\documentclass[12pt,a4paper]{article}
\pdfoutput=1
\usepackage{color}
\usepackage{amssymb,amsmath,bm,bbm}
\usepackage{epsf}
\usepackage{epsfig}
\usepackage{afterpage}
\usepackage{longtable}
\usepackage[dvipsnames]{xcolor}
\usepackage[linktoc=page,bookmarks=false,colorlinks=false,
linkbordercolor=RoyalBlue,citebordercolor=ForestGreen,urlbordercolor=CornflowerBlue]{hyperref}
\usepackage{latexsym,mathrsfs,dsfont}
\usepackage[normalem]{ulem} 
\usepackage[compress]{cite}
\usepackage{graphicx}
\usepackage{url}
\usepackage{paralist}
\usepackage{bbold}

\newcommand{\svs}{\vbox{\vskip 5mm}}

\newcommand{\ord}{\mathcal{O}}

\newcommand{\IM}{{\rm Im}}
\newcommand{\RE}{{\rm Re}}

\newcommand{\tev}{\, {\rm TeV}}
\newcommand{\gev}{\, {\rm GeV}}
\newcommand{\mev}{\, {\rm MeV}}

\newcommand{\vcb}{|V_{cb}|}

\newcommand{\vub}{|V_{ub}|}

\newcommand{\vus}{|V_{us}|}

\newcommand{\bsi}{B_6^{(1/2)}}
\newcommand{\bei}{B_8^{(3/2)}}

\def\epe{\varepsilon'/\varepsilon}
\newcommand{\beq}{\begin{equation}}
\newcommand{\eeq}{\end{equation}}
\newcommand{\be}{\begin{equation}}
\newcommand{\ee}{\end{equation}}
\newcommand{\bi}{\begin{itemize}}
\newcommand{\ei}{\end{itemize}}
\newcommand{\ba}{\begin{array}}
\newcommand{\ea}{\end{array}}
\newcommand{\beqa}{\begin{eqnarray}}
\newcommand{\eeqa}{\end{eqnarray}}
\newcommand{\bea}{\begin{eqnarray}}
\newcommand{\eea}{\end{eqnarray}}
\newcommand{\beqn}{\begin{eqnarray}}
\newcommand{\eeqn}{\end{eqnarray}}

\newcommand{\eps}{\epsilon}

\definecolor{red}{cmyk}{0,1,1,0.4}

\def\kpn{K^+\rightarrow\pi^+\nu\bar\nu}
\def\klpn{K_{L}\rightarrow\pi^0\nu\bar\nu}

\newcommand{\kepe}{\kappa_{\varepsilon^\prime}}
\newcommand{\keps}{\kappa_{\varepsilon}}

\usepackage{fancyhdr}
\advance \headheight by 3.0truept       

\begin{document}


\vspace{-14mm}
\begin{flushright}
        {FLAVOUR(267104)-ERC-115}
\end{flushright}

\vspace{8mm}

\begin{center}
{\Large\bf
\boldmath{New Physics Patterns  in $\epe$ and $\varepsilon_K$ with \\
Implications for Rare Kaon Decays and $\Delta M_K$}}
\\[8mm]
{\large\bf Andrzej~J.~Buras* \\[0.3cm]}
{\small 
      TUM Institute for Advanced Study, Lichtenbergstr.~2a, D-85748 Garching, Germany\\
Physik Department, TU M\"unchen, James-Franck-Stra{\ss}e, D-85748 Garching, Germany\\ * E-mail: aburas@ph.tum.de }
\end{center}

\vspace{4mm}

\begin{abstract}
\noindent
The  Standard Model (SM) prediction for the ratio $\epe$  appears to 
be significantly below the experimental data. Also $\varepsilon_K$ in the SM tends to be below the data. Any new physics (NP) removing 
these anomalies will first of all have impact on flavour observables in the $K$ 
meson system, in particular on rare decays $\kpn$, $\klpn$,
$K_L\to \mu^+\mu^-$ and $K_L\to \pi^0\ell^+\ell^-$ and $\Delta M_K$. Restricting the operators 
contributing to $\epe$ to the SM ones and to the corresponding primed operators, NP contributions to 
$\epe$ are quite generally dominated either by  QCD penguin (QCDP) operators 
$Q_6(Q_6^\prime)$ or  electroweak penguin (EWP) operators $Q_8(Q_8^\prime)$ 
with rather different implications for other flavour observables.
 Our presentation includes general models with tree-level $Z$ and $Z^\prime$ flavour violating exchanges for which we summarize known results and add several new ones. We also briefly discuss few  specific models. The correlations of $\epe$ with 
other flavour observables listed above allow to  differentiate between  models in which $\epe$ can be enhanced.  Various DNA-tables are helpful in this respect. We find that simultaneous enhancements of $\epe$, $\varepsilon_K$, $\mathcal{B}(\klpn)$ and $\mathcal{B}(\kpn)$ in $Z$ scenarios are only possible in the presence of both left-handed  and right-handed flavour-violating couplings. In $Z^\prime$ scenarios  this is not required but the size of NP effects 
 and the correlation between  $\mathcal{B}(\klpn)$ and $\mathcal{B}(\kpn)$ 
depends strongly on whether QCDP or EWP dominate NP 
contributions to $\epe$. In the QCDP case possible enhancements of both
branching ratios are much larger than for EWP scenario and take place only on the branch parallel to the Grossman-Nir bound, which is in the case of EWP dominance  only possible in the absence of NP in $\varepsilon_K$. We point out that 
 QCDP and EWP scenarios of NP in $\epe$ 
can also be uniquely distinguished by the size and the sign of NP contribution to $\Delta M_K$, elevating the importance of the precise calculation of $\Delta M_K$ in the SM.  We emphasize the importance of the theoretical improvements
not only on $\epe$,  $\varepsilon_K$  and $\Delta M_K$   but also on   $K_L\to\mu^+\mu^-$,  $K_L\to \pi^0\ell^+\ell^-$, and the $K\to\pi\pi$ isospin amplitudes ${\rm Re} A_0$ and  ${\rm Re} A_2$ which would in the future enrich our analysis.
\end{abstract}

\setcounter{page}{0}
\thispagestyle{empty}
\newpage

\tableofcontents

\newpage

\section{Introduction}

The ratio
$\epe$  measures the size of the direct CP violation in $K_L\to\pi\pi$ decays
relative to the indirect CP violation described by $\varepsilon_K$ and is rather 
sensitive to new physics (NP). In the Standard Model (SM)
$\varepsilon^\prime$ is governed by QCD penguins (QCDP) but receives also an important destructively interfering  contribution from electroweak penguins (EWP).  
Beyond the SM the structure of NP contributions to $\epe$ is in general different as often only the EWP operators contribute in 
a significant manner. But, one can also construct scenarios in which 
NP contributions from QCDP dominate. This is for instance the case 
of certain $Z^\prime$ models which we will present in detail below. 
Moreover, there exist models in which NP contributions to $\epe$ 
can be  dominated by new operators which can be neglected within the SM.
A prominent example is the chromomagnetic penguin operator in supersymmetric models.

The present status of $\epe$ in the SM has been reviewed recently in \cite{Buras:2015yba,Buras:2015hna}, where references to rich literature can be found.
 After the new results for the hadronic matrix elements of QCDP and 
EWP $(V-A)\otimes (V+A)$ operators from  RBC-UKQCD lattice
collaboration \cite{Blum:2015ywa,Bai:2015nea,Garron:2015vhs} and the extraction of the
corresponding matrix elements of penguin  $(V-A)\otimes (V-A)$ operators  from the CP-conserving 
$K\to\pi\pi$ amplitudes one finds \cite{Buras:2015yba} 
\be\label{LBGJJ}
   (\epe)_\text{SM} = (1.9 \pm 4.5) \times 10^{-4} \,.
\ee
This result differs with $2.9\,\sigma$ significance from 
the experimental world average
from NA48 \cite{Batley:2002gn} and KTeV
\cite{AlaviHarati:2002ye,Abouzaid:2010ny} collaborations, 
\be\label{EXP}
(\epe)_\text{exp}=(16.6\pm 2.3)\times 10^{-4} \,,
\ee
suggesting evidence for NP in $K$ decays. 

 As demonstrated in  \cite{Buras:2015xba}  these new results from lattice-QCD are supported by the large $N$ approach,  which moreover allows to derive upper 
bounds on the matrix elements of the dominant penguin operators. 
This implies  \cite{Buras:2015yba,Buras:2015xba} 
\be\label{BoundBGJJ}
(\epe)_\text{SM}\le (8.6\pm 3.2) \times 10^{-4} \,, \qquad ({\rm large~}N),
\ee
still $2\,\sigma$  below the experimental data. Additional arguments for this bound will be given in Section~\ref{sec:1}.

While, the improvement on the estimate of isospin corrections, final state 
interactions (FSI) \cite{Pallante:1999qf,Pallante:2000hk,Buras:2000kx,Buchler:2001np,Buchler:2001nm,Pallante:2001he} and the inclusion of NNLO QCD corrections could in principle increase $\epe$ 
with respect to the one in (\ref{LBGJJ}), it is rather unlikely that values of
$\epe$ violating  the upper bound in (\ref{BoundBGJJ}) will be found within the SM. After all, until now, lattice QCD confirmed most of earlier results on $K$ meson flavour physics 
obtained in the large $N$ approach (see \cite{Buras:2014maa,Buras:2015hna}).

 In particular a recent analysis of FSI in this approach in 
\cite{Buras:2016fys} gives additional support for these expectations. As stated in this paper, it turns out 
 that  beyond the strict large $N$ limit,  FSI are likely to be important for the $\Delta I=1/2$  rule, in agreement with \cite{Pallante:1999qf,Pallante:2000hk,Buras:2000kx,Buchler:2001np,Buchler:2001nm,Pallante:2001he}, but much less relevant  for $\epe$. 
It appears then that the SM has significant difficulties in explaining the experimental value of $\epe$. This implies that NP models 
in which this ratio can be enhanced with respect to its SM value 
are presently favoured. 

Now, the renormalization group effects play a very important role in the 
analysis of $\epe$. They have been known already
for more than twenty years at the NLO level
\cite{Buras:1991jm,Buras:1992tc,Buras:1992zv,Ciuchini:1992tj,Buras:1993dy,Ciuchini:1993vr} and present technology could extend them to the NNLO level if necessary. First steps in this direction have been taken in \cite{Buras:1999st,Gorbahn:2004my,Brod:2010mj}. The situation with hadronic matrix elements is another story and even if significant progress on their evaluation has been made  over the last 25 years, the present status is clearly not satisfactory. Still, 
both the large $N$ approach and lattice QCD show that  hadronic matrix 
elements of QCD and EWP  $(V-A)\otimes (V+A)$ operators, 
$Q_6$ and $Q_8$ respectively, are by far the largest among those of contributing operators  with the relevant 
matrix element $\langle Q_8\rangle_2$ being larger than $\langle Q_6\rangle_0$ 
in magnitude by roughly a factor of two.

With the Wilson coefficient $y_6$ of $Q_6$ being roughly by a factor of $90$ larger 
than $y_8$ of $Q_8$ (see \cite{Buras:2015yba}) one would  expect  the $Q_6$ operator to be by far the dominant one in
$\epe$. That this does not happen is due to the factor
\be\label{IDI12}
\frac{{\RE} A_2}{{\RE} A_0}=\frac{1}{22.4}
\ee
which in the basic formula for $\epe$ in (\ref{eprime}) suppresses the $Q_6$ contribution 
 relative to the $Q_8$ one. As a result strong cancellation between these 
two dominant contributions to $\epe$ in 
the SM takes place so that contributions of other less important  $(V-A)\otimes (V-A)$ operators matter. A detailed anatomy of such contributions has been presented in \cite{Buras:2015yba}.

Beyond the SM quite often the Wilson coefficients of $Q_6$ and $Q_8$ and 
of the primed operators $Q_6^\prime$ and $Q_8^\prime$\footnote{These operators are obtained from $Q_6$ and $Q_8$ through the interchange of $V-A$ and $V+A$.}
in the NP contribution to $\epe$ are of the same order and then operators $Q_8$ and/or  $Q_8^\prime$  
 win easily this competition because of the suppression of the $Q_6$ and 
$Q_6^\prime$  contributions 
by the factor in (\ref{IDI12}) and the fact that their hadronic matrix elements 
are smaller than the ones of $Q_8$ and $Q_8^\prime$.
Therefore retaining only the latter contributions in the NP part  is a reasonable approximation if one 
wants to make a rough estimate of $\epe$ with the accuracy of $10\%$. Only in the presence of a flavour symmetry 
which assures the flavour universality of diagonal quark
couplings, $Q_6$ and/or  $Q_6^\prime$  win this competition because the contribution of $Q_8 (Q_8^\prime)$ is then either negligible or absent. In such cases 
$Q_6$ and/or  $Q_6^\prime$ are by far the dominant contributions to $\epe$.

This simplification in the renormalization group analysis, pointed out in
\cite{Buras:2014sba}, and present in many  extensions of the SM, 
allows for a quick rough estimate of the size of  NP contributions to $\epe$ 
in a given model. Moreover, the absence of cancellations between QCD and electroweak 
penguin contributions in the NP part makes it subject to much smaller theoretical uncertainties than it is the case within the SM.
Then leading order renormalization group analysis 
is sufficient, in particular, for finding the sign of 
 NP contribution as a function of model parameters, generally couplings 
of NP to quarks. This sign is in 
most cases not unique because of the presence of free parameters represented 
by new couplings in a given model. But requiring that NP enhances $\epe$ relative to its 
SM value, determines the signs of these couplings  with implications 
for other observables in the $K$ meson system. As $\epe$ is only sensitive 
to imaginary couplings, we will simultaneously assume that there is a modest 
anomaly in $\varepsilon_K$, which together with $\epe$ will allow us to determine both imaginary and real flavour violating couplings of $Z$ and $Z^\prime$ 
implied by these anomalies. This in turn will give us predictions for NP 
contributions to 
$\klpn$, $\kpn$, $K_L\to\mu^+\mu^-$ and $\Delta M_K$  implied by these 
anomalies. In certain models the enhancement of 
$\epe$ implies uniquely enhancement or suppression of other observables 
or even eliminates significant NP contributions from them. In this manner even 
patterns of deviations from SM predictions can identify the favoured NP 
models.

This strategy of identifying NP through quark flavour violating processes 
has been proposed in \cite{Buras:2013ooa} and graphically represented in 
terms of DNA-charts. But the case of $\epe$ has not been discussed there in 
this manner and we would like to do it here in the form of DNA-tables, see Tables~\ref{tab:ZPQ6} and \ref{tab:ZPQ8}, concentrating fully on the $K$ meson system. But as we will see this system by itself can already give us a valuable insight into physics beyond the SM. 
The implications for other meson systems require more assumptions on the 
flavour structure of NP and will be considered elsewhere. A recent study of 
the impact of $K$ physics observables on the determination of the Unitarity 
triangle can be found in \cite{Lehner:2015jga}.

Our paper is organized as follows. In Section~\ref{sec:1} we recall the basic 
formula for $\epe$ that is valid in all extensions of the SM and recall the relevant hadronic matrix elements of the operators $Q_6(Q_6^\prime)$ and $Q_8(Q_8^\prime)$. In Section~\ref{sec:strategy} we present our strategy for addressing 
the sizable $\epe$ anomaly and a modest $\varepsilon_K$ anomaly with the hope that it will make our paper more transparent.
In  Section~\ref{sec:2} we discuss models in which NP contributions to 
$\epe$ come dominantly from tree-level $Z$ exchanges and identify  a number of
scenarios for flavour-violating $Z$ couplings  that could provide  the required enhancement of $\epe$ with concrete 
implications for other flavour observables listed above. 
In Section~\ref{sec:3} we generalize this discussion to models with tree-level 
$Z^\prime$ exchanges and discuss briefly the effects of $Z-Z^\prime$ mixing. 
We also consider there the case of $G^\prime$, a colour octet of heavy gauge bosons.
In both sections
 we demonstrate how
these different models can be differentiated with the help of other observables.
Of particular interest is the case of $M_{Z^\prime}$ outside the reach of the LHC if the flavour structure of a given model is such that the suppression by 
$Z^\prime$ propagator is compensated by the increase of flavour-violating couplings. We also stress that for  $M_{Z^\prime}\ge 10\tev$ renormalization group effects imply 
additional significant enhancements of both QCDP and EWP contributions to $\epe$.
In Section~\ref{sec:4} we briefly discuss scenarios in which contributions of 
both $Z$ and $Z^\prime$ are present even in the absence of significant  $Z-Z^\prime$ mixing. 
This is the case of models in which in addition to $Z^\prime$ also new heavy fermions, like vector-like quarks, are present implying through the mixing with 
SM quarks flavour-violating $Z$ couplings.
While our discussion is rather general, in Section~\ref{sec:5} 
we give examples of specific $Z$ 
and $Z^\prime$  models, in which one can reach clear cut conclusions and 
briefly summarize more complicated models. In Section~\ref{DeltaI} we 
discuss possible implications of NP in the $K\to\pi\pi$ isospin amplitudes ${\rm Re} A_0$ and  ${\rm Re} A_2$. In Section~\ref{Vision} we contemplate 
on the implications of the possible discovery of NP in $\kpn$ by NA62 experiment in 2018 in the presence of $\epe$ anomaly, dependently on whether NP in $\varepsilon_K$ is present or absent. Finally in Section~\ref{sec:6} we summarize most important findings and give  a brief outlook for the coming years and list most important open questions.
Several appendices contain a collection of useful formulae.

Our paper differs from other papers on flavour physics in $K$ meson system in 
that we do not obtain the results for $\epe$ and $\varepsilon_K$ as output 
of a complicated analysis but treat them as input parametrizing the size
of NP contributions to them by two parameters of $\ord(1)$: $\kepe$ and 
$\keps$. See Section~\ref{sec:strategy} for the explicit formulation of this
strategy.

Our paper differs also from many papers on rare processes present in the literature in that it does not contain a single plot coming from a sophisticated numerical analysis. The uncertainty in the QCDP contribution to $\epe$ in the 
SM leaves still a very large room for NP in $\epe$ and a detailed numerical analysis would only wash out the pattern of NP  required to enhance $\epe$. The 
absence of sophisticated plots is compensated by numerous simple analytic formulae and DNA-tables that should allow model builders to estimate quickly the pattern of NP in the $K$ meson system in her or his favourite model. Our goal is to 
present the material in such a manner that potential readers  
can follow all steps in detail.

Finally, our paper differs also from the recent literature on flavour physics which is dominated by the anomalies in most recent data for $B\to K(K^*)\ell^+\ell^-$ and $B\to D(D^*)\tau\nu_\tau$ reported by LHCb, BaBar and Belle. The case of $\epe$ 
is different as the data is roughly fifteen years old and the progress is presently  done by theorists, not experimentalists. But as the recent papers 
\cite{Buras:2015yba,Blum:2015ywa,Bai:2015nea,Buras:2015xba} show, $\epe$ after 
rather silent ten years is striking back, in particular in correlation with $\kpn$ and $\klpn$ \cite{Buras:2015yca,Blanke:2015wba} on which the data \cite{Rinella:2014wfa,Romano:2014xda,Shiomi:2014sfa} will improve significantly in the coming years. Moreover, as we will see in the context of our presentation,  theoretical improvements not only on $\epe$ but also on
 $\varepsilon_K$, $\Delta M_K$,  $K_L\to\mu^+\mu^-$,  $K_L\to \pi^0\ell^+\ell^-$, and the $K\to\pi\pi$ isospin amplitudes ${\rm Re} A_0$ and  ${\rm Re} A_2$ 
 will give us new insights in NP at short distance scales.

\begin{boldmath}
\section{Basic formula for $\epe$}\label{sec:1}
\end{boldmath}
The basic formula for $\epe$ reads \cite{Buras:2015yba}
\be\label{eprime}
\frac{\varepsilon'}{\varepsilon} = -\,\frac{\omega_+}{\sqrt{2}\,|\varepsilon_K|}
\left[\, \frac{{\IM} A_0}{{\RE}A_0}\,(1-\hat\Omega_{\rm eff}) -
\frac{1}{a}\frac{{\IM} A_2}{{\RE}A_2} \,\right],
\ee
with $(\omega_+,a)$ and $\hat\Omega_{\rm eff}$ given 
as follows
\be\label{OM+}
\omega_+ =a\,\frac{\RE A_2}{\RE A_0} =(4.53\pm0.02)\times 10^{-2}, \qquad a=1.017, \qquad
\hat\Omega_{\rm eff} = (14.8\pm 8.0)\times 10^{-2}\,.
\ee
Here $a$  and $\hat\Omega_{\rm eff}$ summarize  isospin breaking corrections and include  strong isospin
violation $(m_u\neq m_d)$, the correction to the isospin limit coming from
$\Delta I=5/2$ transitions and  electromagnetic corrections \cite{Cirigliano:2003nn,Cirigliano:2003gt,Bijnens:2004ai}. $\hat\Omega_{\rm eff}$ differs from $\Omega_{\rm eff}$ in 
\cite{Cirigliano:2003nn,Cirigliano:2003gt} which includes
contributions of EWP. Here they are present in  
 ${\IM} A_0$ and of course in ${\IM} A_2.$  Strictly speaking $\epe$ is a 
complex quantity and the expression in (\ref{eprime}) applies to its real 
part but its phase is so small that we can drop the symbol ``Re'' in all
expressions below in order to simplify the notation.

The 
amplitudes ${\RE}A_{0,2}$ are then extracted from the branching ratios on 
$K\to\pi\pi$ decays in the isospin limit. Their values are given by 
\begin{equation}
\RE A_0 = 33.22(1)\times 10^{-8}\gev \, ,
\qquad \qquad
\RE A_2 = 1.479(3)\times 10^{-8}\gev \, .
\label{eq:6.3}
\end{equation}

For the analysis of NP contributions in our paper the only relevant operators 
are the following  QCDP and 
EWP $(V-A)\otimes (V+A)$ operators

{\bf QCD--Penguins:}
\begin{equation}\label{O3}
Q_5 = (\bar s d)_{V-A}\!\!\sum_{q=u,d,s,c,b,t}(\bar qq)_{V+A}~~~~~
Q_6 = (\bar s_{\alpha} d_{\beta})_{V-A}\!\!\sum_{q=u,d,s,c,b,t}
      (\bar q_{\beta} q_{\alpha})_{V+A} 
\end{equation}

{\bf Electroweak Penguins:}
\begin{equation}\label{O4} 
Q_7 = \frac{3}{2}\,(\bar s d)_{V-A}\!\!\sum_{q=u,d,s,c,b,t} e_q\,(\bar qq)_{V+A} 
~~~~~Q_8 = \frac{3}{2}\,(\bar s_{\alpha} d_{\beta})_{V-A}\!\!\sum_{q=u,d,s,c,b,t}
      e_q\,(\bar q_{\beta} q_{\alpha})_{V+A} \,.
\end{equation}

The primed operators $Q_i^\prime$ are obtained from $Q_i$ through the interchange 
of $V-A$ and $V+A$. Summation over colour indices $\alpha$ and $\beta$ is understood. In the case of $Z$ models top quark contribution should be omitted.

Eventually, if we are only interested in signs of  NP contributions to $\epe$ and approximate estimates of their magnitudes, only $Q_6(Q_6^\prime)$ will be relevant for ${\IM} A_0$ and only contribution of $Q_8(Q_8^\prime)$ for 
 ${\IM} A_2$. Thus we only need two hadronic matrix elements:
\begin{eqnarray}\label{eq:Q60}
\langle Q_6(m_c) \rangle_0 &=& -\,4 \sqrt{\frac{3}{2}}
\left[ \frac{m_{\rm K}^2}{m_s(m_c) + m_d(m_c)}\right]^2 (F_K-F_\pi)
\,B_6^{(1/2)} \,=-0.58 \,\bsi \gev^3\,  \\
\label{eq:Q82}
\langle Q_8(m_c) \rangle_2 &=& \sqrt{2}  \sqrt{\frac{3}{2}}
\left[ \frac{m_{\rm K}^2}{m_s(m_c) + m_d(m_c)}\right]^2 F_\pi \,B_8^{(3/2)}\,= 1.06\,\bei \gev^3.
\end{eqnarray}
This approximate treatment would not be justified within the SM because 
of strong cancellations between QCDP and EWP contributions. But as we explained above such cancellations are absent in many extensions of the SM and 
for sure in the models considered by us.

The choice of the scale $\mu=m_c$ is convenient as it is used in analytic 
formulae for $\epe$ in \cite{Buras:2015yba}. But otherwise the precise value of $\mu$ is not relevant as the dominant 
$\mu$ dependence of the Wilson coefficients and of the
matrix elements of $Q_6$ and $Q_8$ operators has a simple structure being 
dominantly governed by the $\mu$ dependence of involved quark masses. As a 
result of this the $\mu$ dependence of the 
parameters $\bsi$ and $\bei$ is negligible for $\mu\ge 1\gev$
 \cite{Buras:1993dy}. The matrix elements of primed operators differ only by sign from the ones given above.
The numerical values in (\ref{eq:Q60}) and (\ref{eq:Q82}) are
given for the central values of
\cite{Agashe:2014kda,Aoki:2013ldr}
\be\label{FpFK}
m_K=497.614\mev, \qquad  F_\pi=130.41(20)\mev,\qquad \frac{F_K}{F_\pi}=1.194(5)\, ,
\ee
\be\label{mc}
m_s(m_c)=109.1(2.8)\mev, \qquad  m_d(m_c)=5.44(19)\mev\,.
\ee
The values of other parameters are collected in Table~\ref{tab:input}.

\begin{table}[!tb]
\renewcommand{\arraystretch}{1.2}
\centering%
\begin{tabular}{|cl|cl|}
\hline\hline
$G_F$ & $1.16637(1)\times 10^{-5}\gev^{-2}$ & $M_W$ & $80.385 \gev$\\
  $\sin^2\theta_W$ & $0.23116(13)$ & $M_Z$ &$ 91.1876 \gev$ \\
$|\eps_K|$ & $2.228(11)\times 10^{-3}$\hfill\cite{Beringer:1900zz} 
&
$m_K $ & $ 0.4976\gev$
\\
$\Delta M_K$ & $ 3.483 \times 10^{-15} \,\gev$\hfill\cite{Beringer:1900zz} 
&
$\hat{B}_K$ & $0.750(15)$ \hfill \cite{Aoki:2013ldr,Buras:2014maa}
\\
$\lambda=|V_{us}|$ & $0.2252(9)$\hfill\cite{Amhis:2012bh} 
&
$\bsi$ & $0.70 $\hfill
\\
$\alpha_s(M_Z)$ & $0.1185(6) $\hfill\cite{Beringer:1900zz}
&
$\bei$ & $ 0.76 $\hfill
\\
$\tilde\kappa_\eps$ & $0.94\pm0.02$ \hfill\cite{Buras:2008nn,Buras:2010pza} 
&
$\eta_{2}$ & $0.5765(65)$\hfill\cite{Buras:1990fn}
\\
$\tilde r(M_Z)$& $1.068$ & ${\rm Re}\lambda_t$  & $-3.0\cdot 10^{-4}$ \\
$\tilde r(3\tev)$& $0.95$ & ${\rm Im}\lambda_t$  & $ 1.4\cdot 10^{-4}$\\
\hline\hline
\end{tabular} 
\caption{\it Values of theoretical and experimental quantities used as input parameters. See also (\ref{FpFK}) and (\ref{mc}).}\label{tab:input}
\end{table}

Recently significant progress on the values of $\bsi$ and $\bei$ has been 
made by the RBC-UKQCD collaboration, who presented new results on 
the relevant hadronic matrix elements of the operators $Q_6$ \cite{Bai:2015nea}
and $Q_8$ \cite{Blum:2015ywa}. These results imply the following values 
for $\bsi$ and $\bei$ \cite{Buras:2015yba,Buras:2015qea}
\be\label{Lbsi}
\bsi=0.57\pm 0.19\,, \qquad \bei= 0.76\pm 0.05\,, \qquad (\mbox{RBC-UKQCD})
\ee
to be compared with their values in the strict large $N$ limit of QCD
\cite{Buras:1985yx,Bardeen:1986vp,Buras:1987wc}
\be\label{LN}
\bsi=\bei=1, \qquad {\rm (large~N~Limit)}\,.
\ee

But, in this analytic, dual approach to QCD, one can demonstrate explicitly the
suppression of both $\bsi$ and $\bei$ below their large-$N$ limit 
and  derive conservative upper bounds on both $\bsi$ and $\bei$ which read
\cite{Buras:2015xba}
\be\label{NBOUND}
\bsi < \bei < 1 \, \qquad (\mbox{\rm large-}N).
\ee
While this approach gives $B_8^{(3/2)}(m_c)=0.80\pm 0.10$, the result for $\bsi$ is less
precise but there is a strong indication that $\bsi < \bei$, with typical values $\bsi\approx 0.5-0.6$ at scales $\ord(1\gev)$, in agreement with
(\ref{Lbsi})\footnote{On the other hand a number of other large $N$ approaches \cite{Bijnens:2000im,Hambye:2003cy,Bijnens:2001ps}
violates strongly the bounds in (\ref{NBOUND}) with $\bsi$ in the ballpark of $3$ and $\bei > 1$ in striking disagreement with lattice results. Similar comment applies to $\bei$ in the dispersive approach \cite{Cirigliano:2001qw,Cirigliano:2002jy}.}

We should emphasize that this suppression of $\bsi$ and $\bei$ results from the meson
evolution of the density-density operators $Q_6$ and $Q_8$ from $\mu\approx 0$  (strict large $N$ limit) to scales $\ord(1\gev)$, where the hadronic matrix elements are multiplied 
by the Wilson coefficients. The scale dependence of both parameters is logarithmic but the one of $\bsi$ is stronger than of $\bei$ implying at scales  $\ord(1\gev)$ the inequalities in (\ref{NBOUND}). This pattern of scale dependence of 
both parameters is consistent with the one for $\mu> 1\gev$  \cite{Buras:1993dy}
 that can be found by usual renormalization group methods. But the scale dependence  for $\mu> 1\gev$  is weaker than for lower scales, as expected.
 For further details, see \cite{Buras:2015xba}.

It is probably useful to recall at this stage that the recent finding of $\epe$ in the SM being
 below its experimental value has been signalled already by early analyses, among them in  \cite{Ciuchini:1995cd,Bosch:1999wr,Buras:2003zz},
 which used $\bsi=\bei=1$. See \cite{Bertolini:1998vd} for an early review.
The new result in (\ref{NBOUND}) tells us that this is an upper bound on 
these two parameters and the recent lattice and large $N$ calculations 
show that these parameters are significantly below this bound making 
$\epe$ in the SM even smaller than previously expected. 

On the other hand, it has been advocated by the chiral perturbation theory 
practitioners \cite{Pallante:1999qf,Pallante:2000hk,Buchler:2001np,Buchler:2001nm,Pallante:2001he} that final state interactions (FSI), not included in the large 
$N$ approach in the leading order, effectively increase the value of $\bsi$ by roughly a factor of $1.5$ and suppress $\bei$ by roughly $10\%$ bringing SM predition for $\epe$ close to the experimental result in (\ref{EXP}). However, the 
recent analysis in \cite{Buras:2016fys}demonstrates that this claim in the case of $\bsi$  cannot be justified. 
 In fact, as  pointed out in that paper, within a pure effective (meson) field approach like {chiral perturbation theory}   the dominant current-current operators governing the $\Delta I=1/2$ rule and the dominant density-density (four-quark) QCD penguin operator $Q_6$ governing $\epe$ cannot be disentangled from each other. Therefore, without an UV completion, that is QCD at short distance scales, the claim that the isospin amplitude ${\rm Re} A_0$ and $\bsi$ are enhanced through FSI in the same manner, as done in \cite{Pallante:1999qf,Pallante:2000hk,Buchler:2001np,Buchler:2001nm,Pallante:2001he} , cannot be justified.
But in the context of a dual QCD approach, which includes both long distance 
dynamics and the QCD at short distance scales,
 such a distinction is possible. One  finds then that  beyond the strict large $N$ limit FSI are likely to be important for the $\Delta I=1/2$ rule but much less relevant for $\epe$ \cite{Buras:2016fys}.

It should also be emphasized that the estimates in \cite{Pallante:1999qf,Pallante:2000hk,Buchler:2001np,Buchler:2001nm,Pallante:2001he} 
omitted the non-factorizable 
contributions to $\bsi$ and $\bei$, represented by meson evolution mentioned above and calculated in  \cite{Buras:2015xba}.
As stressed above, the inclusion of them in the hadronic matrix elements is mandatory in order for the calculations of matrix elements and of Wilson coefficients to be compatible with each other. 

These findings diminish significantly 
hopes that improved 
treatment of FSI within lattice QCD approach and dual QCD approach would 
 bring the SM prediction for $\epe$ to agree with the experimental data,
 opening thereby an arena for important NP contributions to this ratio
and giving strong motivation for the analysis presented in our paper.

Unfortunately, due to cancellations between various contributions, the error on  $\epe$ in the SM remains to be substantial.  From present perspective 
we do not expect that this error can be reduced significantly by 
using large $N$ 
approach or other analytical approaches. Therefore, the efforts to find out the room left for NP contributions in $\epe$ will be led in the coming years by lattice 
QCD. But it would be important to have at least second lattice group, beyond 
RBC-UKQCD, which would take part in these efforts. As this may take still several  years, it is useful to develop some strategies to be able to face NP in $\epe$, if the present results on $\epe$ in the SM from lattice QCD and large $N$ 
approach will be confirmed by more precise lattice calculations. One such strategy is proposed below.

This information is sufficient for our analysis which as the main goal has the 
identification of NP patterns in flavour observables in a number of models 
implied by the desire to  enhance $\epe$ over its SM value in a significant 
manner. In particular those models are of interest 
which can provide a positive shift in $\epe$ by at least $5\times 10^{-4}$. 

\section{Strategy}\label{sec:strategy}
\subsection{Present}
In our paper the central role will be played by $\epe$ and 
$\varepsilon_K$ for which in the presence of NP contributions we have
\be\label{GENERAL}
\frac{\varepsilon'}{\varepsilon}=\left(\frac{\varepsilon'}{\varepsilon}\right)^{\rm SM}+\left(\frac{\varepsilon'}{\varepsilon}\right)^{\rm NP}\,, \qquad \varepsilon_K\equiv 
e^{i\varphi_\eps}\, 
\left[\varepsilon_K^{\rm SM}+\varepsilon^{\rm NP}_K\right] \,.
\ee
In view of uncertainties present in the SM estimates of $\epe$ and to a lesser 
extent in $\varepsilon_K$ we will fully concentrate on NP contributions. Therefore in order to identify the pattern of NP contributions to flavour observables
implied by the $\epe$ anomaly in a transparent manner, we will proceed 
in a given model as follows:

{\bf Step 1:} We assume that NP provides a positive shift in $\epe$: 
\be\label{deltaeps}
\left(\frac{\varepsilon'}{\varepsilon}\right)^{\rm NP}= \kepe\cdot 10^{-3}, \qquad   0.5\le \kepe \le 1.5,
\ee
with the range for $\kepe$ indicating the required size of this contribution. 
But in the formulae below, $\kepe$  will be a free parameter. This step 
will determine the imaginary parts of flavour-violating $Z$ and $Z^\prime$ couplings to quarks as functions of $\kepe$.

{\bf Step 2:} In order to determine the relevant real parts of the couplings 
involved, in the presence of the imaginary part determined from $\epe$, we will assume that in addition to the $\epe$ 
anomaly, NP can also affect the parameter  $\varepsilon_K$.  We will describe 
this effect by the parameter $\keps$  so that now in addition 
to (\ref{deltaeps}) we will study the implications of the shift in $\varepsilon_K$ due to NP
\be \label{DES}
(\varepsilon_K)^{\rm NP}= \keps\cdot 10^{-3},\qquad 0.1\le \keps \le 0.4 \,.
\ee
The positive sign of $\keps$  is motivated by the fact 
that if $\varepsilon_K$ is predicted in the SM 
using CKM parameters extracted from $B$ system observables, its value is found 
typically below the data as first emphasized in \cite{Lunghi:2008aa,Buras:2008nn}. See also \cite{Bona:2006ah,Charles:2015gya}. But it should be stressed that
this depends on whether inclusive or exclusive determinations of $\vub$ and 
$\vcb$ are used and with the inclusive ones SM value of $\varepsilon_K$ agrees well with the data. But then as emphasized in \cite{Blanke:2016bhf} $\Delta M_s$ and $\Delta M_d$ are significantly above the data. Other related discussions 
can be found  in \cite{Buras:2013raa,Buras:2014sba,Bailey:2015frw,Bazavov:2016nty}.

While this possible `` anomaly'' is certainly not as pronounced as the $\epe$ one, it 
is instructive to assume that it is present at the level indicated in 
(\ref{DES}), that is at most $20\%$. 

{\bf Step 3:} In view of the uncertainty in $\kepe$ we set several 
 parameters  to their central values. In particular 
for the SM contributions to rare decays we set the CKM factors to 
\be\label{CKMfixed}
{\rm Re}\lambda_t=-3.0\cdot 10^{-4}, \qquad {\rm Im}\lambda_t=1.4\cdot 10^{-4}
\ee
which are in the ballpark of present  estimates obtained by UTfit \cite{Bona:2006ah} and CKMfitter \cite{Charles:2015gya} collaborations. For this 
choice of CKM parameters the central value of the resulting $\varepsilon_K^{\rm SM}$ is $1.96\cdot 10^{-3}$. With the experimental value of $\varepsilon_K$ in 
 Table~\ref{tab:input}  this  implies $\keps=0.26$.  But we will still vary 
$\keps$ while keeping the values in (\ref{CKMfixed}) as NP contributions 
do not depend on them but are sensitive functions of $\keps$.

{\bf Step 4:} Having fixed the flavour violating couplings of $Z$ or $Z^\prime$ 
in this manner, we will express NP contributions to the branching ratios for $\kpn$, $\klpn$ 
and $K_L\to\mu^+\mu⁻$ and to $\Delta M_K$ in terms of $\kepe$ and $\keps$. This
will allow us to study directly the impact of $\epe$ and $\varepsilon_K$ anomalies in $Z$ and $Z^\prime$ scenarios on these four observables. In  Table~\ref{tab:IMRE} we indicate the dependence of a given observable on the {\it real} and/or the {\it imaginary} $Z$ or $Z^\prime$ flavour violating coupling to quarks. 
 In our strategy imaginary parts depend only on  $\kepe$, while the real parts
on both $\kepe$ and $\keps$. The pattern of flavour violation  depends in a given NP scenario on the relative size of real and imaginary parts of couplings 
and we will see this explicitly later on.

\begin{table}[!htb]
\begin{center}
\begin{tabular}{|c|c|c|c|c|c|c|}
\hline
& $\epe$ & $\varepsilon_K$ & $\klpn$& $\kpn$& $K_L\to\mu^+\mu^-$ & $\Delta M_K$ \\
\hline
${\rm Im}\Delta$ & $*$ & $*$ & $*$ & $*$ &   & $*$\\
${\rm Re}\Delta$ &  & $*$ &  & $*$ &  $*$ & $*$\\
\hline
\end{tabular}
\end{center}
\caption{The dependence of various observables on the  imaginary and/or real 
parts of $Z$ and $Z^\prime$ flavour-violating couplings.
\label{tab:IMRE}}
\end{table}

In the context of our presentation we will see that in $Z$ scenarios with 
only left-handed or right-handed flavour violating couplings the most important 
constraint on the real parts of new couplings comes not from $\varepsilon_K$ or 
$\Delta M_K$ 
but from $K_L\to\mu^+\mu^⁻$. On the other hand, in all $Z^\prime$ scenarios and in the case 
of $Z$ scenarios with 
left-right operators contributing to $\varepsilon_K$, these are always  $\varepsilon_K$ and $\Delta M_K$ and not $K_L\to \mu^+\mu^-$ that are most important 
for the determination of the real parts of the new couplings after the $\epe$ 
constraint has been imposed.
\subsection{Future}
The present strategy above assumes that the progress in the evaluation of $\epe$
 in the SM will be faster than experimental information on $\kpn$. If in 2018 
the situation will be reverse, it will be better to choose as variables 
$\keps$ and $R^{\nu\bar\nu}_+$ defined in (\ref{Rnn+}). In the next sections 
we will provide $R^{\nu\bar\nu}_+$ as a function of $\kepe$ for fixed $\keps$ using the present strategy. But knowing  $R^{\nu\bar\nu}_+$  better than $\epe$ in the SM will allow us to read off from our plots 
the favourite range for $\kepe$ in a given NP scenario for given $\keps$ and 
the diagonal couplings of $Z^\prime$. As these plots will 
be given for $\bsi=0.70$ and $\bei=0.76$, the shift in $\epe$ represented by
$\kepe$ will be given for other values of $\bsi$ and $\bei$ simply by
\be\label{dictionary}
\kepe(\bsi)=\kepe\left[\frac{\bsi}{0.70}\right],\qquad 
\kepe(\bei)=\kepe\left[\frac{\bei}{0.76}\right]\,,
\ee
where $\kepe$ without the argument is the one found in the plots. Even if going
 backwards will require resolution of some sign ambiguities, they should be 
easily resolved. Note that knowing $R^{\nu\bar\nu}_+$ will allow to obtain 
$R^{\nu\bar\nu}_0$, defined in (\ref{Rnn0}) directly from our plots, using the 
value of $\kepe$ extracted from  $R^{\nu\bar\nu}_+$ and $\keps$. The formulae 
in (\ref{dictionary}) are only relevant for predicting $\epe$ in this manner.
Clearly, when $R^{\nu\bar\nu}_0$ will also be know the analysis will be rather
constrained.

\boldmath
\section{$Z$ Models}\label{sec:2}
\unboldmath
\subsection{Preliminaries}
The most recent analyses of $\epe$ in these models can be found in 
\cite{Buras:2014sba,Buras:2015yca} and some results presented below 
are based on these papers. In particular, the relevant renormalization 
group analysis in the spirit of the present paper has been performed in \cite{Buras:2014sba}. We summarize 
 and slightly extend it in  Appendix~\ref{app:X}.

It is straightforward to calculate the values  of the Wilson coefficients  
entering NP part of the $K\to\pi\pi$ Hamiltonian in these models. We define 
these coefficients  by
\be\label{ZprimeA}
\mathcal{H}_{\rm eff}(K\to\pi\pi)(Z)=\sum_{i=3}^{10}(C_i(\mu)Q_i+C_i^\prime(\mu) Q_i^\prime),
\ee
where the primed operators $Q_i^\prime$ are obtained from $Q_i$ by interchanging 
$V-A$ and $V+A$. The operators $Q_i$ are the ones entering the SM contribution
 \cite{Buras:1993dy}
\be\label{basic}
 \mathcal{H}_{\rm eff}(K\to\pi\pi)({\rm SM})=\frac{G_F}{\sqrt{2}}V_{ud}V_{us}^*\sum_{i=1}^{10}(z^{\rm SM}_i(\mu)+\tau y^{\rm SM}_i(\mu))Q_i, \qquad
\tau=-\frac{V_{td}V_{ts}^*}{V_{ud}V_{us}^*}\,.
\ee
Explicit expressions for some of them have been given above and the remaining ones   can be found in \cite{Buras:1993dy}. $Q_{1,2}$ are current-current operators, $Q_3-Q_6$ are QCDP operators and $Q_7-Q_{10}$ EWP operators. Note that whereas $z_i$ and $y_i$ are dimensionless, the coefficients in 
(\ref{ZprimeA}) carry dimension as seen explicitly below.

We define the relevant flavour violating $Z$ couplings $\Delta^{sd}_{L,R}(Z)$ by \cite{Buras:2012jb}
\be\label{Zcouplings}
i\mathcal{L}(Z)=i\left[\Delta_L^{sd}(Z)(\bar s\gamma^\mu P_Ld)+\Delta_R^{sd}(Z)(\bar s\gamma^\mu P_R d)\right] Z_\mu,\qquad P_{L,R}=\frac{1}{2}(1\mp \gamma_5)\,.
\ee
Considering then the simple tree-level $Z$ exchange,
the non-vanishing Wilson coefficients at $\mu=M_{Z}$ are then given at the 
LO as follows \cite{Buras:2014sba}
\begin{align}
\begin{split}
C_3(M_{Z})
& = -\left[\frac{g_2}{6 c_W}\right]\frac{\Delta_L^{s d}(Z)}{4 M^2_Z}, \qquad 
C_5^\prime(M_Z)= -\left[\frac{g_2}{6 c_W}\right]\frac{\Delta_R^{s d}(Z)}{4 M^2_Z}
 \,,\end{split}\label{C3Z}\\
\begin{split}
C_7(M_Z)
& = -\left[\frac{4 g_2 s_W^2}{6 c_W}\right]\frac{\Delta_L^{s d}(Z)}{4 M^2_Z}, 
\quad
C_9^\prime(M_Z) =
 -\left[\frac{4 g_2 s_W^2}{6 c_W}\right]\frac{\Delta_R^{s d}(Z)}{4 M^2_Z}\,
\,,\end{split}\label{C7Z}\\
\begin{split}
C_9(M_Z)
& = \left[\frac{4 g_2 c_W^2}{6 c_W}\right]\frac{\Delta_L^{s d}(Z)}{4 M^2_Z}, 
\qquad
C_7^\prime(M_Z) = \left[\frac{4 g_2 c_W^2}{6 c_W}\right]\frac{\Delta_R^{s d}(Z)}{4 M^2_Z}
\,.\end{split}\label{C9Z}
\end{align}
We have used the known flavour conserving couplings of $Z$ to quarks which 
are collected in the same notation in an appendix in \cite{Buras:2013dea}.
  The $SU(2)_L$ gauge coupling constant $g_2(M_Z)=0.652$.
We note that the values of the coefficients in front of $\Delta_{L,R}$ are 
in the case of $C_9$ and $C_7^\prime$ by a factor of $c_W^2/s_W^2\approx 3.33$ larger 
than for the remaining coefficients.  It should also be stressed that 
these formulae are also valid for new $Z$ penguins which provide  one loop contributions to the  couplings $\Delta_{L,R}^{s d}(Z)$.

We also notice that in contrast to the SM the contributions of current-current 
operators $Q_{1,2}$ are absent and they cannot be generated through renormalization group
effects from penguin operators\footnote{If new heavy charged gauge bosons are 
present in a given model new contributions to Wilson coefficients of current-current operators would be 
generated and in turn also the coefficients of penguin operators would be 
modified through renormalization group effects. But these effects are expected to be significantly smaller than the
ones considered here.}.  Moreover, whereas the QCDP operator 
coefficients in the SM are enhanced by more than an order of magnitude over the EWP coefficients due to the factor $\alpha_s/\alpha_{\rm em}$, this 
enhancement is absent here.

In  Appendix~\ref{app:X} we demonstrate that after performing the renormalization 
group evolution from $M_Z$ down to $m_c$ and considering the size of 
hadronic matrix elements  it is sufficient to keep only contributions of  $Q_6$ and $Q_6^\prime$ generated from  $Q_5$ and $Q_5^\prime$ or contributions of $Q_8$ and $Q_8^\prime$, generated from  $Q_7$ and $Q_7^\prime$,  if we want to identify the sign of NP contribution to $\epe$ and do not aim for high precision. But, in $Z$ scenarios, the known structure of flavour  diagonal $Z$ couplings to 
quarks implies that only EWP $Q_8$ and $Q_8^\prime$ matter.

\subsection{Left-handed Scenario (LHS)}\label{LHSZ}
\boldmath
\subsubsection{$\epe$}
\unboldmath
In this scenario only LH flavour-violating couplings are non-vanishing and 
the pair $(Q_7,Q_8)$ has to be considered. 
Even if  at $\mu=M_Z$ the Wilson coefficient of the EWP 
operator $Q_8$ vanishes in the leading order,  its large mixing with $Q_7$ operator, its large anomalous dimension and enhanced hadronic $K\to \pi \pi$ 
matrix elements  make it the dominant EWP operator 
in $\epe$. It  leaves behind the $Q_7$ operator whose Wilson coefficient, as seen in (\ref{C7Z}), does not vanish at $\mu=M_Z$.  We find then \cite{Buras:2014sba}
\be\label{ZfinalL}
\left(\frac{\varepsilon'}{\varepsilon}\right)^L_{Z}=\frac{1}{a}
\frac{\omega_+}{|\varepsilon_K|\sqrt{2}}\frac{{\IM} [A_2^{\rm NP}]^L}{{\RE}A_2}=
0.96\times 10^9\,\left[\frac{{\IM} [A_2^{\rm NP}]^L}{\gev}\right]
\ee
with 
\be\label{LENPZ}
{\IM} [A_2^{\rm NP}]^L= {\IM} C_8(m_c)\langle Q_8(m_c)\rangle_2
\ee
and 
\be\label{C8}
 C_8(m_c)= 0.76\, C_7(M_Z)= -0.76 \,\left[\frac{4 g_2 s_W^2}{6 c_W}\right]\frac{\Delta_L^{s d}(Z)}{4 M^2_Z} = -2.62 \times 10^{-6} 
\left[\frac{\Delta_L^{s d}(Z)}{\gev^2}\right]\,.
\ee
Here $g_2=g_2(M_Z)=0.652$ is the $SU(2)_L$ gauge coupling and the factor $0.76$ is the outcome of the RG evolution summarized in  Appendix~\ref{app:X}.
For our purposes most important is the sign in this result and that the RG factor is $\ord(1)$.  $\langle Q_8(m_c)\rangle_2$ is given in (\ref{eq:Q82}).

Collecting all these results we find
\be\label{ZfinalL1}
\left(\frac{\varepsilon'}{\varepsilon}\right)^L_{Z}=\, -2.64
\times 10^3\,\bei\, {\IM}\Delta_L^{s d}(Z)\,.
\ee

While for our purposes this result is sufficient, in this scenario, in which 
the RG running starts at the electroweak scale,  it is straightforward  to proceed in a different manner by 
including  NP effects  
 through particular shifts in the 
 functions $X$, $Y$ and $Z$ entering the 
analytic formula for $\epe$ in  \cite{Buras:2015yba}. 
These shifts read 
\cite{Buras:2014sba}
\be\label{eprimeshifts}
\Delta X=\Delta Y =\Delta Z= c_W\frac{8\pi^2}{g_2^3}\frac{{\rm Im}\Delta_L^{sd}(Z)}{{\rm Im}\lambda_t} = 1.78\times 10^6\left[\frac{1.4\cdot 10^{-4}}{{\rm Im}\lambda_t}\right] {\rm Im}\Delta_L^{sd}(Z)\,.
\ee 
In doing this we include in fact NLO QCD corrections and all operators whose Wilson coefficients 
are affected by NP and this allows us to confirm that  only the modification in the contribution of the operator $Q_8$ really matters if we do not aim for 
high precision.
 Indeed, inserting these shifts into the analytic formula for $\epe$ 
in \cite{Buras:2015yba} we reproduce the result in (\ref{ZfinalL1}) within roughly   $10\%$ and similar accuracy is expected for other estimates of NP 
contributions to $\epe$ below. Compared to the present uncertainty in the 
SM prediction for $\epe$, this accuracy is certainly sufficient, but can be
increased in the future if necessary.

The final formula for $\epe$ in LHS scenario is then given by
\be\label{epsLHS}
\left(\frac{\varepsilon'}{\varepsilon}\right)_\text{LHS}=
\left(\frac{\varepsilon'}{\varepsilon}\right)_\text{SM}+
\left(\frac{\varepsilon'}{\varepsilon}\right)^L_{Z}
\ee
where the second term stands for the contribution in (\ref{ZfinalL1}) and 
if one aims for higher accuracy it originates in the 
modification related to the 
shifts in (\ref{eprimeshifts}). 

In order to see the  implications of the $\epe$ anomaly in this NP scenario we assume that NP provides a positive shift in $\epe$, as defined in (\ref{deltaeps}),  keeping  $\kepe$  as a  free {\it positive} definite parameter. In accordance 
with our strategy  we set other parameters  to their central values. In particular 
for the SM contributions to rare decays we set the CKM factors to the 
values in (\ref{CKMfixed}).

From (\ref{ZfinalL1}) and (\ref{deltaeps}) we find first
\be\label{IMSD}
 {\rm Im} \Delta_L^{s d}(Z)= -5.0\,\kepe \left[\frac{0.76}{\bei}\right]\cdot 10^{-7}\,.
\ee
The sign is fixed through the requirement of the enhancement of $\epe$. In order to simplify the formulae below we set $\bei=0.76$ but having (\ref{IMSD}) 
it is straightforward to find out what happens for or other values of $\bei$.
Moreover, as seen in (\ref{Lbsi}), $\bei$ is already rather precisely known.

\boldmath
\subsubsection{$\varepsilon_K$, $\Delta M_K$ and $K_L\to \mu^+\mu^-$}
\unboldmath
For $\kpn$ we will also need ${\rm Re} \Delta_L^{s d}(Z)$. To this end 
using the formulae of  Appendix~\ref{DF2} we find the shifts in $\varepsilon_K$ and $\Delta M_K$ to be
\be\label{eps}
(\varepsilon_K)^Z_{\rm VLL}=-4.26\times 10^7\, {\rm Im} \Delta_L^{s d}(Z){\rm Re}\Delta_L^{s d}(Z)  \,
\ee
and 
\be\label{DM2a}
\frac{(\Delta M_K)^Z_{\rm VLL}}{(\Delta M_K)_\text{exp}} =6.43\times 10^7\, 
[({\rm Re} \Delta_{L}^{s d}(Z))^2 -({\rm Im} \Delta_{L}^{s d}(Z))^2]\,.
\ee

From (\ref{DES}), (\ref{IMSD}) and (\ref{eps}) we determine 
${\rm Re}\Delta_L^{s d}(Z)$ to be
\be\label{RE1}
{\rm Re}\Delta_L^{s d}(Z)=4.7\, \left[\frac{\keps}{\kepe}\right] \left[\frac{\bei}{0.76}\right] \cdot 10^{-5}\,.
\ee

However, the strongest constraint for ${\rm Re} \Delta_L^{s d}(Z)$ in this scenario 
comes from the 
$K_L\to\mu^+\mu^-$ bound in (\ref{eq:KLmm-bound}) which implies 
the allowed range 
\be\label{RESD}
-1.19 \cdot 10^{-6}\le {\rm Re} \Delta_L^{s d}(Z)\le 3.96\cdot 10^{-6}\,
\ee
and consequently using (\ref{RE1})
\be
\keps\le 0.084\,\kepe\,\left[\frac{0.76}{\bei}\right]\,.
\ee

Inserting the values of the couplings in (\ref{IMSD}) and  (\ref{RESD})
into (\ref{eps}) and (\ref{DM2a})
we find that
the shift in $\varepsilon_K$ is very small, at the level of $4\%$ at most
\be
-2.7\,\kepe \cdot 10^{-5}\le (\varepsilon_K)^Z_{\rm VLL}\le 8.4\, \kepe \cdot 10^{-5}
\ee
with the sign following the one of ${\rm Re} \Delta_L^{s d}(Z)$.
The shift in $\Delta M_K$ is fully negligible.

Thus in this NP scenario SM must  describe well the data on  $\varepsilon_K$ and $\Delta M_K$
unless NP generating flavour-violating $Z$ couplings can provide 
significant one-loop contributions to $\varepsilon_K$ and $\Delta M_K$. Such a possibility is encountered in  models with heavy vector-like quarks in 
\cite{Ishiwata:2015cga}, provided their masses are above $5\tev$.

\boldmath
\subsubsection{$\kpn$ and $\klpn$}
\unboldmath
All formulae  for these decays that are relevant for us have been collected 
in  Appendix~\ref{KPNN}.
In the case of $\klpn$ we get a unique prediction:
\be\label{Rnn0}
R^{\nu\bar\nu}_0\equiv \frac{\mathcal{B}(\klpn)}{\mathcal{B}(\klpn)_\text{SM}}=(1-0.6\,\kepe)^2
\ee
which for $\kepe=1.0$ amounts to a suppression of the SM prediction by a factor of $6.3$. 

The corresponding branching ratio for $\kpn$ is suppressed through 
the suppression of ${\rm Im} X_{\rm eff}$ governing $\klpn$ and also through suppression 
of ${\rm Re} X_{\rm eff}$ for positive values of ${\rm Re} \Delta_L^{s d}(Z) $.
But for sufficiently negative values of ${\rm Re} \Delta_L^{s d}(Z) $ in 
(\ref{RESD}) it can be enhanced. Using the formulae in  Appendix~\ref{KPNN}
we find then 
\be\label{Rnn+}
R^{\nu\bar\nu}_+\equiv\frac{\mathcal{B}(\kpn)}{\mathcal{B}(\kpn)_\text{SM}}\le 1.94\,.
\ee
This upper limit practically does not  depend 
on $\kepe$ as the NP contribution 
to the dominant part of  $R^{\nu\bar\nu}_+$ coming from the modification of 
 ${\rm Re} X_{\rm eff}$ is independent of  $\kepe$ and is  
directly bounded by $K_L\to\mu^+\mu^-$ and not by the combination of $\epe$ and 
$\varepsilon_K$.

In Fig.~\ref{R1} we show $R^{\nu\bar\nu}_0$  as a function of $\kepe$. For
the chosen values of the CKM parameters in (\ref{CKMfixed}) one has 
\be\label{AJBSM}
\mathcal{B}(\kpn)_\text{SM}=7.7\cdot 10^{-11}\,,  \qquad 
\mathcal{B}(\klpn)_\text{SM}=2.8\cdot 10^{-11}\,
\ee
to be compared with the present SM estimates that include uncertainties 
in the tree-level determinations of CKM parameters \cite{Buras:2015qea}
\be
\mathcal{B}(\kpn)_\text{SM}=(8.4\pm 1.0)\cdot 10^{-11}\,,  \qquad 
\mathcal{B}(\klpn)_\text{SM}=(3.4\pm 0.6)\cdot 10^{-11}\,.
\ee
We will use the values in (\ref{AJBSM}) in all formulae below.

\subsection{Right-handed Scenario (RHS)} 
\boldmath
\subsubsection{$\epe$}
\unboldmath
In this case the operator $Q_8^\prime$ dominates. But its mixing with $Q_7^\prime$ is the same as the one between $Q_8$ and $Q_7$. Only the value of 
 $C^\prime_7(M_Z)$ is different and the matrix element of $Q_8^\prime$ differs 
from the one of $Q_8$ only by sign. Using (\ref{C9Z}) we then find 
\be\label{ZfinalR}
\left(\frac{\varepsilon'}{\varepsilon}\right)^R_{Z}=\frac{1}{a}
\frac{\omega_+}{|\varepsilon_K|\sqrt{2}}\frac{{\IM} [A_2^{\rm NP}]^R}{{\RE}A_2}=
0.96\times 10^9\,\left[\frac{{\IM} [A_2^{\rm NP}]^R}{\gev}\right]
\ee
with 
\be\label{RENPZ}
{\IM} [A_2^{\rm NP}]^R= {\IM} C^\prime_8(m_c)\langle Q^\prime_8(m_c)\rangle_2 \,,\qquad \langle Q^\prime_8(m_c)\rangle_{2}=-\langle Q_8(m_c)\rangle_{2}\,,
\ee
where
\be\label{C8P}
 C^\prime_8(m_c)= 0.76\,  C^\prime_7(M_Z)= 0.76 \,\left[\frac{4 g_2 c_W^2}{6 c_W}\right]\frac{\Delta_R^{s d}(Z)}{4 M^2_Z}\,= 8.71 \times 10^{-6} 
\left[\frac{\Delta_R^{s d}(Z)}{\gev^2}\right]\,.
\ee

Collecting all these results we find now
\be\label{ZfinalR1}
\left(\frac{\varepsilon'}{\varepsilon}\right)^R_{Z}=\, -8.79 \times 10^3\, \bei\, {\IM}\Delta_R^{s d}(Z)
\ee 
and note that the numerical factor on the r.h.s is by a factor $c_W^2/s_W^2=3.33$ 
larger than in (\ref{ZfinalL1}) but the sign is the same.

Thus
\be\label{epsRHS}
\left(\frac{\varepsilon'}{\varepsilon}\right)_\text{RHS}=
\left(\frac{\varepsilon'}{\varepsilon}\right)_\text{SM}+
\left(\frac{\varepsilon'}{\varepsilon}\right)^R_{Z}
\ee
with the last term given in (\ref{ZfinalR1}). 

From (\ref{ZfinalR1}) and (\ref{deltaeps}) we find now
\be\label{IMSDR}
 {\rm Im} \Delta_R^{s d}(Z)= -1.50\,\kepe \left[\frac{0.76}{\bei}\right] \cdot 10^{-7}\,.
\ee
The sign is fixed through the requirement of the enhancement of $\epe$. 
For a given $\kepe$ the magnitude of the required coupling can be smaller than in LHS because the relevant Wilson coefficient contains the  additional factor  $3.33$. This also means that it is easier to enhance $\epe$ in this scenario 
while satisfying other constraints. This difference relative to LHS changes 
the implications for other observables.

\boldmath
\subsubsection{$\varepsilon_K$, $\Delta M_K$ and $K_L\to\mu^+\mu^-$}
\unboldmath
The strongest constraint for ${\rm Re} \Delta_L^{s d}(Z)$ in this scenario 
comes again from the 
$K_L\to\mu^+\mu^-$ bound in (\ref{eq:KLmm-bound}) which implies this time
the allowed range 
\be\label{RESDR}
-3.96 \cdot 10^{-6}\le {\rm Re} \Delta_R^{s d}(Z)\le 1.19\cdot 10^{-6}\,,
\ee
 simply the flip of the sign due to the flip of the sign in 
(\ref{c4}).

Using the formulae of Appendix~\ref{DF2} we find the shifts in $\varepsilon_K$ and $\Delta M_K$ to be even smaller than in LHS. Thus also in this NP scenario SM must describe the data on $\varepsilon_K$ and $\Delta M_K$ well unless loop contributions could be significant. 
On the other hand 
the results for $\kpn$ and $\klpn$ are more interesting.

\boldmath
\subsubsection{$\kpn$ and $\klpn$}
\begin{figure}[!tb]
 \centering
\includegraphics[width = 0.60\textwidth]{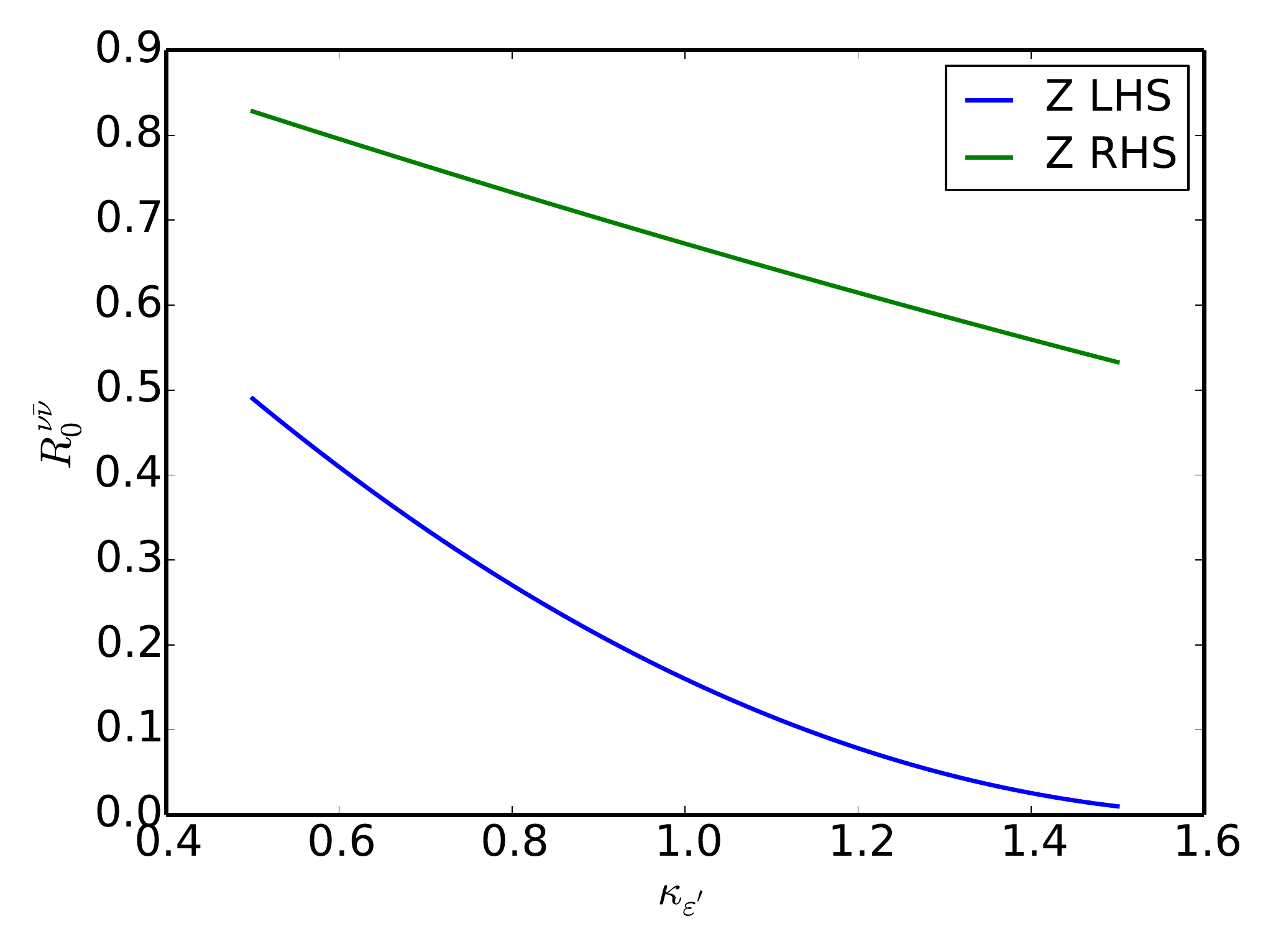}
\caption{ \it $R^{\nu\bar\nu}_0$   as a function of $\kepe$ for LHS and RHS $Z$ 
scenarios.
}\label{R1}~\\[-2mm]\hrule
\end{figure}

\unboldmath
We again obtain a unique prediction:
\be
R^{\nu\bar\nu}_0=(1-0.18\,\kepe)^2,
\ee
but this time the suppression of $R^{\nu\bar\nu}_0$ is smaller.
For $\kepe=1.0$ it amounts to a suppression by a factor of $1.5$. 
In Fig.~\ref{R1} we show $R^{\nu\bar\nu}_0$   as a function of $\kepe$ in this scenario.

The corresponding branching ratio for $\kpn$ is suppressed through 
the suppression of ${\rm Im} X_{\rm eff}$ and also through suppression 
of ${\rm Re} X_{\rm eff}$ for positive values of ${\rm Re} \Delta_R^{s d}(Z) $.
But for sufficiently negative values of ${\rm Re} \Delta_R^{s d}(Z) $ in 
(\ref{RESDR}) it can be enhanced. As the allowed magnitude in the latter case 
is larger than in LHS, the upper bound on the branching ratio is weaker. 
The dependence of this upper bound on $\kepe$ is even weaker than in LHS as 
 ${\rm Re} X_{\rm eff}$, which is  independent of it, is dominantly responsible for the modification of the $\kpn$ rate. We find
\be
R^{\nu\bar\nu}_+ \le 5.7\,.
\ee
Certainly such a large enhancement is very unlikely but it shows that in this 
scenario large enhancements of $\mathcal{B}(\kpn)$ are possible.
The fact that in RHS
the bound on $\kpn$ from $K_L\to\mu^+\mu^-$ is much weaker than in LHS has been pointed out in the context of the analysis of the Randall-Sundrum model with custodial protection, where rare decays are governed by tree-level $Z$ exchanges with RH flavour violating couplings \cite{Blanke:2008yr}.

\boldmath
\subsection{General $Z$ Scenarios} 
\unboldmath
\boldmath
\subsubsection{$\epe$}
\unboldmath
When both $\Delta_L^{sd}(Z)$ and $\Delta_R^{sd}(Z)$ are present
the general formula for $\epe$ is given as follows
\be\label{epsgeneral}
\left(\frac{\varepsilon'}{\varepsilon}\right)_Z=
\left(\frac{\varepsilon'}{\varepsilon}\right)_\text{SM}+
\left(\frac{\varepsilon'}{\varepsilon}\right)^L_{Z}+
\left(\frac{\varepsilon'}{\varepsilon}\right)^R_{Z}
\ee
with the last two terms representing LHS and RHS contributions discussed 
above.  As the operators $Q_i$ and $Q_i^\prime$ do not mix under renormalization we can just add these two contributions to the SM part independently of each other. 

 The $\epe$ constraint now reads
\be\label{c1}
{\rm Im} \Delta_{L}^{s d}(Z)+3.33\, {\rm Im} \Delta_{R}^{s d}(Z)= -5.0 \, \kepe\, \left[\frac{0.76}{\bei}\right]\cdot 10^{-7} \,.
\ee
The presence of two couplings allows now for more possibilities as we will 
see soon. We set $\bei=0.76$ in what follows.

\boldmath
\subsubsection{$\varepsilon_K$, $\Delta M_K$ and $K_L\to\mu^+\mu^-$}
\unboldmath
This time also LR operators contribute to $\varepsilon_K$ and $\Delta M_K$ 
and quite generally constitute by far the dominant contributions to these 
quantities so that 
we can approximate the shifts in $\varepsilon_K$ and  $\Delta M_K$ by 
keeping only LR contributions
\be\label{V1}
(\varepsilon_K)^Z\approx 
2.07\cdot 10^9 [({\rm Im} \Delta_{L}^{s d}(Z){\rm Re} \Delta_{R}^{s d}(Z)+{\rm Im} \Delta_{R}^{s d}(Z){\rm Re} \Delta_{L}^{s d}(Z)]
\ee
and 
\be\label{V2}
R^Z_{\Delta M}\equiv\frac{(\Delta M_K)^Z}{(\Delta M_K)_\text{exp}}\approx 
-6.21\cdot 10^9
[({\rm Re} \Delta_{L}^{s d}(Z){\rm Re} \Delta_{R}^{s d}(Z)-{\rm Im} \Delta_{L}^{s d}(Z){\rm Im} \Delta_{R}^{s d}(Z)].
\ee
The large size of LR contribution with respect to VLL and 
VRR contributions is not only related to enhanced hadronic matrix elements of
LR operators
but also to larger Wilson coefficients at $\mu=m_c$ that are enhanced through 
renormalization group effects \cite{Buras:2001ra}. The ones of VLL and VRR operators are 
suppressed slightly by these effects. 

The presence of LR operators has a very important consequence. While in LHS and RHS the $K_L\to\mu^+\mu^-$ bound provided by far the strongest constraint on ${\rm Re} \Delta_{L,R}^{s d}(Z)$, now also  $\varepsilon_K$ plays a role and 
$\keps$ will enter the game. However, as we will see in the first example below, for $\keps\ge 0.3$
and $\kepe\le 0.6$  the $K_L\to\mu^+\mu^-$ will again bound the 
rate for $\kpn$.

In this context it should be remarked that in principle
 it is possible to eliminate LR contributions by choosing 
properly the real and imaginary parts of LH and RH couplings.  It is also possible to use LR contributions to $\Delta M_K$ or $\varepsilon_K$ to eliminate completely NP contributions to them by 
cancelling the contributions from VLL and VRR operators \cite{Buras:2014sba,Buras:2014zga}.
This is only possible in the presence of suitable hierarchy between LH and RH 
couplings. In what follows we will assume that such fine-tuned situations do not take place.

While, the presence of LR operators is regarded often as a problem, it 
should be realized that in the case of possible anomalies in $\varepsilon_K$ 
and $\Delta M_K$ they could be welcome in the $Z$ case, where in LHS and RHS
NP contributions to  $\varepsilon_K$  and $\Delta M_K$ turned out to be small.
In order to illustrate this we will assume, as announced in Section~\ref{sec:strategy}, that in addition to the $\epe$ 
anomaly, the data show also $\varepsilon_K$ anomaly  parametrized by 
$\keps$ in (\ref{DES}).

\boldmath
\subsubsection{$\kpn$ and $\klpn$}
\unboldmath
For $\kpn$ and $\klpn$ the relevant expressions are collected in Appendix~\ref{KPNN}. In particular (\ref{c3}) implies that
in $\klpn$ the enhancement of its branching  
ratio requires the sum of the imaginary parts of the couplings to be {\it positive}.  This enhances also $\kpn$ but as seen in (\ref{c2}) could be compensated by the 
decrease of ${\rm Re}\,X_{\rm eff}$ unless the sum of the corresponding real parts is {\it negative}. 
For $K_L\to\mu^+\mu^-$ the relevant expressions are given in Appendix~\ref{sec:KLmm}. In particular in (\ref{c4}).

It is clear that with more parameters involved 
there are many possibilities in this NP scenario and which one is realized in nature 
will be only known through precise confrontation of the SM predictions for $\epe$, $\mathcal{B}(\kpn)$, $\mathcal{B}(\klpn)$, $\varepsilon_K$ and $\Delta M_K$ with future data. Indeed, presently it is not excluded that NP contributes to all of these quantities so that some enhancements and/or suppressions will be 
required.

Now among the  five quantities in question only $\epe$ and to a lesser extent $\varepsilon_K$ exhibit
some anomaly and NP models providing enhancements of both of them appear to be favoured. How 
much enhancement is needed in $\epe$ will strongly depend on the future value of $\bsi$. In the 
case of $\varepsilon_K$ this depends on the values of the CKM parameters, in particular on the value of $\vcb$.

 It would also be favourable, in particular for experimentalists, if the nature required the  enhancements of both $\mathcal{B}(\kpn)$ and $\mathcal{B}(\klpn)$  relative to SM predictions, simply, because then these branching ratios would be 
easier to measure and one could achieve a higher experimental precision on them.
But, we have seen in LHS and RHS  that enhancement of $\epe$ implied 
automatically suppression of $\mathcal{B}(\klpn)$, while $\mathcal{B}(\kpn)$ 
could be both enhanced and suppressed. NP contributions 
to $\varepsilon_K$ and $\Delta M_K$ were found 
at the level of a few percent at most after the $\epe$ and $K_L\to\mu^+\mu^-$ constraints
have been imposed. Therefore these scenarios while being 
in principle able to remove $\epe$ anomaly, cannot simultaneously solve possible
$\varepsilon_K$ anomaly. In fact, as already observed in \cite{Buras:2014sba}, in these scenarios a $10-20\%$ NP contribution 
to $\varepsilon_K$ would give significantly larger shift in $\epe$ than it is 
allowed by the data.

The question then arises whether it is possible in a general $Z$ scenario 
to remove the  $\epe$ anomaly through the shift in (\ref{deltaeps}), 
enhance $\varepsilon_K$ by a shift in (\ref{DES})
and simultaneously enhance $\mathcal{B}(\klpn)$ and $\mathcal{B}(\kpn)$ while 
satisfying  the $K_L\to\mu^+\mu^-$ and $\Delta M_K$ constraints.
The inspection of the formulae in  Appendices~\ref{DF2}--\ref{sec:KLmm}
 shows that this is indeed 
possible.

\subsubsection{Phenomenology} 
In order to exhibit this possibility in explicit terms and investigate the 
interplay between various quantities we introduce two real 
parameters $r_1$ and $r_2$ through
\be\label{REL1}
{\rm Im} \Delta_{L}^{s d}(Z)=-r_1\, {\rm Im} \Delta_{R}^{s d}(Z), \qquad
{\rm Re} \Delta_{L}^{s d}(Z)=r_2\, {\rm Re} \Delta_{R}^{s d}(Z)\,.
\ee

Using (\ref{V1}) we find then
\be\label{V1a}
(\varepsilon_K)^Z\approx 
2.07\cdot 10^9\, (r_2-r_1) {\rm Im} \Delta_{R}^{s d}(Z){\rm Re} \Delta_{R}^{s d}(Z)\,.
\ee

Imposing the shifts  in (\ref{deltaeps})  and
(\ref{DES}) we can determine:
\be\label{REL2}
{\rm Im} \Delta_{R}^{s d}(Z)=\frac{5.0}{(r_1-3.33)} \kepe\cdot 10^{-7}, \qquad 
{\rm Re} \Delta_{R}^{s d}(Z)=0.97\,\frac{(r_1-3.33)}{(r_2-r_1)}\frac{\keps}{\kepe}\cdot 10^{-6}\,.
\ee
Formulae (\ref{REL1}) and (\ref{REL2}) inserted in the expressions in 
Appendices~\ref{DF2}--\ref{sec:KLmm} allow to express the branching ratios 
for $\kpn$, $\klpn$ and $K_L\to\mu^+\mu^-$ and $\Delta M_K$ in terms of 
$\kepe$, $\keps$, $r_1$ and $r_2$.

In particular in order to see the signs of NP effects we find first
\begin{align}
{\rm Re}\,X_{\rm eff}(Z)&= -4.44\cdot 10^{-4}+2.51\cdot 10^2\,(1+r_2)\, {\rm Re} \Delta_{R}^{s d}(Z)\,,\label{c2b}\\
\label{c3b}
{\rm Im}\,X_{\rm eff}(Z)&=2.07\cdot 10^{-4}+2.51\cdot 10^2\,(1-r_1)\, {\rm Im} \Delta_{R}^{s d}(Z)\,, \\
\label{c4b}
{\rm Re}\,Y_{\rm eff}(Z)&= -2.83\cdot 10^{-4}+2.51\cdot 10^2\,(r_2-1)\, {\rm Re} \Delta_{R}^{s d}(Z)\,
\end{align}
and 
\be\label{V2a}
R^Z_{\Delta M}\approx 
-6.21\cdot 10^9
\left[r_2\,({\rm Re} \Delta_{R}^{s d}(Z))^2+r_1\,({\rm Im} \Delta_{R}^{s d}(Z))^2\right].
\ee

\begin{figure}[!tb]
 \centering
\includegraphics[width = 0.60\textwidth]{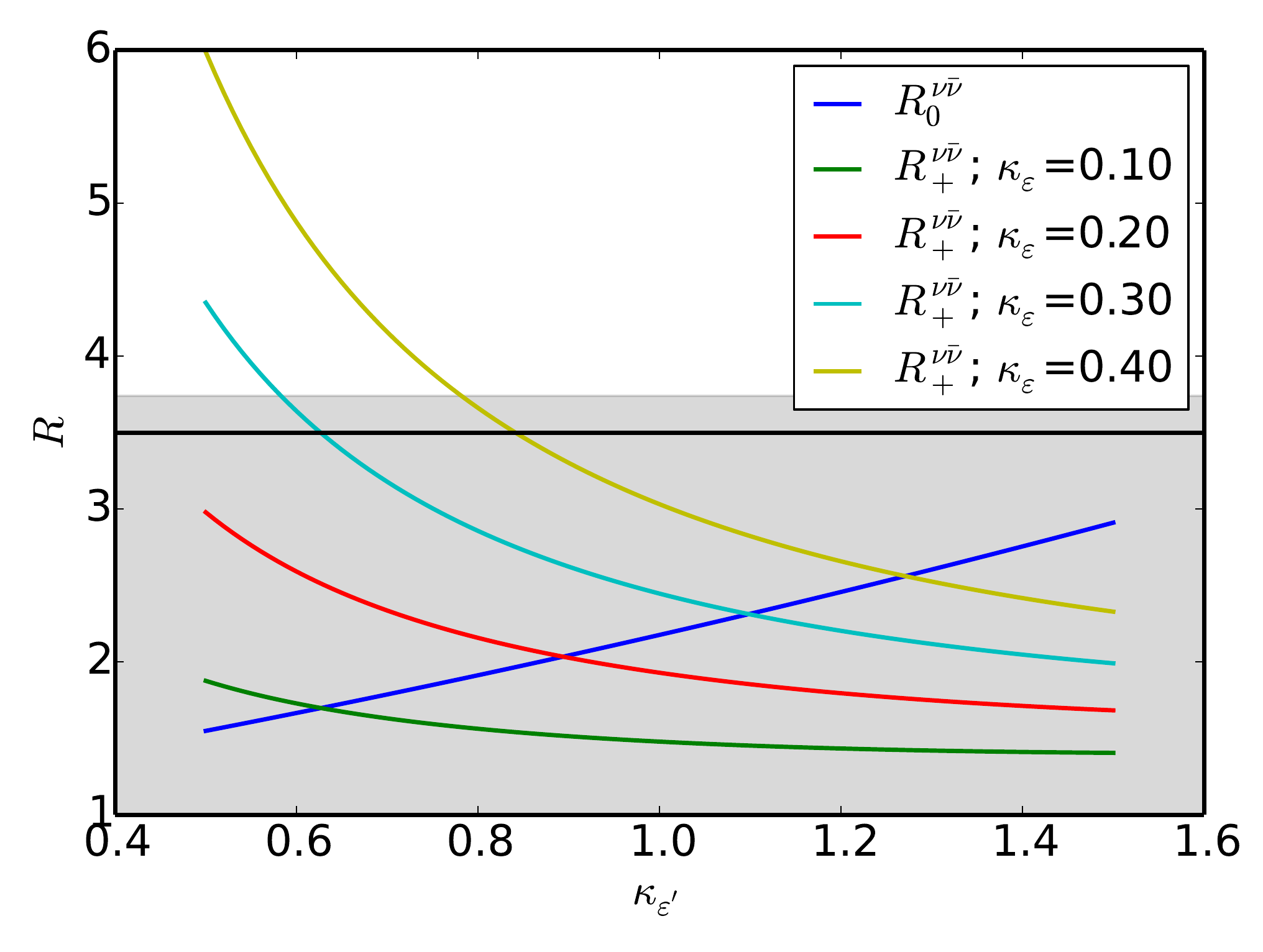}
\caption{ \it $R_0^{\nu\bar\nu}$ and  $R_+^{\nu\bar\nu}$, as functions 
of $\kepe$ for $\keps=0.1,\,0.2,\, 0.3,\, 0.4$ for the example 1. The horizontal 
{\it black} line corresponds to the upper bound in (\ref{KLKP}). The experimental $1\sigma$ range for  $R_+^{\nu\bar\nu}$ in (\ref{EXP1}) is displayed by the grey band.
}\label{R2a}~\\[-2mm]\hrule
\end{figure}

With $\kepe$  being positive we find then that 
$\epe$  and $\mathcal{B}(\klpn)$, with the latter governed by ${\rm Im}\,X_{\rm eff}(Z)$,  can be simultaneously enhanced provided
\be 
{\rm Im} \Delta_{R}^{s d}(Z) < 0 \,, \qquad
1.0 < r_1 < 3.33\,.
\ee
If in addition $\varepsilon_K$ and $\mathcal{B}(\kpn)$ should be enhanced
 $r_2$ has to satisfy\footnote{We assume here the enhancement of the magnitude of ${\rm Re} X_\text{eff}$.}
\be
 r_2 > r_1\,, \qquad   {\rm Re} \Delta_{R}^{s d}(Z) < 0 \qquad 
{\rm or} \qquad
r_2 < -1 \,, \qquad   {\rm Re} \Delta_{R}^{s d}(Z) > 0\,.
\ee

We illustrate the implications of these findings with two examples:

{\bf Example: 1 } We fix $r_1=2$ and $r_2=3$ to get
\be\label{E1}
{\rm Im} \Delta_{R}^{s d}(Z)=-3.76\, \kepe\cdot 10^{-7}, \qquad 
{\rm Re} \Delta_{R}^{s d}(Z)=-1.33\,\frac{\keps}{\kepe}\cdot 10^{-6}\,.
\ee
This is in fact the case considered already in  \cite{Buras:2015yca} but
here we present it in more explicit terms. In particular we include $\kpn$ and 
$K_L\to\mu^+\mu^-$ in this discussion and not only $\klpn$ as done in that paper.
The inspection of formulae for ${\rm Re}\,X_{\rm eff}(Z)$ and ${\rm Re}\,Y_{\rm eff}(Z)$ above accompanied by numerical analysis show that in this example
\be\label{KLKP}
R_+^{\nu\bar\nu}\approx R_L^{\mu\bar\mu}=\frac{\mathcal{B}(K_L\to\mu^+\mu^-)}{\mathcal{B}(K_L\to\mu^+\mu^-)_\text{SM}} \le 3.5
\ee
with the latter bound resulting from the bound in (\ref{eq:KLmm-bound}). 
On the other hand $\Delta M_K$ does not play any essential role with
$|R^Z_{\Delta M}|\le 0.04$. Here only short distance contributions to $K_L\to\mu^+\mu^-$ are involved.

In Fig.~\ref{R2a} we show $R_0^{\nu\bar\nu}$ and  $R_+^{\nu\bar\nu}$, as functions 
of $\kepe$ for $\keps=0.1,\,0.2,\, 0.3,\, 0.4$\footnote{In principle while varying $\keps$ we should also modify our CKM parameters as they correspond to 
$\keps=0.26$. But the dominant dependence on CKM parameters cancels in the 
ratios considered and keeping CKM fixed exposes better the dependence on 
 $\keps$ in the plots.} represented in the case of  $R_+^{\nu\bar\nu}$ by different colours
\be\label{coding}
\keps=0.1~{(\rm green)}, \qquad \keps=0.2~{(\rm red)}, \qquad 
\keps=0.3~{(\rm cyan)}, \qquad \keps=0.4~{(\rm yellow)}\,.
\ee
 $R_0^{\nu\bar\nu}$ is given by {\it blue} line and 
the upper bound in (\ref{KLKP}) is indicated by a black horizontal 
line. 

We observe that with increasing $\kepe$ the enhancement of $R_0^{\nu\bar\nu}$ slowly increases. On the other hand for a given $\keps$ the ratio $R_+^{\nu\bar\nu}$
decreases with increasing  $\kepe$. Both properties can  easily be understood 
from the formulae in (\ref{c2b}), (\ref{c3b}) and (\ref{E1}). We note that for a given $\kepe$ the upper bound  in (\ref{KLKP}) implies and upper bound on $\keps$ which becomes weaker with  
increasing $\kepe$.
Most interesting appear the values $\kepe \ge 1.0$ and $\keps\approx 0.25$ 
for which both $\epe$ and $\varepsilon_K$ anomalies can be solved in 
agreement with the $K_L\to\mu^+\mu^-$ bound and both $\kpn$ and $\klpn$ are 
significantly enhanced over their SM values.

\begin{figure}[!tb]
 \centering
\includegraphics[width = 0.60\textwidth]{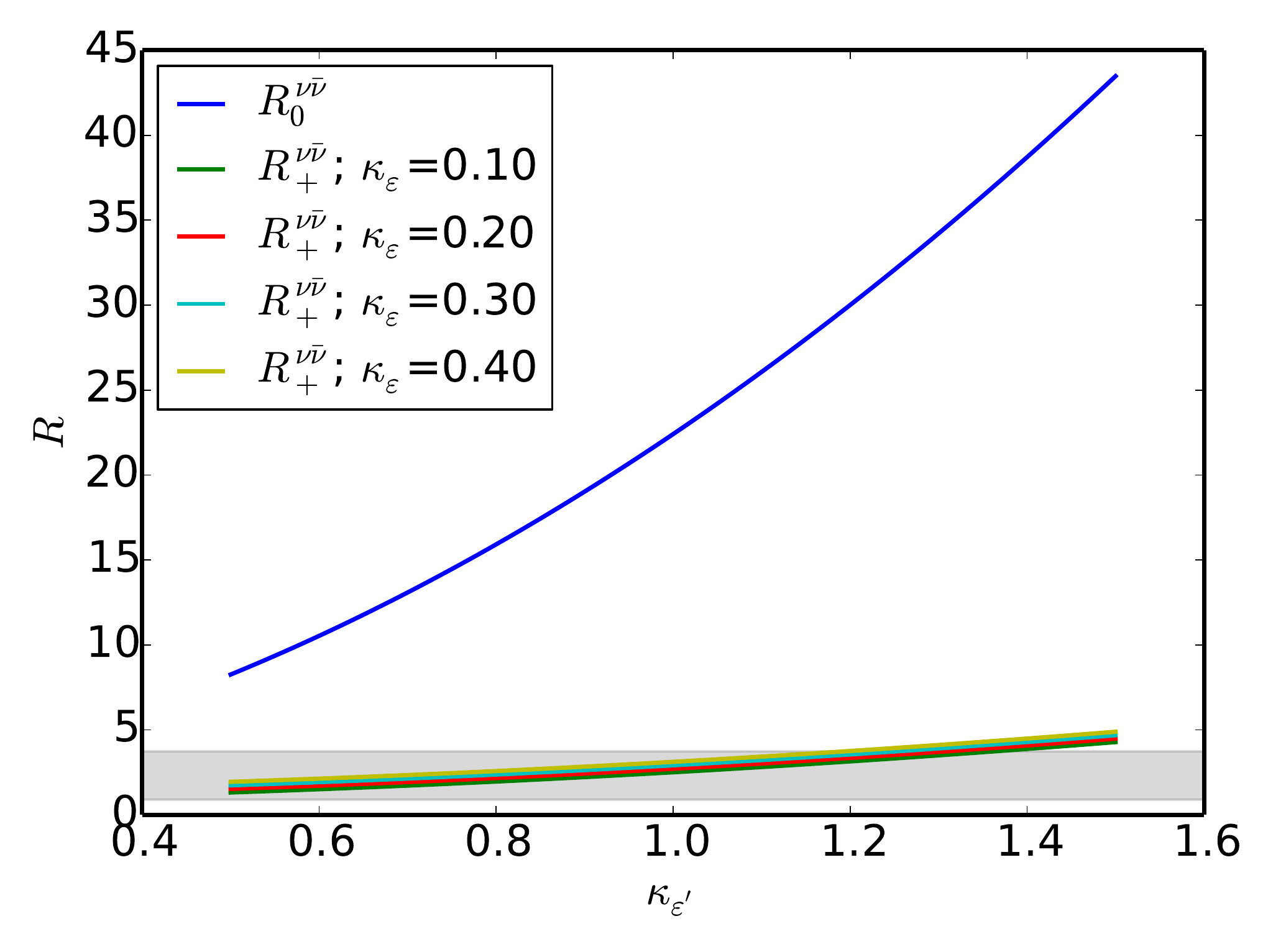}
\caption{ \it $R_0^{\nu\bar\nu}$ and  $R_+^{\nu\bar\nu}$, as functions 
of $\kepe$ for the example 2.  $R_0^{\nu\bar\nu}$ is independent of $\keps$ and the 
dependence of  $R_+^{\nu\bar\nu}$ on  $\keps$ is negligible. The experimental $1\sigma$ range for  $R_+^{\nu\bar\nu}$ in (\ref{EXP1}) is displayed by the grey band.
}\label{R3a}~\\[-2mm]\hrule
\end{figure}

{\bf Example: 2 } We fix $r_1=3$ and $r_2=-2$ to get
\be\label{E2}
{\rm Im} \Delta_{R}^{s d}(Z)=-1.52\, \kepe\cdot 10^{-6}, \qquad 
{\rm Re} \Delta_{R}^{s d}(Z)=6.6\,\frac{\keps}{\kepe}\cdot 10^{-8}\,.
\ee
Note that now imaginary parts of the couplings are larger than  the real parts with interesting consequences. 
In Fig.~\ref{R3a} we show for this case  
 $R_0^{\nu\bar\nu}$ and  $R_+^{\nu\bar\nu}$, 
 as functions of $\kepe$ again for  $\keps=0.1,\,0.2,\, 0.3,\, 0.4$.  Now
the relation (\ref{KLKP}) is no longer valid and  the bound from 
$K_L\to\mu^+\mu^-$ is irrelevant because the real parts of the couplings
are much smaller than in the previous example. 
We observe basically no dependence  of $R_+^{\nu\bar\nu}$ on $\keps$ as this 
parameter affects only the real parts of the couplings which are small in this
example. Again  $\Delta M_K$ does not play any essenial role with
$|R^Z_{\Delta M}|\le 0.05$.

\begin{figure}[!tb]
 \centering
\includegraphics[width = 0.45\textwidth]{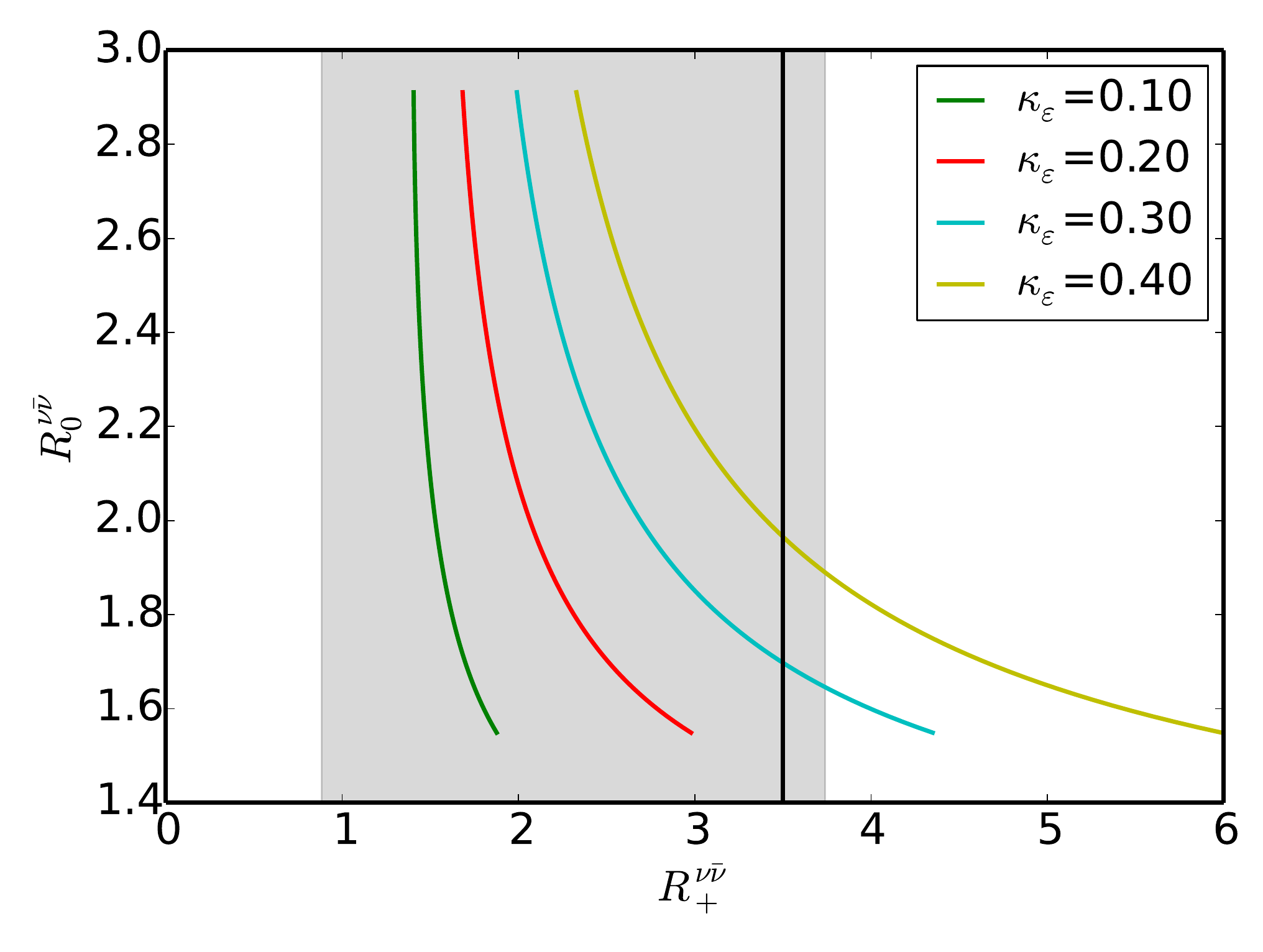}
\includegraphics[width = 0.45\textwidth]{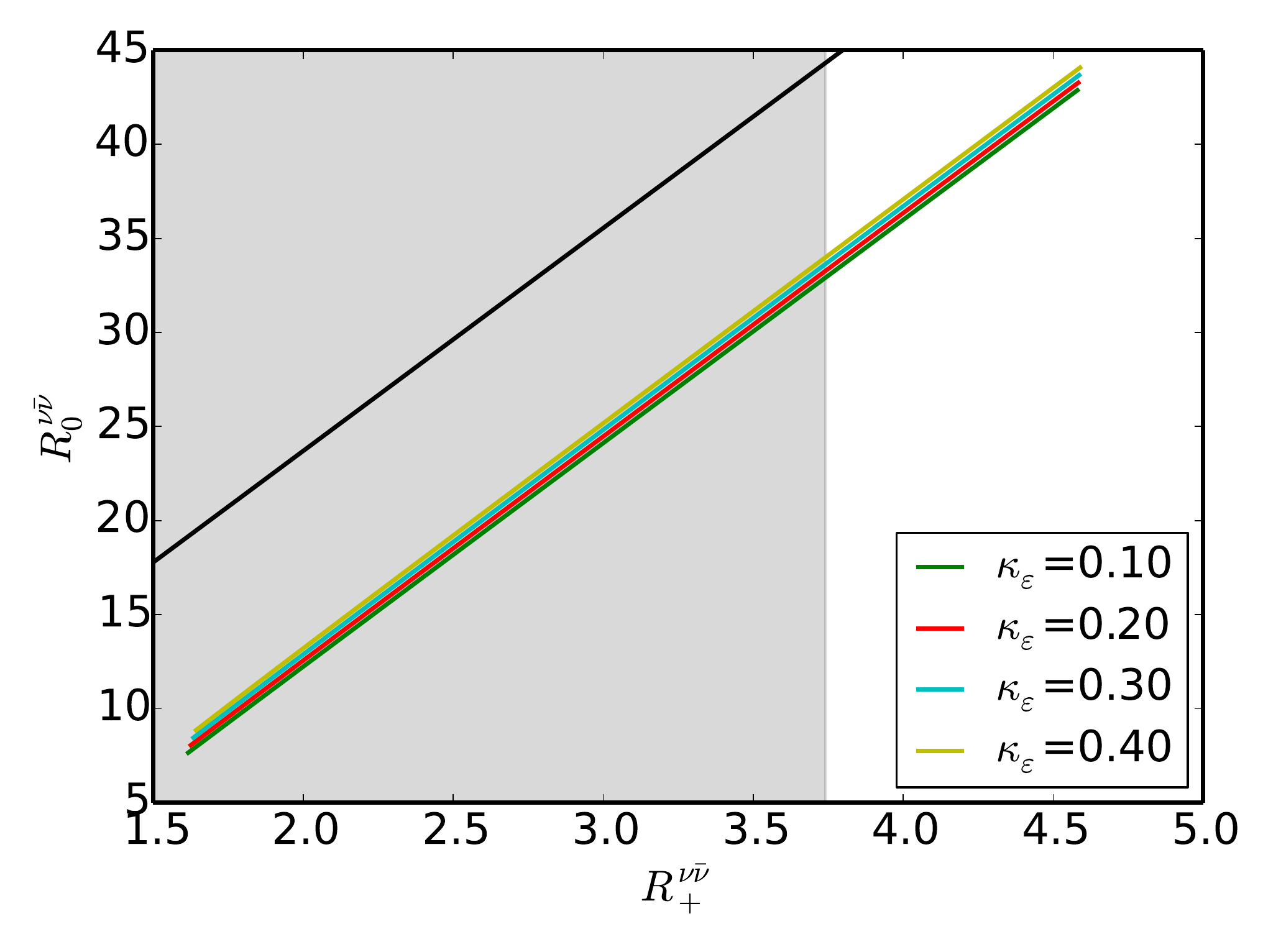}
\caption{ \it $R_0^{\nu\bar\nu}$ vs  $R_+^{\nu\bar\nu}$  
for $\keps=0.1,\,0.2,\, 0.3,\, 0.4$ for the example 1 (left panel) and the
example 2 (right panel) varying $0.5\le\kepe\le 1.5$. The vertical 
{\it black} line in the left panel corresponds to the upper bound in (\ref{KLKP}). The dependence on $\keps$ in the right panel is negligible and the {\it black} line 
represents the GN bound in (\ref{GNa}). The experimental $1\sigma$ range for  $R_+^{\nu\bar\nu}$ in (\ref{EXP1}) is displayed by the grey band.
}\label{R4ab}~\\[-2mm]\hrule
\end{figure}

We observe a very strong enhancement of {\it both} branching ratios which increases with increasing $\kepe$. This should be contrasted with the previous example
in which for a given $\keps$ the two branching ratios were anticorrelated. This 
is best seen in  Fig.~\ref{R4ab} where 
we show  in the left panel $R_0^{\nu\bar\nu}$ vs  $R_+^{\nu\bar\nu}$ for the  example 1 and in the right panel the corresponding plot 
 for the example 2. A given line in the left panel, on which the ratios are anticorrelated, corresponds 
to a fixed value of $\keps$ and the range on each line results from the 
variation of $\kepe$ in the range $0.5\le \kepe \le 1.5$. We impose
the constraint from   $\mathcal{B}(K_L\to\mu^+\mu^-)$. In the right panel the 
value of $\keps$ does not matter and the range for the values of both branching
ratios corresponds to  $0.5\le \kepe \le 1.5$ with largest enhancements for
largest $\kepe$. Moreover, 
 the two ratios increase in a correlated manner on the line parallel to the 
GN bound in (\ref{GN}) which
expressed through
the ratios  $R_0^{\nu\bar\nu}$ and  $R_+^{\nu\bar\nu}$ reads
\be\label{GNa}
 R_0^{\nu\bar\nu}\le 11.85\, R_+^{\nu\bar\nu}\,.
\ee
We indicate this bound by a black line. Such a correlation between $\kpn$ and 
$\klpn$ is characteristic for cases in which  only imaginary parts in the new
couplings matter and both branching ratios are affected only by the modification
of ${\rm Im} X_{\rm eff}$. For a general discussion see  \cite{Blanke:2009pq}.

\boldmath
\subsection{Summary of NP  Patterns in $Z$ Scenarios}
\unboldmath
The lessons from these four exercises are as follows:
\begin{itemize}
\item
In the LHS , a given request for the enhancement of $\epe$ determines 
the coupling ${\rm Im} \Delta_{L}^{s d}(Z)$.
\item
This result has direct unique implications on $\klpn$: {\it suppression} of
  $\mathcal{B}(\klpn)$. This property is known from NP scenarios in which 
 NP to $\klpn$ and $\epe$ enters dominantly through the modification of 
$Z$-penguins.
\item
The imposition of the $K_L\to\mu^+\mu^-$ constraint determines the range for
${\rm Re} \Delta_{L}^{s d}(Z)$ which with the already fixed ${\rm Im} \Delta_{L}^{s d}(Z)$ allows to calculate the shifts in $\varepsilon_K$ and $\Delta M_K$. 
These shifts turn out to be very small for $\varepsilon_K$ and negligible for 
 $\Delta M_K$. Therefore unless loop contributions from physics generating 
$ \Delta_{L}^{s d}(Z)$ play significant role in both quantities, the SM predictions
 for  $\varepsilon_K$ and $\Delta M_K$ must agree well with data for this NP scenario to survive.
\item
Finally, with fixed ${\rm Im} \Delta_{L}^{s d}(Z)$ and the allowed range 
for ${\rm Re} \Delta_{L}^{s d}(Z)$, the range for  $\mathcal{B}(\kpn)$ 
can be obtained. But in view of uncertainties in the  $K_L\to\mu^+\mu^-$ constraint both enhancement and suppressions of  $\mathcal{B}(\kpn)$ are possible 
and no specific pattern of correlation between  $\mathcal{B}(\klpn)$ and 
 $\mathcal{B}(\kpn)$ is found. In the absence of a relevant $\varepsilon_K$ 
constraint this is consistent with the general analysis in \cite{Blanke:2009pq}.
$\mathcal{B}(\kpn)$  can be enhanced by a factor of $2$ at most.
\item
Analogous pattern is found in RHS, although the numerics is different. First  due the modification of the initial conditions for the Wilson coefficients 
the suppression of $\mathcal{B}(\klpn)$ for a given $\kepe$ is smaller.
Moreover,  the flip of the sign in NP contribution to $K_L\to \mu^+\mu^-$  allows 
for larger enhancement of  $\mathcal{B}(\kpn)$, a property known from our
previous analyses. An enhancement of $\mathcal{B}(\kpn)$  up to a factor of  $5.7$ 
is possible.
\item
In a general $Z$ scenario the pattern of NP effects 
changes because of the appearance of 
LR operators dominating NP contributions to  $\varepsilon_K$ and $\Delta M_K$. 
Consequently for large range of parameters  
these two quantities, in particular  $\varepsilon_K$, provide stronger 
constraint on ${\rm Re} \Delta_{L,R}^{s d}(Z)$ than $K_L\to\mu^+\mu^-$. But 
the main virtue of the general scenario is the possibility of enhancing 
simultaneously $\epe$, $\varepsilon_K$, $\mathcal{B}(\kpn)$ and $\mathcal{B}(\klpn)$ which is not possible in LHS and RHS. Thus the presence of both 
LH and RH flavour-violating currents is essential for obtaining
simultaneously the enhancements in question.
\item
We have illustrated this on two examples with the results shown in Figs.~\ref{R2a}--\ref{R4ab} for which as seen in Fig.~\ref{R4ab} the correlation 
between branching ratios for $\kpn$ and $\klpn$ are strikingly different. In 
particular in the second example in which the imaginary parts in the couplings dominate the correlation takes place along the line parallel to the line representing GN bound.
\end{itemize}

We will now turn our attention to $Z^\prime$ models which, as we will see, 
exhibit quite different pattern of NP effects in the $K$ meson system than the 
LH and RH $Z$ scenarios. In particular we will find  that at the qualitative 
level  $Z^\prime$ models with only LH or RH flavour-violating couplings can 
generate very naturally the patterns found in the two examples 
in Figs.~\ref{R2a}--\ref{R4ab} that in $Z$ scenario required the presence 
of both LH and RH couplings. In fact the pattern of correlation between 
$\kpn$ and $\klpn$ found in the example 1 
will also be found in $Z^\prime$ scenario in which NP in $\epe$ is dominated 
by EWP operator $Q_8$. On the other hand QCDP operator $Q_6$ generated by 
$Z^\prime$ exchange implies a pattern of correlation between $\klpn$ and $\kpn$
found in the example 2. But the implication for 
NP effects in $\Delta M_K$ will turn out to be more interesting than found in 
the latter example.

\boldmath
\section{$Z^\prime$ Models}\label{sec:3}
\unboldmath
\subsection{Preliminaries}
Also in this case the operators $Q_8$ and $Q_8^\prime$ dominate NP 
contribution to $\epe$
in several models and we will recall some of them below. 
However, this time flavour diagonal $Z^\prime$ couplings to quarks are model dependent, which allows to construct models in which the QCDP operator $Q_6$ or the operator $Q_6^\prime$ dominates NP contribution to $\epe$. As this case cannot be realized in $Z$ scenarios it is instructive 
to discuss this scenario first. In particular it will turn out  that in this case it is much easier to reach our goal of enhancing simultaneously $\epe$, $\varepsilon_K$, $\kpn$ and $\klpn$. Moreover, the presence of flavour-violating right-handed currents is not required. 

In order for $Q_6$ or $Q_6^\prime$ to dominate the scene
 the diagonal RH or LH quark couplings 
must be flavour universal which with the normalization of Wilson coefficients 
in (\ref{ZprimeA}) implies \cite{Buras:2014sba}
\begin{align}
\begin{split}
C_3(M_{Z^\prime})
& = \frac{\Delta_L^{s d}(Z^\prime)\Delta_L^{q q}(Z^\prime)}{4 M^2_{Z^\prime}}, \qquad 
C_3^\prime(M_{Z^\prime})
 = \frac{\Delta_R^{s d}(Z^\prime)\Delta_R^{q q}(Z^\prime)}{4 M^2_{Z^\prime}}
 \,,\end{split}\label{C3}\\
\begin{split}
C_5(M_{Z^\prime})
& = \frac{\Delta_L^{s d}(Z^\prime)\Delta_R^{q q}(Z^\prime)}{4 M^2_{Z^\prime}}, \qquad
C_5^\prime(M_{Z^\prime})
  = \frac{\Delta_R^{s d}(Z^\prime)\Delta_L^{q q}(Z^\prime)}{4 M^2_{Z^\prime}}
\,.\end{split}\label{C6}
\end{align}
The couplings $\Delta_{L,R}^{s d}(Z^\prime)$ are defined by (\ref{Zcouplings}) with $Z$ replaced by $Z^\prime$.
$\Delta_{L,R}^{q q}(Z^\prime)$ are flavour universal quark couplings which 
are assumed to be real. It should be noted that EWP are 
absent here. Moreover, they cannot be generated from QCDP through 
QCD renormalization group effects so that their contributions to NP part of 
$\epe$ can be neglected.
This should be contrasted with the SM, where they are generated by electroweak 
interactions from the mixing with current-current operators that have much larger Wilson coefficients than QCDP.

Now as briefly discussed in Section~\ref{sec:5} there exist models in which 
only LH or RH couplings are present. In that case the Wilson coefficients 
$C_5(M_{Z^\prime})$ and $C^\prime_5(M_{Z^\prime})$ vanish in the leading order. 
Non-vanishing contribution of $Q_6$ and $Q_6^\prime$ can still be generated 
through their mixing with $(V\mp A)\times(V\mp A)$ operators                     $Q_3$ and $Q_3^\prime$, respectively. But this mixing 
is significantly smaller than between  $Q_6$ and $Q_5$ and between
$Q_6^\prime$ and $Q_5^\prime$ leading to much smaller Wilson coefficients of 
 $Q_6$ and $Q_6^\prime$ at $\mu=m_c$ than it is possible when the Wilson 
coefficients $C_5(M_{Z^\prime})$ and $C^\prime_5(M_{Z^\prime})$ do not vanish. We 
will therefore consider only the latter case but the former case of only LH 
or RH couplings implies similar phenomenology to the one presented below 
except that NP effects in $\epe$ are significantly smaller than the ones
discussed by us.

In this context we also note that without a specific model there is a considerable freedom in the values of the diagonal quark 
and lepton couplings of $Z^\prime$, although one must make sure that they 
are consistent with LEP II and LHC bounds.
Concerning LHC bounds, the study in  \cite{deVries:2014apa} implies
\be \label{LHCbound}
\left|\Delta_R^{q \bar q}(Z') \right| \leq 1.0 \left[ \frac{M_{Z'}}{3\tev}
 \right] \left[ 1 + \left(\frac{1.3\tev}{M_{Z'}}\right)^2\right]\,.
\ee
On the other hand bounds on the leptonic $Z^\prime$ couplings can be 
extracted from the final analysis of the LEP-II data  \cite{Schael:2013ita}, 
although there is still a considerable freedom as the bounds are for 
products of electron and other lepton couplings. Therefore the allowed coupling
 $\Delta^{\nu\bar\nu}_L(Z^\prime)$ can be increased by lowering  $\Delta^{e\bar e}_L(Z^\prime)$ coupling.

 As an example for our nominal value $M_{Z^\prime}=3\tev$ the choices
\be\label{COUP}
\Delta^{q \bar q}_R(Z^\prime)=1, \qquad
\Delta^{q \bar q}_L(Z^\prime)=-1, \qquad \Delta^{\nu\bar\nu}_L(Z^\prime)=\Delta^{\mu\bar\mu}_L(Z^\prime)=0.5
\ee
are consistent with these bounds\footnote{The relation between leptonic couplings follows from $SU(2)_L$ gauge invariance}. Yet, it should be kept in mind 
that these couplings can in principle be larger or  smaller. 
For larger (smaller) $\Delta^{\nu\bar\nu}_L(Z^\prime)$ NP contributions to the branching ratios  
for $\kpn$ and $\klpn$ will be larger (smaller), but in a correlated manner. The implications of 
the change of the couplings $\Delta^{q \bar q}_{L,R}(Z^\prime)$ are more profound as we will see in the context of our presentation. But, for the time being we will
 assume 
that $\Delta^{q \bar q}_{L,R}(Z^\prime)$  are $\ord(1)$. 

\boldmath
\subsection{$Z^\prime$ with QCD Penguin Dominance (LHS)}\label{Q6LHC}
\subsubsection{$\epe$}
\unboldmath
We begin with NP scenario with purely LH flavour-violating quark 
couplings and flavour universal RH flavour diagonal couplings.
In this case the operator $Q_6$ is dominant. It mixes with the operator 
$Q_5$ and the LO RG analysis gives \cite{Buras:2014sba} (see  Appendix~\ref{app:X})
\be\label{LOC6}
C_6(m_c)= 1.13\frac{\Delta_L^{s d}(Z^\prime)\Delta_R^{q q}(Z^\prime)}{4 M^2_{Z^\prime}}=3.14\times 10^{-8} \left[\frac{\Delta_L^{s d}(Z^\prime)\Delta_R^{q q}(Z^\prime)}{\gev^2}\right]\left[\frac{3\tev}{M_{Z^\prime}}\right]^2
\ee
with $1.13$ resulting from RG evolution from $M_{Z^\prime}=3\tev$ down to 
$\mu=m_c$. With increasing $M_{Z^\prime}$ this factor increases logarithmically but $C_6(m_c)$ decreases much faster because of the last factor. Still as we 
will discuss later
if the flavour structure of a given model is such that the suppression by 
$Z^\prime$ propagator is compensated by the increase of flavour-violating couplings, for  $M_{Z^\prime}\ge 10\tev$ the RG effects above $M_{Z^\prime}=3\tev$ begin to play some 
role  implying additional enhancements of both QCDP and EWP contributions to $\epe$. The contribution 
of $Q_5$ can be neglected because of its strongly colour suppressed matrix element. Moreover, relative importance of $Q_5$ decreases with increasing $M_{Z^\prime}$ again due to RG effects. See  Appendix~\ref{app:X} for details.

We then find 
\be\label{eprimeZprime}
\left(\frac{\varepsilon'}{\varepsilon}\right)^L_{Z^\prime}=
-\frac{{\IM} [A_0^{\rm NP}]^L}{{\RE}A_0}\left[\frac{\omega_+}{|\varepsilon_K|\sqrt{2}}\right](1-\hat\Omega_{\rm eff})=-3.69\times 10^7\,\left[\frac{{\IM} [A_0^{\rm NP}]^L}{\gev}\right]
\ee
where we set all relevant quantities at their central values and
\be\label{A0ZP}
[A_0^{\rm NP}]^L= C_6(m_c)\langle Q_6(m_c)\rangle_0
\ee
with $\langle Q_6(\mu)\rangle_0$ given in (\ref{eq:Q60}).

Collecting all these results we find
\be\label{eprimeZprime1}
\left(\frac{\varepsilon'}{\varepsilon}\right)^L_{Z^\prime}= 0.67 \,\bsi \,\left[\frac{3\tev}{M_{Z^\prime}}\right]^2 {\IM}(\Delta_L^{s d}(Z^\prime))\Delta_R^{q q}(Z^\prime)\,.
\ee
It should be noted that due to a large value of $M_{Z^\prime}$ and the suppression 
factors of the $Q_6$ contribution to $\epe$ mentioned before, the overall numerical factor 
in this result is for $\Delta_R^{q\bar q}(Z^\prime)=\ord(1)$ by more than three orders of magnitude smaller than in 
the case of the corresponding $Z$ scenario. See (\ref{ZfinalL1}). 

We next request the enhancement of $\epe$ as given
in (\ref{deltaeps}) and set the values of CKM factors to the ones in 
(\ref{CKMfixed}). Setting $\bsi=0.7$, a typical value consistent with lattice and large $N$ 
results,  we find  from  (\ref{eprimeZprime1}) and  (\ref{deltaeps})
\be\label{IMSDPrime}
 {\rm Im} \Delta_L^{s d}(Z^\prime)= 2.1\, \left[\frac{\kepe}{\Delta_R^{q\bar q}(Z^\prime)} \right]\left[\frac{0.70}{\bsi}\right] \left[\frac{M_{Z^\prime}}{3\tev}\right]^2\cdot 10^{-3}\,.
\ee
The sign is fixed through the requirement of the enhancement of $\epe$ in 
(\ref{deltaeps}) and 
the sign of  $\Delta^{q \bar q}_R(Z^\prime)$ in (\ref{COUP}). The large difference
between the values in (\ref{IMSD}) and  (\ref{IMSDPrime}) is striking.
The strong suppression of NP contribution to $\epe$ by a large 
$Z^\prime$ mass,  suppressed matrix element of $Q_6$ relative to the one of $Q_8$ and  the inverse ``$\Delta I=1/2$'' factor in (\ref{IDI12}) have 
to be compensated by increasing $ {\rm Im} \Delta_L^{s d}(Z^\prime)$. 
This will 
have interesting consequences.

\boldmath
\subsubsection{$\varepsilon_K$, $\Delta M_K$ and $K_L\to\mu^+\mu^-$}
\unboldmath
Because of the increased value of ${\rm Im} \Delta_L^{s d}(Z^\prime)$,
not $K_L\to\mu^+\mu^-$ bound 
(\ref{eq:KLmm-bound}), as in the corresponding $Z$ scenario, 
but $\varepsilon_K$ and $\Delta M_K$
put the  strongest constraints on ${\rm Re} \Delta_L^{s d}(Z^\prime)$.

Using the instructions at the end of Appendix~\ref{DF2} we find
\be\label{epsZprime}
(\varepsilon_K)^{Z^\prime}_{\rm VLL}=-3.51\times 10^4\, \left[\frac{3\tev}{M_{Z^\prime}}\right]^2 
{\rm Im} \Delta_{L}^{s d}(Z^\prime){\rm Re} \Delta_{L}^{s d}(Z^\prime)
\ee
and
\be\label{DMZprime}
R^{Z^\prime}_{\Delta M}= \frac{(\Delta M_K)^{Z^\prime}_{\rm VLL}}{(\Delta M_K)_\text{exp}}=
5.29\times 10^4\, \left[\frac{3\tev}{M_{Z^\prime}}\right]^2 
[({\rm Re} \Delta_{L}^{s d}(Z^\prime))^2 -({\rm Im} \Delta_{L}^{s d}(Z^\prime))^2]\
\ee

 Requiring 
the enhancement of $\varepsilon_K$ as in (\ref{DES}) and using (\ref{IMSDPrime}) we find 
\be\label{RESDPrime}
{\rm Re} \Delta_L^{s d}(Z^\prime)=-1.4 \, \keps\,
\left[\frac{\Delta_R^{q \bar q}(Z^\prime)}{\kepe}\right] \left[\frac{\bsi}{0.70}\right]  \cdot 10^{-5}\,.
\ee
It should be noted that this result is independent of the value of $M_{Z^\prime}$.
Moreover, there is again a striking difference from the $Z$ case as now 
${\rm Re} \Delta_L^{s d}(Z^\prime)$ is much smaller than ${\rm Im} \Delta_L^{s d}(Z^\prime)$ making the coupling $\Delta_L^{s d}(Z^\prime)$ to an excellent approximation imaginary with two first interesting consequences:
\begin{itemize}
\item
The $K_L\to\mu^+\mu^-$ constraint is easily satisfied.
\item
$\Delta M_K$ is uniquely {\it suppressed} with the suppression increasing with 
increasing $\kepe$ and $M_{Z^\prime}$: 
\be\label{SUPQ6}
R^{Z^\prime}_{\Delta M}({\rm QCDP})\equiv\frac{(\Delta M_K)^{Z^\prime}_{\rm VLL}}{(\Delta M_K)_\text{exp}}=-0.23\, 
\left[\frac{\kepe}{\Delta_R^{q \bar q}(Z^\prime)}\right]^2
\left[\frac{M_{Z^\prime}}{3\tev}\right]^2\,\left[\frac{0.70}{\bsi}\right]^2 .
\ee
\end{itemize}
Whether this suppression is consistent with the data cannot be answered at 
present because of large uncertainties in the evaluation of $\Delta M_K$ 
within the SM.

Indeed the present SM estimate without the inclusion of long distance 
effects reads \cite{Brod:2011ty}
\be\label{SMDMK}
R^{\rm SM}_{\Delta M}= 0.89 \pm 0.34 \,.
\ee
Large N approach  \cite{Buras:2014maa} indicates that long distance contributions enhance this ratio by roughly $20\%$. First lattice calculations  
\cite{Bai:2014cva} are still subject to large uncertainties 
and also the large error in (\ref{SMDMK}) precludes any definite conclusions 
at present whether NP should enhance or suppress this ratio.

\boldmath
\subsubsection{$\kpn$ and $\klpn$}
\unboldmath
But the most interesting implications of the $\epe$ anomaly in this scenario are the ones for 
$\kpn$ and $\klpn$. Inserting the couplings in  (\ref{IMSDPrime}) and 
(\ref{RESDPrime}) into (\ref{c7}) and (\ref{c8})we find that
the branching ratios $\mathcal{B}(\klpn)$ and $\mathcal{B}(\kpn)$ are to an excellent approximation 
affected only through the shift in ${\rm Im}\,X_{\rm eff}$. Therefore,
there is a  strict correlation 
between  $\mathcal{B}(\klpn)$ and  $\mathcal{B}(\kpn)$  which in the plane 
of these two branching ratios takes place on the branch parallel to the 
Grossman-Nir bound \cite{Grossman:1997sk} in (\ref{GNa}). This is a very striking difference from $Z$ scenarios LHS and RHS which to our knowledge has not been noticed before. On the other hand there are some similarities to the example 2 
in the general $Z$ scenario in which the imaginary parts of the couplings 
dominate and the value of the parameter $\keps$ does not play any role 
for $\kpn$ and $\klpn$.

\begin{figure}[!tb]
 \centering
\includegraphics[width = 0.45\textwidth]{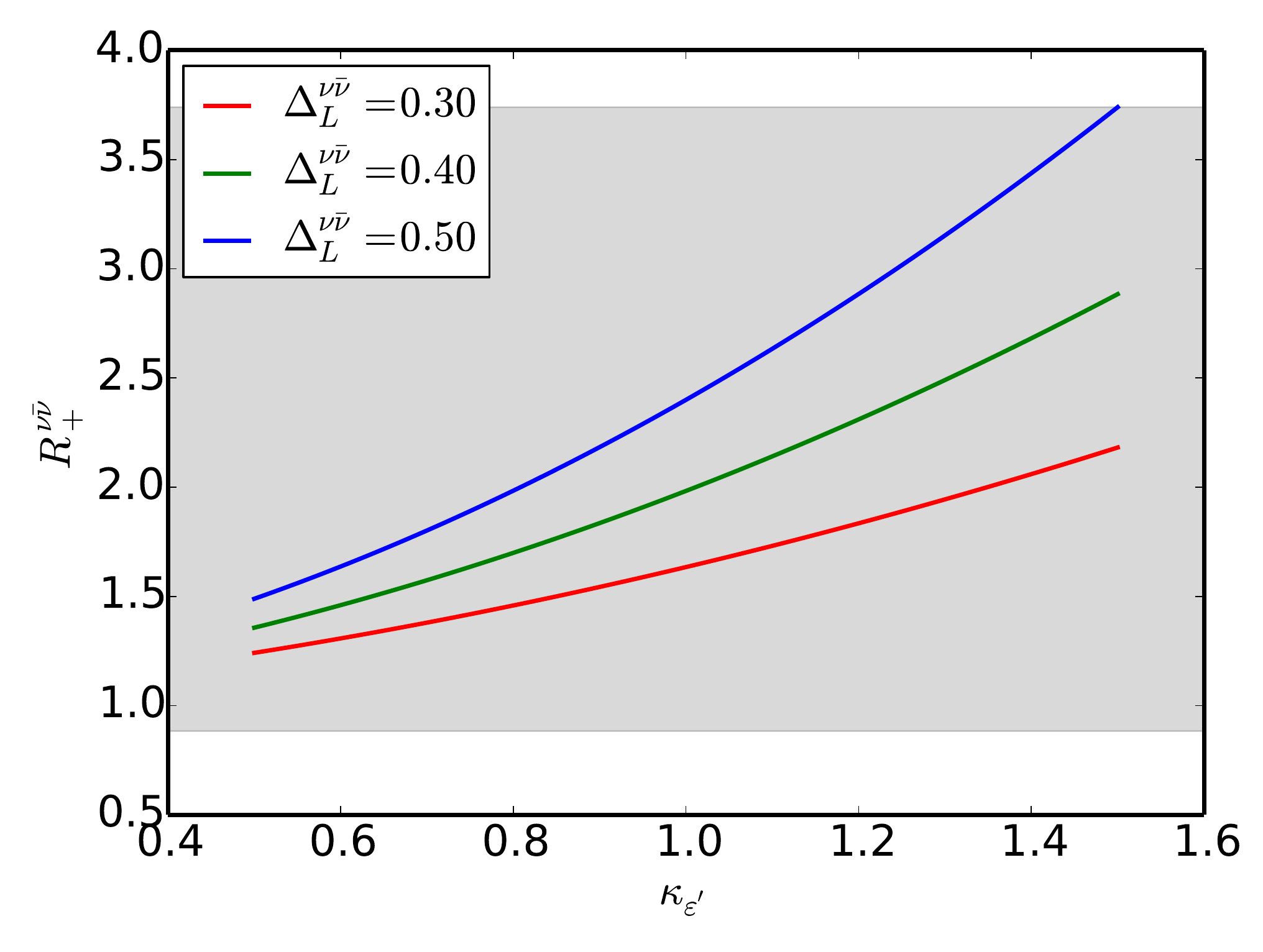}
\includegraphics[width = 0.45\textwidth]{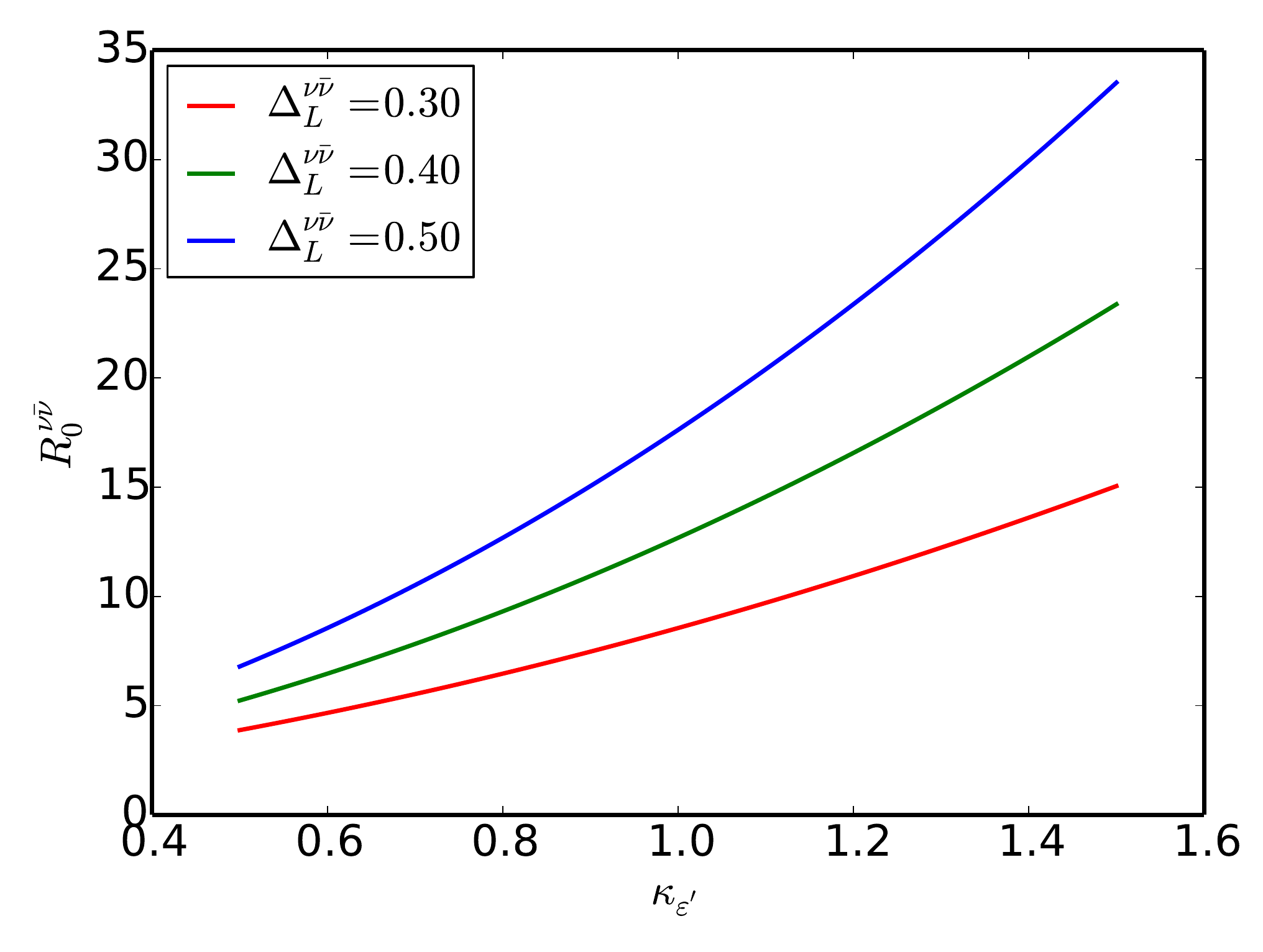}
\includegraphics[width = 0.45\textwidth]{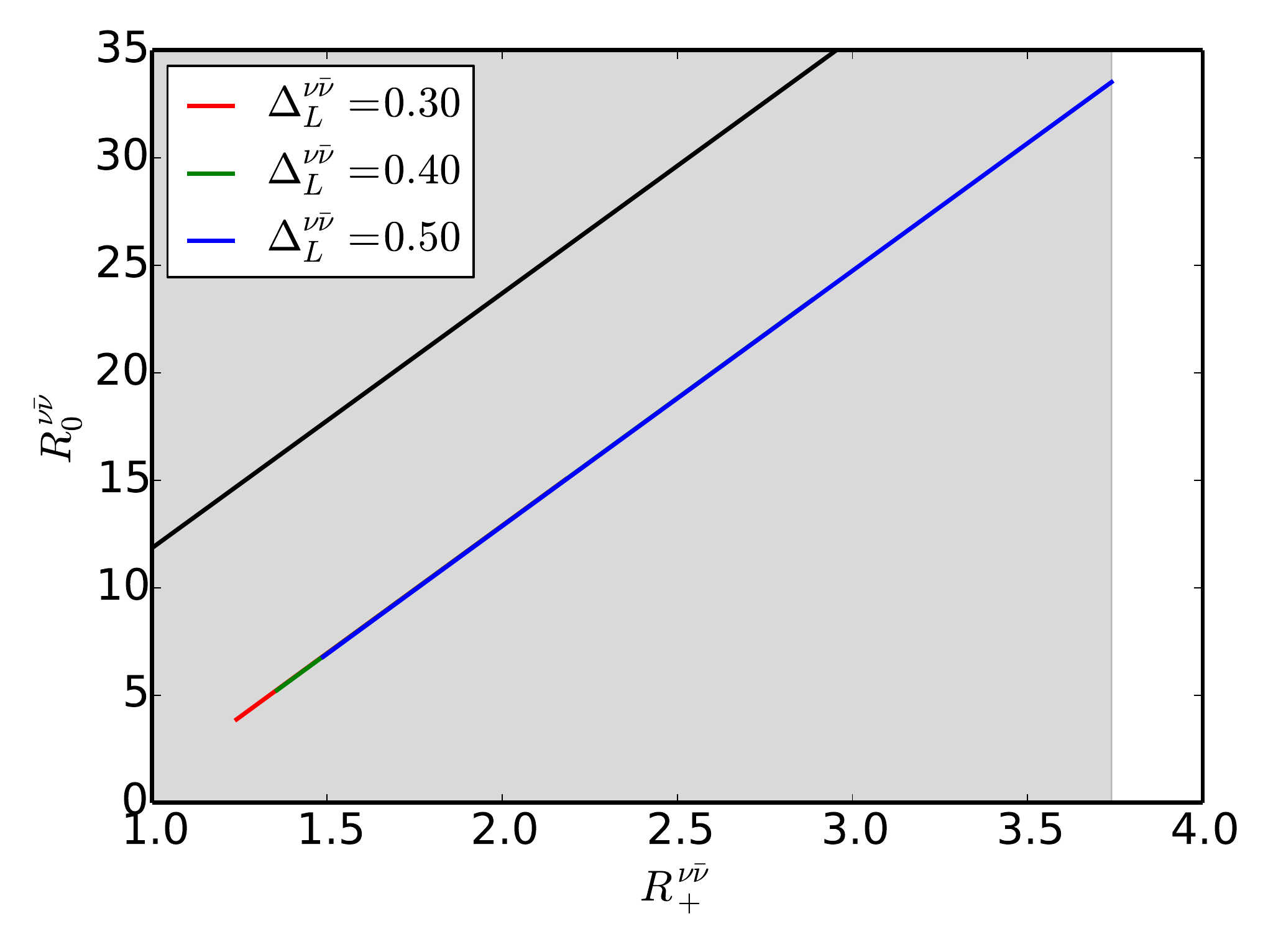}
\includegraphics[width = 0.45\textwidth]{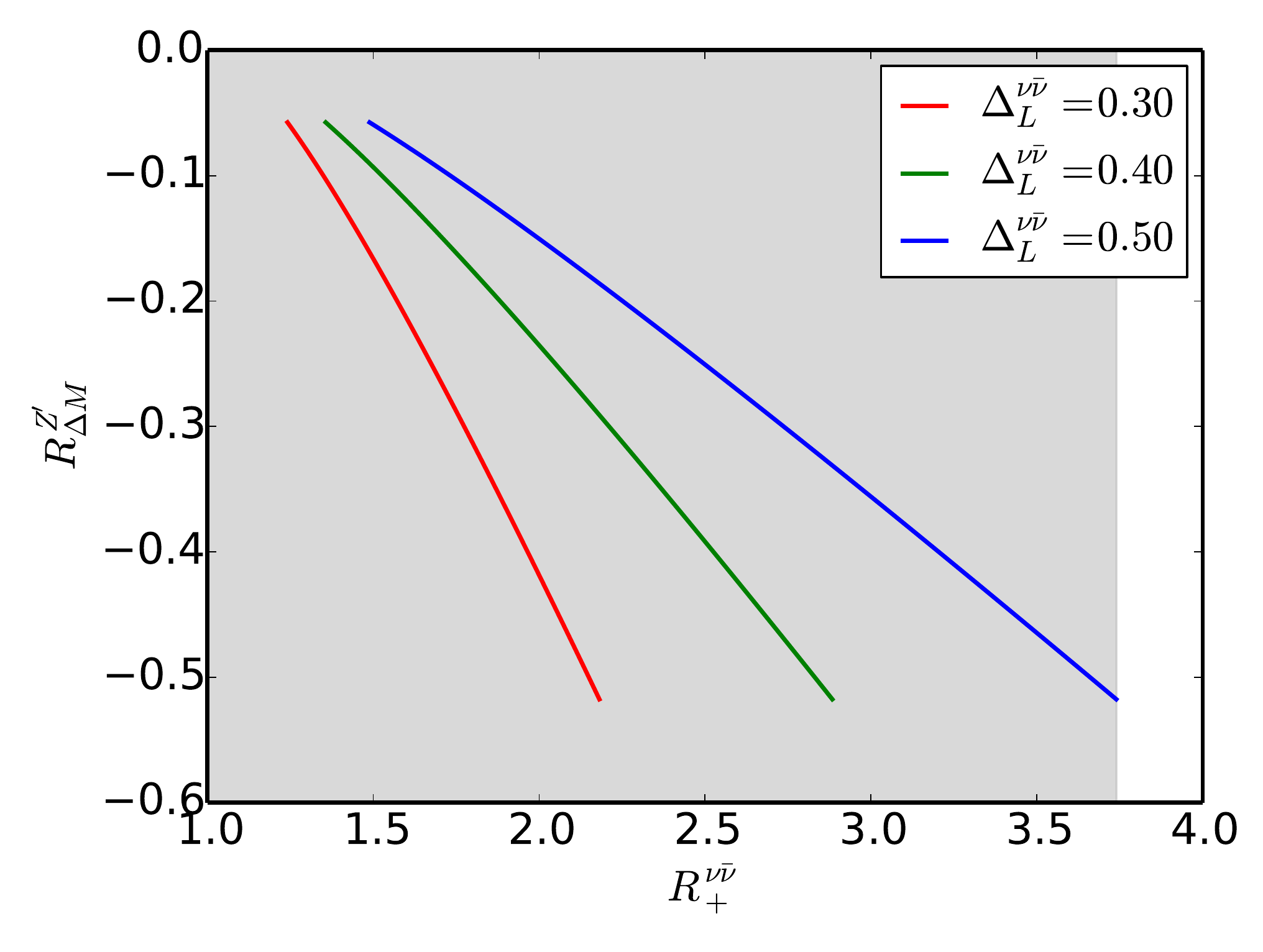}
\caption{ \it $R_+^{\nu\bar\nu}$ and  $R_0^{\nu\bar\nu}$, as functions 
of $\kepe$ for $\Delta^{\nu\bar\nu}_L(Z^\prime)=0.3,\,0.4,\, 0.5$ for {\rm QCDP} scenario. $M_{Z^\prime}=3\tev$. The dependence on $\keps$ is negligible. The upper 
black line in the lower left panel is the GN bound. In the fourth panel correlation of $R^{Z^\prime}_{\Delta M}$ with  $R_+^{\nu\bar\nu}$ is given. The experimental $1\sigma$ range for  $R_+^{\nu\bar\nu}$ in (\ref{EXP1}) is displayed by the grey band.
}\label{R5}~\\[-2mm]\hrule
\end{figure}

In Fig.~\ref{R5} we show $R_0^{\nu\bar\nu}$ and $R_+^{\nu\bar\nu}$ 
as functions of $\kepe$ and different 
values of $\Delta^{\nu\bar\nu}_L(Z^\prime)$ with the colour coding:
\be\label{codinga}
\Delta^{\nu\bar\nu}_L(Z^\prime)=0.3~{(\rm red)}, \qquad 
\Delta^{\nu\bar\nu}_L(Z^\prime)=0.4~{(\rm green)}, \qquad \Delta^{\nu\bar\nu}_L(Z^\prime)= 0.5~{(\rm blue)}\,.
\ee
We keep the diagonal quark coupling $\Delta_R^{q\bar q}(Z^\prime)=1$ but as seen 
in (\ref{IMSDPrime}) the results depend only on the ratio 
$\kepe/\Delta_R^{q\bar q}(Z^\prime)$ and it is straightforward to find out what 
happens for other values of $\Delta_R^{q\bar q}(Z^\prime)$. As the real parts 
of flavour violating couplings are small the parameter $\keps$ has no impact 
on this plot. In the third panel we show $R_0^{\nu\bar\nu}$ vs $R_+^{\nu\bar\nu}$ 
with the lower straight line representing the strict correlation between both ratios 
mentioned before and the upper line is the GN upper bound. In the
 fourth panel we show  the correlation of $R^{Z^\prime}_{\Delta M}$ with 
  $R_+^{\nu\bar\nu}$ for different values of $\Delta^{\nu\bar\nu}_L(Z^\prime)$.

 We observe that for $\Delta^{\nu\bar\nu}_L(Z^\prime)=0.5$ and 
$\kepe=1.0$ the branching ratio $\mathcal{B}(\klpn)$ is enhanced by a factor of $17.6$ and  $\mathcal{B}(\kpn)$ by  a factor of $2.4$ with respect to the SM values. Moreover $\Delta M_K$ is suppressed by roughly $25\%$.
These results are for $M_{Z^\prime}=3\tev$. Larger values of $M_{Z^\prime}$
will be considered in 
Section~\ref{BLHC}.

The NP effects for largest $\kepe$ are spectacular but probably unrealistic. There are various 
means to decrease them as can be deduced from the plots in Fig.~\ref{R5}. We give two examples
\begin{itemize}
\item
$\kepe$ in (\ref{deltaeps}) could turn out to be moderate, say $\kepe=0.5$, so that  ${\rm Im} \Delta_L^{s d}(Z^\prime)$ is smaller by a factor of two relative 
to the $\kepe=1.0$ case. 
The enhancements of $\mathcal{B}(\klpn)$ and  $\mathcal{B}(\kpn)$ will then
decrease approximately to $6.8$ and $1.5$, respectively. Moreover the suppression 
of $\Delta M_K$ will only be by $6\%$. The enhancement of $\varepsilon_K$  in (\ref{DES}) can still be kept by increasing ${\rm Re} \Delta_L^{s d}(Z^\prime)$ by 
a factor of $2$ without any visible consequences for other observables.
\item
The enhancements of  $\mathcal{B}(\klpn)$ and  $\mathcal{B}(\kpn)$ can 
be decreased by making $\Delta^{\nu\bar\nu}_L(Z^\prime)$ smaller. In fact 
this will be the only option if $\kepe$ will be required to be close 
to unity. Note, however, that modifying $\Delta^{\nu\bar\nu}_L(Z^\prime)$ will
affect the two branching ratios in a correlated manner.
\end{itemize}
It should also be kept in mind that an increase of  $\Delta^{q \bar q}_R(Z^\prime)$ to obtain larger 
enhancement of $\epe$ and smaller ${\rm Im} \Delta_L^{s d}(Z^\prime)$ is bounded by the LHC data in (\ref{LHCbound}).

Clearly, the result that the branching ratios   $\mathcal{B}(\klpn)$ and  $\mathcal{B}(\kpn)$ are enhanced because $\epe$ is enhanced is related to the 
choice of the signs of flavour diagonal quark and neutrino couplings in (\ref{COUP}). If the sign of one of these couplings is reversed but still the enhancement of $\epe$ is required, both branching ratios are suppressed along the branch 
parallel to the GN bound. But $\Delta M_K$ being governed by the square of 
the imaginary couplings is always suppressed. We summarize all cases in 
Table~\ref{tab:ZPQ6}. It should also be noticed that this pattern would not 
change if it turned out that $\varepsilon_K$ should be suppressed ($\keps < 0$), which would reverse the sign of ${\rm Re} \Delta_L^{s d}(Z^\prime)$.
Simply, because ${\rm Re} \Delta_L^{s d}(Z^\prime)$ is so much smaller than 
${\rm Im} \Delta_L^{s d}(Z^\prime)$ that its sign does not matter.

\begin{table}[!htb]
\begin{center}
\begin{tabular}{|c|c|c|c|c|c|}
\hline
$\Delta^{q \bar q}_R(Z^\prime)$ & $\Delta^{\nu\bar\nu}_L(Z^\prime)$ & $\epe$& $\mathcal{B}(\klpn)$& $\mathcal{B}(\kpn)$ & $\Delta M_K$ \\
\hline
$+$ & $+$ & $+$ & $+$ & $+$ &  $-$ \\
$-$ & $+$ & $+$ & $-$ & $-$ &  $-$ \\
$+$ & $-$ & $+$ & $-$ & $-$ &  $-$ \\
$-$ & $-$ & $+$ & $+$ & $+$ &  $-$ \\
\hline
\end{tabular}
\end{center}
\caption{Pattern of correlated enhancements ($+$) and suppressions ($-$) in $Z^\prime$ scenarios in which NP in $\epe$ is dominated by QCDP operator $Q_6$.
\label{tab:ZPQ6}}
\end{table}

In summary the two striking predictions of this scenario is the simultaneous 
enhancement or simultaneous suppression of the branching ratios for $\kpn$ and 
$\klpn$ accompanied always by the suppression of $\Delta M_K$. Finding 
the enhancement of $\kpn$ and suppression of $\klpn$ or vice versa at NA62 and 
KOPIO experiments and/or the need for an enhancement of $\Delta M_K$ by 
NP would rule out this scenario independently of what will happen with $\varepsilon_K$.

\boldmath
\subsection{$Z^\prime$ with QCD Penguin Dominance (RHS)}
\unboldmath
In the case of LHS the flavour symmetry on all diagonal RH 
quark couplings 
has to be imposed. But in the RHS the flavour diagonal couplings are left-handed and the ones in an $SU(2)_L$ doublet  must be equal to each other  due to  $SU(2)_L$ gauge 
symmetry which is still unbroken for $Z^\prime$ masses larger than few
$\tev$. Thus it is more natural in this case to generate only QCDP operators than in LHS.

We find this time 
\be\label{LOC6P}
C^\prime_6(m_c)= 1.13\frac{\Delta_R^{s d}(Z^\prime)\Delta_L^{q q}(Z^\prime)}{4 M^2_{Z^\prime}} =3.14\times 10^{-8} \left[\frac{\Delta_R^{s d}(Z^\prime)\Delta_L^{q q}(Z^\prime)}{\gev^2}\right]\left[\frac{3\tev}{M_{Z^\prime}}\right]^2
\ee
$\epe$ is again given by (\ref{eprimeZprime}) but this time
\be\label{A0ZPP}
[A_0^{\rm NP}]^R= C^\prime_6(\mu)\langle Q^\prime_6(\mu)\rangle_0, \qquad 
\langle Q^\prime_6(\mu)\rangle_0=-\langle Q_6(\mu)\rangle_0
\ee

Collecting all these results we find
\be\label{eprimeZprimeR}
\left(\frac{\varepsilon'}{\varepsilon}\right)^R_{Z^\prime}= -0.67 \,\bsi \,\left[\frac{3\tev}{M_{Z^\prime}}\right]^2 {\IM}(\Delta_R^{s d}(Z^\prime))\Delta_L^{q q}(Z^\prime)\,.
\ee
The difference in sign from (\ref{eprimeZprimeR}) is only relevant in a model
in which the flavour diagonal couplings are known or can be measured somewhere.
With the choice of the quark flavour diagonal couplings in (\ref{COUP}) there is no change in the values of flavour violating couplings except that now these are right-handed couplings 
instead of left-handed ones. Even if NP contribution to $K_L\to\mu^+\mu^-$ 
changes sign, this change is too small to be relevant because the real parts of NP couplings are small. For other choices 
of signs of flavour diagonal  couplings a DNA-Table analogous to Table~\ref{tab:ZPQ6} can be constructed by just reversing the signs of $\Delta^{q \bar q}_R(Z^\prime)$ 
and replacing it by $\Delta^{q \bar q}_L(Z^\prime)$.

\boldmath
\subsection{$Z^\prime$ with QCD Penguin Dominance (General)}
\unboldmath
\boldmath
\subsubsection{$\epe$}
\unboldmath
We will next consider scenario in which both LH and RH 
flavour violating $Z^\prime$ couplings are present.
From (\ref{eprimeZprime1}) and (\ref{eprimeZprimeR}) we find
\be\label{eprimeZprimeLR}
\left(\frac{\varepsilon'}{\varepsilon}\right)_{Z^\prime}= 0.67 \,\bsi \,\left[\frac{3\tev}{M_{Z^\prime}}\right]^2 \left[{\IM}(\Delta_L^{s d}(Z^\prime))\Delta_R^{q q}(Z^\prime)  -{\IM}(\Delta_R^{s d}(Z^\prime))\Delta_L^{q q}(Z^\prime)\right]\,.
\ee
This result is interesting in itself. If $Z^\prime$ couplings to quarks are left-right symmetric there is, similar to $K_L\to\mu^+\mu^-$, no NP contribution to 
$\epe$. In view of strong indication for $\kepe\not=0$ left-right symmetry in the $Z^\prime$ couplings to quarks has to be broken.

But there is still another reason that such a situation cannot be 
realized as  either the coupling $\Delta_L^{q q}(Z^\prime)$ or the 
coupling $\Delta_R^{qq}(Z^\prime)$ 
can be flavour universal. They cannot be both flavour universal as then 
it would not be possible to generate large flavour violating couplings in the mass eigenstate basis for any of the terms in (\ref{eprimeZprimeLR}).
But one could consider e.g. 
 $\Delta_R^{qq}(Z^\prime)$ to be flavour universal to a high degree still allowing for a strongly suppressed but non-vanishing coupling $\Delta_R^{s d}(Z^\prime)$. In any case for these reasons only one term in (\ref{eprimeZprimeLR}) will 
be important allowing in principle the solution to the $\epe$ anomaly. 
But the presence of both LH and RH flavour-violating couplings, even if one is much smaller than the other, changes the $\varepsilon_K$ and $\Delta M_K$ 
constraints through LR operators, as we have seen in the general $Z$ case. While in the latter scenario this allowed us to obtain interesting results for 
 rare decays, in $Z^\prime$ scenarios the requirement of much larger couplings 
than in the $Z$ case for solving the $\epe$ anomaly makes the  $\varepsilon_K$ and $\Delta M_K$  constraints problematic as we will discuss briefly now.

\boldmath
\subsubsection{$\varepsilon_K$ and $\Delta M_K$}
\unboldmath
We have now
\be
(\varepsilon_K)^{\rm NP}= (\varepsilon_K)^{Z^\prime}_{\rm VLL}+(\varepsilon_K)^{Z^\prime}_{\rm VRR}+(\varepsilon_K)^{Z^\prime}_{\rm LR}
\ee
where 
\be
(\varepsilon_K)^{Z^\prime}_{\rm LR}=-3.39 \times 10^6\,
\left[\frac{3\tev}{M_{Z^\prime}}\right]^2 {\rm Im}\left[\Delta_L^{s d}(Z^\prime)\Delta_R^{s d}(Z^\prime)\right]^* \,
\ee

and
\be
R^{Z^\prime}_{\Delta M}= \frac{(\Delta M_K)^{Z^\prime}_{\rm VLL}}{(\Delta M_K)_\text{exp}}+ \frac{(\Delta M_K)^{Z^\prime}_{\rm VRR}}{(\Delta M_K)_\text{exp}}+\frac{(\Delta M_K)^{Z^\prime}_{\rm LR}}{(\Delta M_K)_\text{exp}}\,
\ee
with
\be
\frac{(\Delta M_K)^{Z^\prime}_{\rm LR}}{(\Delta M_K)_\text{exp}}=-1.02 \times 10^7\, 
\left[\frac{3\tev}{M_{Z^\prime}}\right]^2
{\rm Re}\left[\Delta_L^{s d}(Z^\prime)\Delta_R^{s d}(Z^\prime)\right]^* \,.
\ee

\subsubsection{Implications} 
In view of the large coupling  ${\rm Im} \Delta_L^{s d}(Z^\prime)$ or 
 ${\rm Im} \Delta_R^{s d}(Z^\prime)$ required to solve the $\epe$ anomaly, 
NP contributions to $\varepsilon_K$ and $\Delta M_K$ in the presence of 
both LH and RH currents are very large. 
The only solution would be a very fine-tuned scenario in which the 
four couplings  ${\rm Im} \Delta_{L,R}^{s d}(Z^\prime)$ and  ${\rm Re} \Delta_{L,R}^{s d}(Z^\prime)$ take very particular values. But eventually in order to 
get significant shift in $\epe$  and satisfy $\Delta M_K$ and 
$\varepsilon_K$ constraints either RH or LH couplings 
would have to be very small bringing us back to the LHS or RHS scenario, 
respectively. 

We conclude therefore that the solution to the $\epe$ anomaly in $Z^\prime$ 
scenarios through the QCDP is only possible in the LHS or RHS if 
one wants to avoid fine-tuning of couplings.  
Then also the branching ratios for $\kpn$ and $\klpn$ can be enhanced 
in a correlated manner and $\varepsilon_K$ enhanced as favoured by the data.

This is different from  the $Z$ case, where the four enhancements in question 
could only be simultaneously obtained in the presence of LH and 
RH couplings without fine-tuning of parameters. 
\boldmath
\subsection{A heavy $G^\prime$} 
\unboldmath
We have just seen that the removal of $\epe$ anomaly in $Q_6$ scenario implies for $\Delta^{\nu\bar\nu}_L(Z^\prime)=\ord(1)$ 
large  NP effects in 
$\kpn$ and $\klpn$. It is possible that the $\epe$ anomaly will remain but no NP
will be found in $\kpn$ and $\klpn$. The simplest solution would be to set
$\Delta^{\nu\bar\nu}_L(Z^\prime)=0$.
But another possibility would 
be the presence of a heavy $G^\prime$ which does not couple to neutrinos. 
 One of the prominent examples of this type are Kaluza-Klein gluons in Randall-Sundrum scenarios that belong to the adjoint representation of the 
colour $SU(3)_c$. But here we want to consider a simplified scenario 
that  has been considered in 
the context of NP contribution to the $\Delta I=1/2$ rule in \cite{Buras:2014sba} and some of the results obtained there can be used in the case of $\epe$ 
here.

 Following \cite{Buras:2014sba}  
 we will then assume that these gauge bosons carry a common mass $M_{G^\prime}$ and being in the octet representation of $SU(3)_c$
couple to fermions in the same manner as gluons do. However, we will allow 
for different values of their left-handed and right-handed couplings. Therefore  up to the colour matrix $t^a$, the couplings to quarks will 
be again parametrized by:
\be\label{couplingsA}
\Delta_L^{s d}(G^\prime),\qquad \Delta_R^{s d}(G^\prime), \qquad \Delta_L^{q q}(G^\prime),\qquad \Delta_R^{qq}(G^\prime)\,.
\ee

As $G^\prime$ carries colour, the RG analysis is modified through the 
change of the initial conditions  at $\mu=M_{G^\prime}$ that read now \cite{Buras:2014sba}  
\begin{align}
\begin{split}
C_3(M_{G^\prime})
& = \left[-\frac{1}{6}\right]\frac{\Delta_L^{s d}(G^\prime)\Delta_L^{q q}(G^\prime)}{4 M^2_{G^\prime}}, \qquad 
C_3^\prime(M_{G^\prime})
 = \left[-\frac{1}{6}\right]\frac{\Delta_R^{s d}(G^\prime)\Delta_R^{q q}(G^\prime)}{4 M^2_{G^\prime}}
 \,,\end{split}\label{C3A}\\
\begin{split}
C_4(M_{G^\prime})
& = \left[\frac{1}{2}\right]\frac{\Delta_L^{s d}(G^\prime)\Delta_L^{q q}(G^\prime)}{4 M^2_{G^\prime}}, \qquad 
C_4^\prime(M_{G^\prime})
 = \left[\frac{1}{2}\right]\frac{\Delta_R^{s d}(G^\prime)\Delta_R^{q q}(G^\prime)}{4 M^2_{G^\prime}}
 \,,\end{split}\label{C4A}\\
\begin{split}
C_5(M_{G^\prime})
& = \left[-\frac{1}{6}\right]\frac{\Delta_L^{s d}(G^\prime)\Delta_R^{q q}(G^\prime)}{4 M^2_{G^\prime}}, \qquad 
C_5^\prime(M_{G^\prime})
 = \left[-\frac{1}{6}\right]\frac{\Delta_R^{s d}(G^\prime)\Delta_L^{q q}(G^\prime)}{4 M^2_{G^\prime}}
 \,,\end{split}\label{C5A}\\
\begin{split}
C_6(M_{G^\prime})
& = \left[\frac{1}{2}\right]\frac{\Delta_L^{s d}(G^\prime)\Delta_R^{q q}(G^\prime)}{4 M^2_{G^\prime}}, \qquad 
C_6^\prime(M_{G^\prime})
 = \left[\frac{1}{2}\right]\frac{\Delta_R^{s d}(G^\prime)\Delta_L^{q q}(G^\prime)}{4 M^2_{G^\prime}}
 \,.\end{split}\label{C6A}
\end{align}

In the LHS scenario the contributions 
of primed operators are absent. Moreover, due the non-vanishing 
value of $C_6(M_{G^\prime})$ the dominance of the operator $Q_6$  
is this time even more pronounced than in the case of a colourless $Z^\prime$. 
See Appendix~\ref{app:X}. One finds then in the LHS \cite{Buras:2014sba}
\be\label{LOC6GP}
C_6(m_c)= 1.61\frac{\Delta_L^{s d}(G^\prime)\Delta_R^{q q}(G^\prime)}{4 M^2_{G^\prime}}
\ee
with $1.61$ resulting from RG evolution from $M_{G^\prime}=3.0\tev$ down to
$m_c$.

We find then 
\be\label{eprimeGprime}
\left(\frac{\varepsilon'}{\varepsilon}\right)^L_{G^\prime}= 0.70 \,\bsi \,\left[\frac{3.5\tev}{M_{G^\prime}}\right]^2 {\IM}(\Delta_L^{s d}(G^\prime))\Delta_R^{q q}(G^\prime)\,,
\ee
where the difference in the RG factor for $M_{G^\prime}=3.0\tev$ and $M_{G^\prime}=3.5\tev$ can be neglected.

Now the upper bound on $\Delta_R^{q q}(G^\prime)$ from LHC reads 
 \cite{deVries:2014apa}
\be \left|\Delta_R^{q \bar q}(G') \right| \leq 2.0 \left[ \frac{M_{G'}}{3.5\tev}
 \right] \left[ 1 + \left(\frac{1.4\tev}{M_{G'}}\right)^2\right]\,.
\ee
Taking $\bsi=0.7$, $\Delta_R^{q q}(G^\prime)=2.0$ and $M_{G'}=3.5\tev$ 
we find then 
\be\label{eprimeGprime1}
\left(\frac{\varepsilon'}{\varepsilon}\right)^L_{G^\prime}= 0.98 \,{\IM}\Delta_L^{s d}(G^\prime)
\ee
and consequently the removal of $\epe$ anomaly  requires now
\be\label{IMGP}
{\IM}\Delta_L^{s d}(G^\prime)= 1.02 \,\kepe\left[\frac{2.0}{\Delta_R^{q \bar q}(G^\prime)}\right]\, 10^{-3},
\ee
which is by a factor of two lower than in the case of $Z^\prime$. 

As shown in \cite{Buras:2014sba} NP contributions to $\varepsilon_K$ and $\Delta M_K$ are for $M_{G^\prime}=M_{Z^\prime}$ suppressed by a colour factor of three relative to $Z^\prime$ case, but also
in this case the removal of the $\varepsilon_K$ tension together with 
(\ref{IMGP})  implies that the 
coupling $\Delta_L^{s d}(G^\prime)$ is nearly imaginary. Therefore, also in this 
case the unique prediction is the suppression of $\Delta M_K$ below it 
SM value. Yet, this suppression is  smaller relative to $Z^\prime$ case by 
roughly a factor of $17$ due to smaller value of 
${\IM}\Delta_L^{s d}(G^\prime)$, the colour factor $1/3$ in NP contribution to 
$\Delta M_K$ and the higher mass 
of $G^\prime$. Thus in contrast to the $Z^\prime$ case, NP effects in $\Delta M_K$ are fully negligible in this scenario.

While, this scenario of NP is not very exciting, we cannot exclude it at present. It should also be remarked that NP contributions to $\Delta M_K$ could 
be obtained also with $G^\prime$ by making $\Delta_R^{sd}(G^\prime)$ 
non-vanishing.

\boldmath
\subsection{$Z^\prime$ with Electroweak Penguin Dominance}\label{Q8Zprime}
\unboldmath
\boldmath
\subsubsection{The case of $\Delta_R^{q q}(Z^\prime)=\ord(1)$}
\unboldmath
We will next consider the case of a $Z^\prime$ model of the LHS type 
in which NP contribution to $\epe$ is governed by the $Q_8$ operator. The 
331 models discussed briefly in Section~\ref{sec:331} are  specific models 
belonging to this class of models.
It should be noted that as far as $\kpn$, $\klpn$, $\varepsilon_K$, $\Delta M_K$
and $K_L\to \mu^+\mu^-$ are concerned the formulae of the LH scenario in which $Q_6$ dominated NP 
in $\epe$ remain unchanged. On the other hand the formula for $\epe$ is 
modified in a very significant matter which will imply striking differences 
from QCDP scenario.

Generalizing the analysis of 331 models in \cite{Buras:2014yna} to a 
$Z^\prime$ model with arbitrary diagonal couplings we find 
\be\label{LOC8ZP}
C_8(m_c)= 1.35\, C_7(M_{Z^\prime})=1.35\,\frac{\Delta_L^{s d}(Z^\prime)\Delta_R^{q q}(Z^\prime)}{4 M^2_{Z^\prime}}
\ee
with $1.35$ resulting from RG evolution from $M_{Z^\prime}=3.0\tev$ down to
$m_c$. Here, in order to simplify the notation we denoted the RH flavour diagonal quark coupling simply by $\Delta_R^{q \bar q}(Z^\prime)$\footnote{In reality it
is a proper linear combination of diagonal up-quark and down-quark couplings that enters the  $Q_7$ and $Q_8$ penguin operators. We denote this combination 
 simply by $\Delta_R^{q \bar q}(Z^\prime)$. }.

Proceeding as in LHS Z scenario in Section~\ref{LHSZ} and replacing 
$C_8(m_c)$ in (\ref{C8}) by (\ref{LOC8ZP}) we find
instead of (\ref{eprimeZprime1}) 
\be\label{eprimeZprimeQ8}
\left(\frac{\varepsilon'}{\varepsilon}\right)^L_{Z^\prime}= 38.0 \,\bei \,\left[\frac{3\tev}{M_{Z^\prime}}\right]^2 {\IM}(\Delta_L^{s d}(Z^\prime))\Delta_R^{q \bar q}(Z^\prime)\,.
\ee

Compared to  (\ref{eprimeZprime1}) the larger overall coefficient implies 
a smaller ${\IM}\Delta_L^{s d}(Z^\prime)$ required to solve the $\epe$ anomaly.
On the other hand compared to (\ref{ZfinalL1}) in the LHS Z scenario, the sign of the model dependent 
$\Delta_R^{q q}(Z^\prime)$ can be chosen in such a manner that one can enhance 
simultaneously $\epe$ and  $\mathcal{B}(\klpn)$. This  was not possible 
in the LH Z scenario in which the diagonal quark couplings were fixed.

Setting $\bei=0.76$ we find the required couplings for the solution of $\epe$ 
 and $\varepsilon_K$ anomalies through the shifts in (\ref{deltaeps}) and (\ref{DES})
to be:
\begin{align}
 {\rm Im} \Delta_L^{s d}(Z^\prime)&= 3.5\,\left[\frac{\kepe}{\Delta_R^{q \bar q}(Z^\prime)}\right]  \left[\frac{0.76}{\bei}\right]  \left[\frac{M_{Z^\prime}}{3\tev}\right]^2 \cdot 10^{-5}\,, \label{IMSDPrime8}\\
\label{IMSDPrime8a}
{\rm Re} \Delta_L^{s d}(Z^\prime)&= -8.2\, \keps\left[\frac{\Delta_R^{q \bar q}(Z^\prime)}{\kepe}\right] \left[\frac{\bei}{0.76}\right] \cdot 10^{-4}
\end{align}
 which in view of a large $M_{Z^\prime}$ can be made consistent with the $K_L\to \mu^+\mu^-$ bound for 
$\Delta_A^{\mu\bar\mu}(Z^\prime)=\ord(1)$. Note that ${\rm Re} \Delta_L^{s d}(Z^\prime)$ is independent of  $M_{Z^\prime}$.

We observe that the signs in (\ref{IMSDPrime8}) and (\ref{IMSDPrime8a}) are the same as in 
(\ref{IMSDPrime}) and (\ref{RESDPrime}), respectively implying that also now 
$\mathcal{B}(\klpn)$ and $\mathcal{B}(\kpn)$ will be enhanced over 
their SM values but the correlation between these enhancements is
 different due to the fact that the real part of $\Delta_L^{s d}(Z^\prime)$
is larger than its 
imaginary part. Moreover NP effects implied in these decays by the $\epe$ and 
$\varepsilon_K$ anomalies turn out to be significantly   smaller than in the QCDP scenario.

\begin{figure}[!tb]
 \centering
\includegraphics[width = 0.45\textwidth]{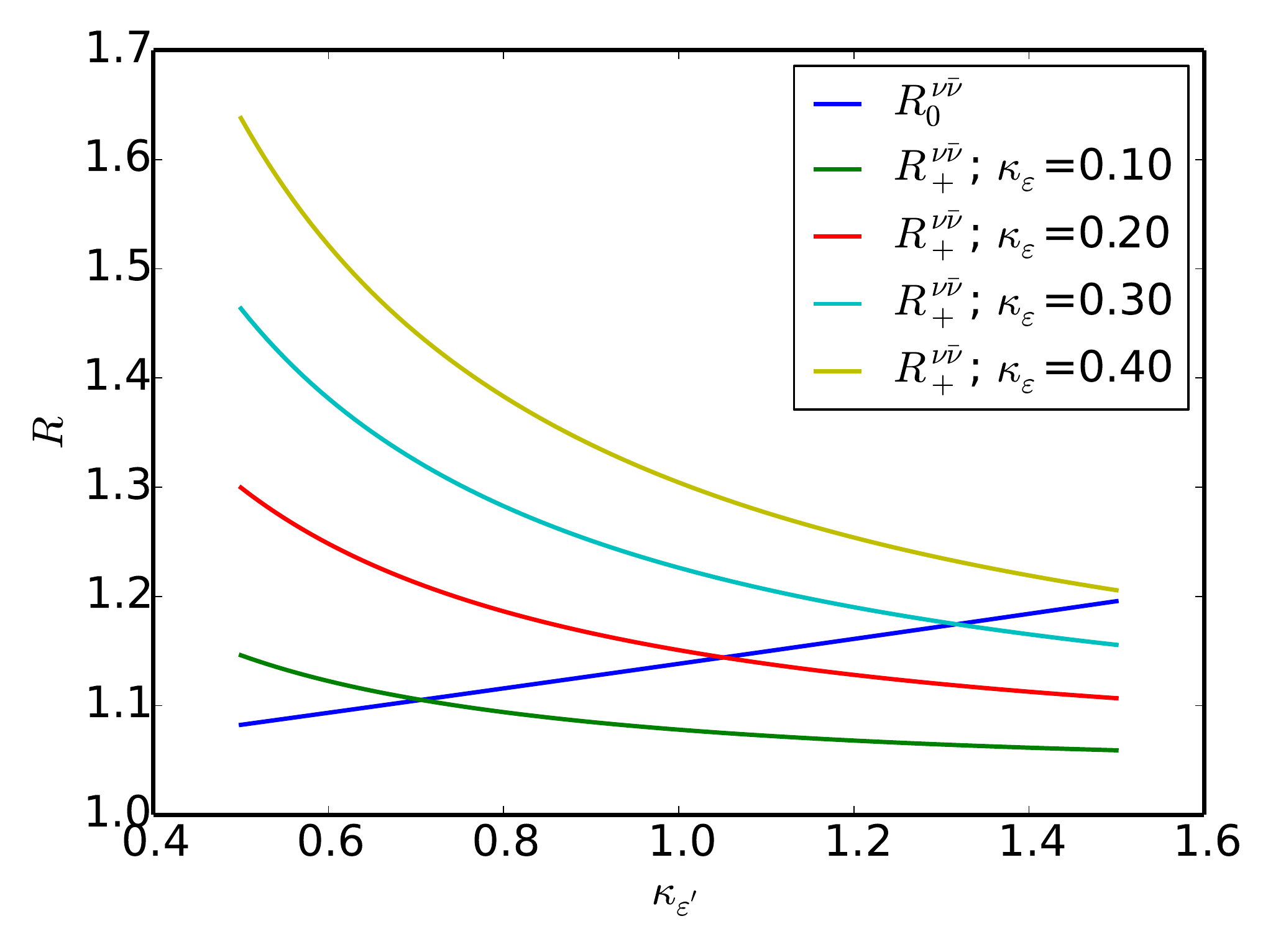}
\includegraphics[width = 0.45\textwidth]{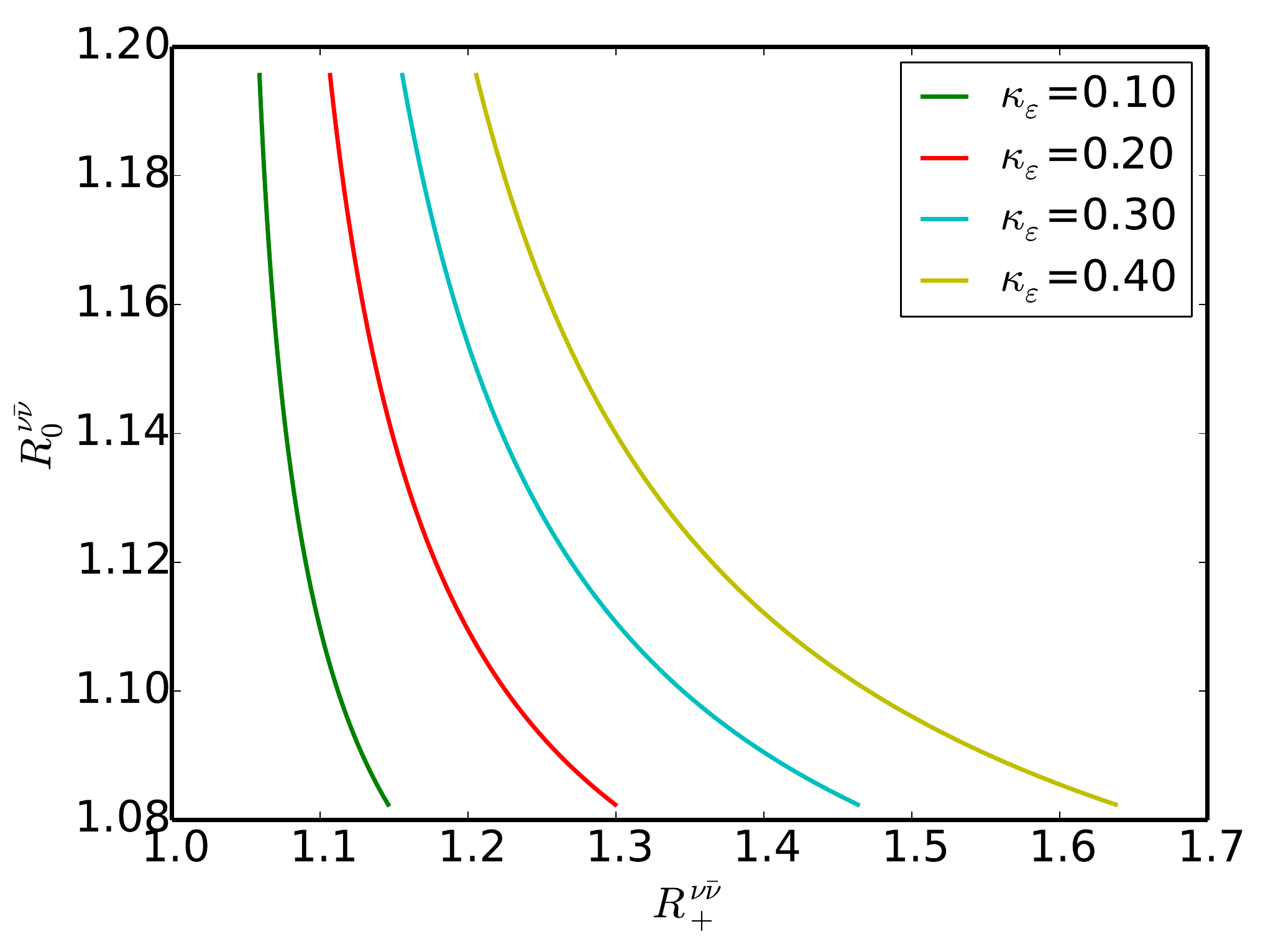}
\caption{ \it $R_0^{\nu\bar\nu}$ and  $R_+^{\nu\bar\nu}$, as functions 
of $\kepe$ for $\keps=0.1,\,0.2,\, 0.3,\, 0.4$ for EWP scenario.
}\label{R6}~\\[-2mm]\hrule
\end{figure}

In the first panel in Fig.~\ref{R6} we show $R_0^{\nu\bar\nu}$ and $R_+^{\nu\bar\nu}$  as functions of $\kepe$ and different values of $\keps$ with the colour coding in (\ref{coding}). $R_0^{\nu\bar\nu}$ is given by the {\it blue} line. Due 
 to smaller values of imaginary parts required for a given $\kepe$ to fit
the data on $\epe$ the implied
NP effects in both ratios are smaller than in the QCDP case and therefore 
we set this time 
$\Delta_L^{\nu\bar\nu}(Z^\prime)=0.5$. On the other hand in contrast to QCDP 
case, where there is no dependence on $\keps$, the enhancement of 
$\mathcal{B}(\kpn)$  in EWP scenario strongly depends on the ratio $\keps/\kepe$.  This is also seen in the second panel in which we present the results 
of the first panel as $R_0^{\nu\bar\nu}$ vs $R_+^{\nu\bar\nu}$. This result has 
a pattern similar to the first $Z$ example in Fig.~\ref{R4ab} but NP effects 
are now much smaller.

\begin{table}[!htb]
\begin{center}
\begin{tabular}{|c|c|c|c|c|c|c|}
\hline
$\Delta^{q \bar q}_R(Z^\prime)$ & $\Delta^{\nu\bar\nu}_L(Z^\prime)$ & $\epe$& $|\varepsilon_K|$ &
 $\mathcal{B}(\klpn)$& $\mathcal{B}(\kpn)$ & $\Delta M_K$ \\
\hline
$+$ & $+$ & $+$ & $+$ &  $+$ & $+$ &  $+$ \\
$+$ & $+$ & $+$ & $-$ &  $+$ & $-$ &  $+$ \\
\hline
$-$ & $+$ & $+$ & $+$ &  $-$ & $-$ &  $+$ \\
$-$ & $+$ & $+$ & $-$ &  $-$ & $+$ &  $+$ \\
\hline
$+$ & $-$ & $+$ & $+$ &  $-$ & $-$ &  $+$ \\
$+$ & $-$ & $+$ & $-$ &  $-$ &  $+$ &   $+$ \\
\hline
$-$ & $-$ & $+$ & $+$ &  $+$ & $+$ &  $+$ \\
$-$ & $-$ & $+$ & $-$ &  $+$ & $-$ &  $+$ \\
\hline
\end{tabular}
\end{center}
\caption{Pattern of correlated enhancements ($+$) and suppressions ($-$) in $Z^\prime$ scenarios in which NP in $\epe$ is dominated by EWP operator $Q_8$.
\label{tab:ZPQ8}}
\end{table}

Interestingly, we find that $\Delta M_K$ is  exclusively {\it enhanced} as opposed 
to its suppression in QCDP scenario as seen in (\ref{SUPQ6}). This 
time we have
\be\label{SUPQ8}
R^{Z^\prime}_{\Delta M}({\rm EWP})\equiv\frac{(\Delta M_K)^{Z^\prime}_{\rm VLL}}{(\Delta M_K)_\text{exp}}=3.6\cdot 10^{-2}\, \keps^2 \left[\frac{\Delta_R^{q \bar q}(Z^\prime)}{\kepe}\right]^2 \left[\frac{0.76}{\bei}\right]^2\left[\frac{3\tev}{M_{Z^\prime}}\right]^2 
\,.
\ee
We note that dependence on $\kepe$ and $\Delta_R^{q \bar q}(Z^\prime)$ is different 
than in  (\ref{SUPQ6}) and the enhancement depends on $\keps$. But the striking 
difference is in the size of the effect and its  $M_{Z^\prime}$ dependence. NP contribution to 
$\Delta M_K$ is now in the ballpark of a few percent only and  decreases with increasing $M_{Z^\prime}$ as opposed to the QCD 
penguin case, where it is sizable and increases with increasing $M_{Z^\prime}$ thereby significantly suppressing $\Delta M_K$. See Fig.~\ref{R5}.

Clearly, similar to the case of the $Q_6$ dominance, the result that the branching ratios   $\mathcal{B}(\klpn)$ and  $\mathcal{B}(\kpn)$ are enhanced because $\epe$ is enhanced is related to the 
choice of the signs of flavour diagonal quark and neutrino couplings in (\ref{COUP}). If the sign of one of these couplings is reversed but still the enhancement of $\epe$ is required, both branching ratios are suppressed. But if 
in addition we require that $\varepsilon_K$ is suppressed then $\kpn$ is enhanced again but $\klpn$ suppressed. We show various possibilities in Table~\ref{tab:ZPQ8}. This table differs from Table~\ref{tab:ZPQ6} because the flip of 
the sign of ${\rm Re} \Delta_L^{s d}(Z^\prime)$, caused by the flip of the sign of NP contribution to $\varepsilon_K$, now matters as  ${\rm Re} \Delta_L^{s d}(Z^\prime)$  is much larger than in the QCDP case. 
This has no impact on $\klpn$ but
changes enhancement of $\kpn$ into its suppression and vice versa.
On the other hand $\Delta M_K$ being governed this time by the square of 
the real couplings is always enhanced as opposed to the QCDP case.

The striking prediction of this scenario is also the prediction that in the 
case 
of a negative shift of $\varepsilon_K$ by NP one of the branching ratios 
must be enhanced with respect to the SM and the other suppressed, a feature 
which is not possible in the QCDP scenario.

In view of these rather different results it should be possible to distinguish the QCDP 
and EWP mechanisms in $Z^\prime$ scenarios when the situation with $\epe$ and  $\varepsilon_K$ anomalies will be clarified and the data 
on $\mathcal{B}(\klpn)$ and $\mathcal{B}(\kpn)$ will be available. 
The improved  knowledge of $\Delta M_K$ will be important  in this distinction due to the different signs and sizes of NP contributions to $\Delta M_K$ in these two scenarios.

\boldmath
\subsubsection{The case of $\Delta_R^{q q}(Z^\prime)\ll 1$}
\unboldmath
The pattern just discussed is modified if  $\Delta_R^{q q}(Z^\prime)$ is strongly 
suppressed for some dynamical reason. For instance choosing $\Delta_R^{q q}(Z^\prime)=0.01$
we find
\be\label{IMSDPrime8b}
 {\rm Im} \Delta_L^{s d}(Z^\prime)= 3.5\,\kepe \, 10^{-3}\,, \qquad 
{\rm Re} \Delta_L^{s d}(Z^\prime)= -8.2\, \frac{\keps}{\kepe}\, 10^{-6}
\ee
which as seen in (\ref{IMSDPrime} ) and (\ref{RESDPrime}) is rather similar 
to the case of the QCDP so that enhancements of $\kpn$ and $\klpn$ 
are correlated on a branch parallel to the GN bound. Yet, it should be 
emphasized that in the EWP case this can only be obtained 
by choosing the coupling $\Delta_R^{q q}(Z^\prime)$ to be very small, while in 
the case of QCDP one obtains this result automatically as in order to satisfy 
 all flavour bounds while solving the $\epe$ anomaly $\Delta_R^{q q}(Z^\prime)$ must be $\ord(1)$.

We will not consider the cases of RHS and of a general scenario.
Due to the arbitrary values of diagonal couplings not much new can be learned 
relative to the cases already considered. But such scenarios could be of interest in specific models.

\boldmath
\subsection{The Impact of  $Z-Z^\prime$ mixing}
\unboldmath
Generally, in a $Z^\prime$ scenario, the  $Z-Z^\prime$ mixing will generate
in the process of electroweak symmetry breaking flavour-violating tree-level 
$Z$ contributions. As an example a non-vanishing coupling 
\be\label{ZZmix}
\Delta^{sd}_L(Z)=\sin\xi \, \Delta^{sd}_L(Z^\prime)
\ee
will be generated with $\xi$ being the mixing angle. This mixing is bounded 
by LEP data to be $\ord(10^{-3})$ and has the structure
\be
\sin\xi= c_\text{mix}\frac{M_Z^2}{M_{Z^\prime}^2}
\ee
with $c_\text{mix}$ being a model dependent factor. Inserting (\ref{ZZmix}) 
into (\ref{C7Z}) and performing RG evolution from $M_Z$ to $m_c$ we find the 
$Z$ contribution to $C_8$ generated by this mixing:
\be
C_8(m_c)
 = - 0.76\,c_\text{mix}\left[\frac{4 g_2 s_W^2}{6 c_W}\right]
\frac{\Delta_L^{s d}(Z^\prime)}{4 M^2_{Z^\prime}}\,. 
\ee

Comparing with (\ref{LOC8ZP}) we observe that $Z$ contribution has eventually 
the same dependence on $M_{Z^\prime}$ as $Z^\prime$ contribution. Which of these 
contributions is larger depends on the model dependent values of $c_\text{mix}$ 
and $\Delta_R^{q\bar q}$ which govern $Z^\prime$ contribution to $\epe$.

A simple class of models that illustrates these effects are 331 
models in which $c_\text{mix}$ 
and $\Delta_R^{q\bar q}$ are given in terms of fundamental parameters of these 
models. A detailed  analysis of the impact of $Z-Z^\prime$ mixing on flavour observables in 331 models, including $\epe$, 
can be found in \cite{Buras:2014yna} and in a recent update in \cite{Buras:2015kwd}.
One finds after taking electroweak precision constraints into account, that in most of these models for a large range of parameters 
$Z^\prime$ contributions dominate but if one aims for precision the effects 
of $Z$ contributions cannot be neglected. A brief summary of the analysis in 
\cite{Buras:2015kwd} is given in Section~\ref{sec:331}.

\boldmath
\subsection{$Z^\prime$ Outside the Reach of the LHC}\label{BLHC}
\unboldmath
\subsubsection{QCD Penguin Dominance}
Our discussion in Section~\ref{Q6LHC} has revealed  interesting $M_{Z^\prime}$ 
dependence of flavour observables when the $\epe$ and $\varepsilon_K$ constraints in (\ref{deltaeps}) and (\ref{DES}) are imposed. They originate in the fact that these constraints taken together
 require the following  $M_{Z^\prime}$ dependence of the $Z^\prime$ couplings
\begin{itemize}
\item
${\rm Im} \Delta_L^{s d}(Z^\prime)$ must increase as $M^2_{Z^\prime}$,
\item
${\rm Re} \Delta_L^{s d}(Z^\prime)$ must be independent of  $M_{Z^\prime}$\,.
\end{itemize}
Therefore the increase of  $M_{Z^\prime}$ assures the dominance of imaginary couplings. This should be no surprise as both quantities are CP-violating and the 
imaginary couplings have to be larger in order to explain the anomalies 
in $\epe$ and $\varepsilon_K$ at larger $M_{Z^\prime}$.

As a consequence of this $M_{Z^\prime}$ dependence
\begin{itemize}
\item
$\mathcal{B}(\klpn)$ is independent of  $M_{Z^\prime}$ because ${\rm Im}X_\text{eff}$ is independent of it. The suppression by $1/M^2_{Z^\prime}$ is cancelled by 
the increase of ${\rm Im} \Delta_L^{s d}(Z^\prime)$.
\item
But  ${\rm Re}X_\text{eff}$ decreases with increasing $M_{Z^\prime}$ and consequently in principle $\mathcal{B}(\kpn)$ will  decrease. But this effect is so small 
in QCDP scenario that similar to $\mathcal{B}(\klpn)$ also this branching ratio will be 
independent of  $M_{Z^\prime}$ with NP contributing only through ${\rm Im}X_\text{eff}$.
\item
On the other hand the branching ratio for $K_L\to \mu^+\mu^-$ decreases with increasing  $M_{Z^\prime}$ as it depends only on real parts of the couplings.
\end{itemize}

As a result of this pattern the correlation between $\mathcal{B}(\klpn)$ and $\mathcal{B}(\kpn)$
will be confined to the line parallel to the GN bound. But what is interesting 
is that this correlation will depend only on $\kepe$ and is independent of 
$M_{Z^\prime}$. Comparing (\ref{IMSDPrime}) with (\ref{RESDPrime}) we find that 
the real parts are comparable with imaginary ones only for  $M_{Z^\prime}< 500\gev$ which is clearly excluded by the LHC. Therefore, for fixed $\kepe$ and 
$\keps$ nothing will change as far as $\kpn$ and $\klpn$ are concerned when 
 $M_{Z^\prime}$ is increased but the constraint from $K_L\to\mu^+\mu^-$ will be 
weaker.

Yet, these  {\it scaling laws} cannot be true forever as for sufficiently 
large $M_{Z^\prime}$ the couplings will enter non-perturbative regime and our
calculations will no longer apply. Moreover, 
these {\it scaling laws}  did not yet take into account the bound on NP contributions 
to $\Delta M_K$. Indeed as seen in (\ref{SUPQ6}) this contribution increases 
in QCDP scenario with increasing $M_{Z^\prime}$ and suppresses $\Delta M_K$ that 
is positive in the SM.  At some value 
of $M_{Z^\prime}$ this NP effect will be too large for the theory to agree with 
experiment. The rescue could come from increased value of $\Delta_R^{q \bar q}(Z^\prime)$ 
 or decreased value of $\kepe$. This simply means that when $\Delta M_K$ 
constraint is taken into account there is an upper bound on $\kepe$ which 
becomes stronger with increasing $M_{Z^\prime}$. Or in other words at 
sufficiently 
high values of  $M_{Z^\prime}$ it will not be possible to explain the anomalies 
in question and with further increase of  $M_{Z^\prime}$  NP will decouple. 

At this stage one should emphasize that for more precise calculations,
  when going to much higher values of  $M_{Z^\prime}$, well
 above  the LHC scales, RG effects represented by numerical factors like 
$1.13$, $1.61$ and $1.35$ for QCDP, $G^\prime$ and EWP contributions to $\epe$
 valid for $M_{Z^\prime}=3\tev$ have to be modified as collected in Table~\ref{RG} in  Appendix~\ref{app:X}. For  $M_{Z^\prime}=100\tev$ they are increased typically by a factor of $1.3-1.5$ relative to  $M_{Z^\prime}=3\tev$.

\begin{figure}[!tb]
 \centering
\includegraphics[width = 0.60\textwidth]{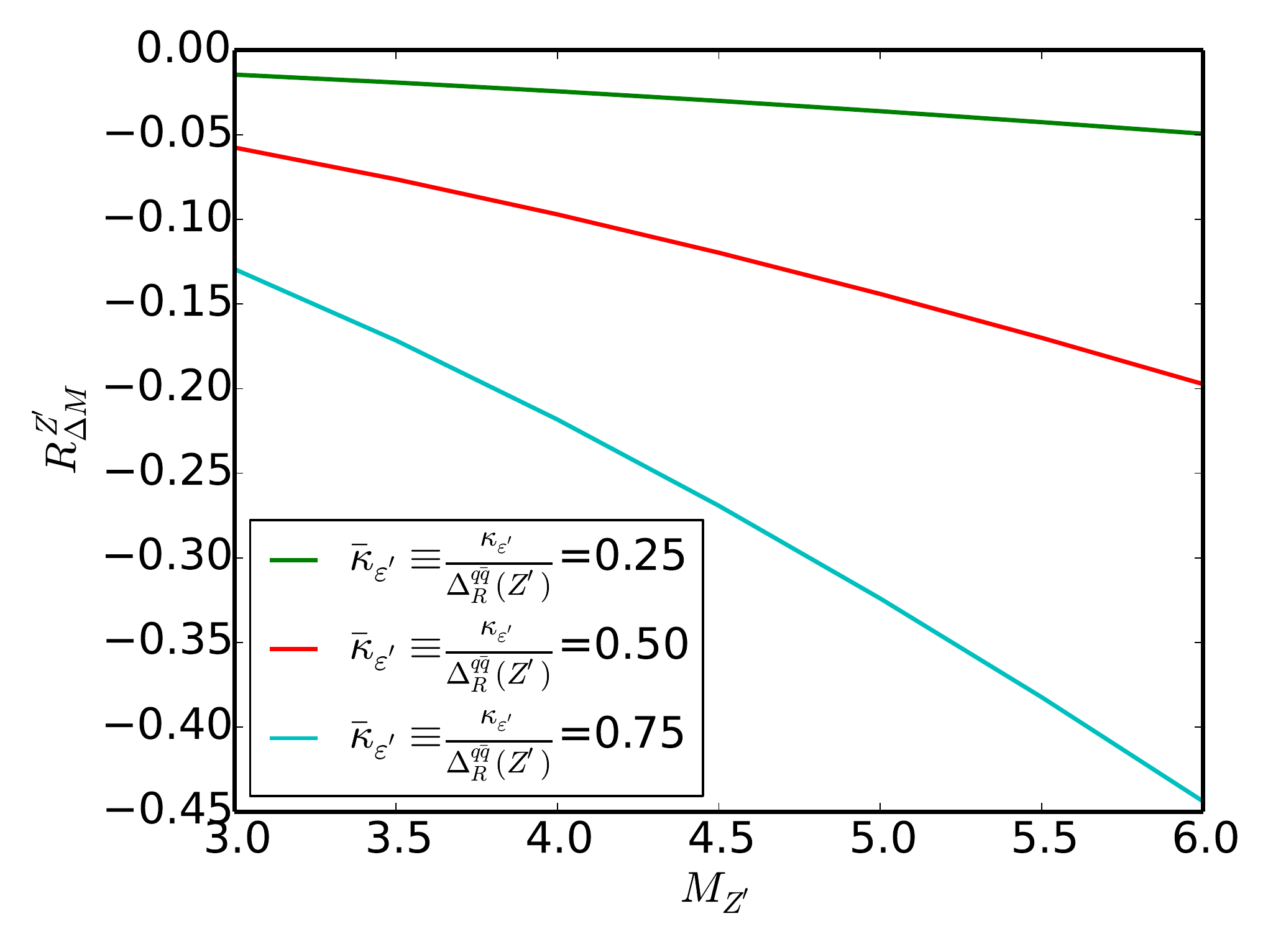}
\caption{ \it $R^{Z^\prime}_{\Delta M}({\rm QCDP})$ as a function of $M_{Z^\prime}$ for different values of $\bar\kappa_{\varepsilon^\prime}$.}\label{R7}~\\[-2mm]\hrule
\end{figure}

Formula (\ref{SUPQ6}) generalized to include RG corrections for 
 $M_{Z^\prime}\ge 3\tev$ reads
\be\label{SUPQ6LARGE}
R^{Z^\prime}_{\Delta M}({\rm QCDP})=-0.23\, \left[\frac{1.13}{r_{65}(M_{Z^\prime})}\right]^2
\left[\frac{\kepe}{\Delta_R^{q \bar q}(Z^\prime)}\right]^2
\left[\frac{M_{Z^\prime}}{3\tev}\right]^2\,\left[\frac{0.70}{\bsi}\right]^2\,,
\ee
with $r_{65}(M_{Z^\prime})$ given in Table~\ref{RG}. In Fig~\ref{R7} we show 
$R^{Z^\prime}_{\Delta M}({\rm QCDP})$ as a function of $M_{Z^\prime}$ for 
different values of the ratio
\be
\bar\kappa_{\varepsilon^\prime}\equiv \frac{\kepe}{\Delta_R^{q \bar q}(Z^\prime)}\,.
\ee
We observe that already for  $M_{Z^\prime}=6\tev$ the shift in $\Delta M_K$ is 
large unless $\kepe$ is at most $0.5$ or $\Delta_R^{q \bar q}(Z^\prime)> 1.0$. 
As seen  in (\ref{LHCbound}) for  $M_{Z^\prime}=6\tev$ the choice 
$\Delta_R^{q \bar q}(Z^\prime)= 2.0$ is still consistent with LHC bounds.

The bound on  $\Delta M_K$ in question can be avoided  to some extent by going to the general $Z^\prime$ 
scenario which contains also $\Delta^{sd}_R(Z^\prime)$. This allows, as 
suggested in \cite{Buras:2014sba},  to weaken with some fine-tuning
$\Delta M_K$ constraint while solving $\epe$ anomaly. But, in 
order to perform a meaningful analysis the value of $\Delta M_K$ in the SM  must be known significantly better than it is the case now. In particular if 
suppressions of $\Delta M_K$ are not allowed one will have to abandon 
this scenario. Then, as we will discuss soon, the EWP scenario 
would be favoured.

It should also be emphasized that in a concrete model additional constraints 
could come from other observables, in particular from observables like 
the $B^0_{s,d}-\bar B^0_{s,d}$ mass differences $\Delta M_{s,d}$ and CP asymmetries  $S_{\psi K_S}$ and $S_{\psi\phi}$  which 
could further change the scaling laws. We refer to \cite{Buras:2015kwd} for 
scaling laws found in the context of 331 models.

\subsubsection{Electroweak  Penguin Dominance}
The main difference in this scenario is the finding that for $\Delta_R^{q \bar q}=\ord(1)$ and $M_{Z^\prime}=3\tev$
\be
{\rm Re}\Delta^{sd}_R(Z^\prime)\gg  {\rm Im}\Delta^{sd}_R(Z^\prime)\,.
\ee
With increasing $M_{Z^\prime}$ this hierarchy becomes for fixed $(\kepe,\keps)$ smaller as ${\rm Im}\Delta^{sd}_R(Z^\prime)$ increases with $M_{Z^\prime}$  and 
${\rm Re}\Delta^{sd}_R(Z^\prime)$ is independent of it. By comparing 
(\ref{IMSDPrime8}) and (\ref{IMSDPrime8a}) we learn that the magnitudes
of both couplings are equal for
\be\label{EQUAL}
M_{Z^\prime}=14.5\,\sqrt{\keps}\, \left[\frac{\Delta_R^{qq}}{\kepe}\right]\,\tev
\ee

 But even for these values of $M_{Z^\prime}$
\begin{itemize}
\item
The correlation between $\kpn$ and $\klpn$ is away from the branch parallel
to the GN bound.
\item
NP contribution to $\Delta M_K$ has opposite sign to the one in QCDP scenario and $\Delta M_K$ is 
enhanced and not suppressed relative to its SM value. Moreover this enhancement is at the level of a few percent only and  decreases with increasing $M_{Z^\prime}$ so that possible problems with $\Delta M_K$ constraint 
encountered in QCDP scenario are absent here unless future precise estimates 
of $\Delta M_K$ in the SM will require sizable contribution from NP.
\end{itemize}

Clearly a precise value of $\Delta M_K$ in the SM  will be crucial in order to 
see whether the enhancement of $\Delta M_K$ predicted here is consistent 
with the data. In particular if an
 enhancement of $\Delta M_K$ is not allowed,  one will have to abandon 
this scenario.

\boldmath
\subsection{Summary of NP  Patterns in $Z^\prime$ Scenarios}
\unboldmath
The striking difference from $Z$ scenarios, known already from our previous 
studies, is the increased importance of the constraints from $\Delta F=2$ 
observables. This has two virtues in the presence of the $\epe$ constraint:
\begin{itemize}
\item
The real parts of the couplings are determined for not too a large $\keps$ from the $\varepsilon_K$ constraint, which is 
theoretically cleaner than the $K_L\to\mu^+\mu^-$ constraint that was more important in LHS and RHS Z scenarios.
\item
There is a large hierarchy between real and imaginary parts of the  flavour 
violating couplings implied by anomalies in  both $Q_6$ and $Q_8$ scenarios.
But as seen in   (\ref{IMSDPrime}) and (\ref{RESDPrime}) in the case of $Q_6$ and in (\ref{IMSDPrime8}) and  (\ref{IMSDPrime8a})  in the case of $Q_8$ this hierarchy is different unless
the $\varepsilon_K$ anomaly is absent.
\end{itemize}

Because of a significant difference in the manner QCDP and electroweak 
penguins enter $\epe$, there are  striking differences in the implications 
for the correlation between $\kpn$ and $\klpn$ in these
two NP  scenarios if significant NP contributions to $\epe$ are required:
\begin{itemize}
\item
In the case of QCDP scenario the correlation between 
$\mathcal{B}(\klpn)$ and $\mathcal{B}(\kpn)$ takes place along the branch 
parallel to the GN bound. Moreover, this feature is independent of $M_{Z^\prime}$.
\item
In the EWP scenario this correlation  proceeds away from  this branch for diagonal couplings 
$\ord(1)$ if 
NP in $\varepsilon_K$ is present with the departure from this branch increasing 
with the increased NP effect in $\varepsilon_K$. But with increasing  $M_{Z^\prime}$ this branch will be approached although it is reached for $M_{Z^\prime}$ well 
beyond the LHC scales unless $\keps$ is very small. See (\ref{EQUAL}).
\item
For fixed values of the neutrino and  diagonal quark couplings the 
predicted enhancements of $\mathcal{B}(\klpn)$ and $\mathcal{B}(\kpn)$ 
are much larger when NP in QCDP is required to remove the 
$\epe$ anomaly. This is simply related to the fact that QCDP operators 
are less effective in enhancing $\epe$ than EWP operators and 
consequently the imaginary parts of the flavour violating couplings 
are required to be larger.
\item
Finally, a striking difference is the manner in which NP affects $\Delta M_K$ in 
these two scenarios. In QCDP scenario  $\Delta M_K$ is {\it suppressed} and this 
effect increases with increasing  $M_{Z^\prime}$ whereas in the EWP scenario 
 $\Delta M_K$ is {\it enhanced} and this effect decreases with increasing
 $M_{Z^\prime}$ as long as real couplings dominate.  Already on the basis of this property one could differentiate between 
these two scenarios when the SM prediction for $\Delta M_K$ improves.
\end{itemize}

The plots in Figs.~\ref{R5} and \ref{R6} show clearly the differences between QCDP and EWP scenarios.

\boldmath
\section{Hybrid Scenarios: $Z$ and $Z^\prime$}\label{sec:4}
\unboldmath
Similar to flavour non-universal $Z^\prime$ couplings to quarks in the flavour 
basis, leading to flavour-violating $Z^\prime$ couplings to quarks in the  mass eigenstate basis, also flavour-violating $Z$ couplings can be generated. As an example in Randall-Sundrum scenario such couplings result from the
breakdown of flavour universality of $Z$ couplings to quarks in the flavour basis.
But such couplings are also generated in the presence of new heavy fermions with
different transformation properties under the SM gauge group than the ordinary 
quarks and leptons. The mixing of these new fermions with the ordinary fermions 
generates flavour-violating $Z$ couplings in the mass eigenstate basis. In order to avoid anomalies the most natural here are vector-like fermions. 

In the presence of both $Z$ and $Z^\prime$ contributions, independently of 
the dynamics behind their origin, the formulae for all observables discussed 
by us can be straightforwardly generalized using the formulae of previous 
sections. We find then
\be\label{H1}
\left(\frac{\varepsilon'}{\varepsilon}\right)_{\rm NP}=
\left(\frac{\varepsilon'}{\varepsilon}\right)_Z+
\left(\frac{\varepsilon'}{\varepsilon}\right)_{Z^\prime}\,.
\ee
$Z$  contribution is given in the case of the LHS 
 in (\ref{ZfinalL1}). $Z^\prime$ contribution in the QCDP scenario is given in
 (\ref{eprimeZprime1}) and the one for EWP in (\ref{eprimeZprimeQ8}). 

Similar we have 
\be\label{epsZZprime}
(\varepsilon_K)^{\rm NP}_{\rm VLL}= (\varepsilon_K)^Z_{\rm VLL}+(\varepsilon_K)^{Z^\prime}_{\rm VLL}
\ee
with the two contributions given in (\ref{eps}) and  (\ref{epsZprime}), respectively.
Next 
\be\label{DM2aZZprime}
\frac{(\Delta M_K)^{\rm NP}_{\rm VLL}}{(\Delta M_K)_\text{exp}} =
\frac{(\Delta M_K)^{Z}_{\rm VLL}}{(\Delta M_K)_\text{exp}} +\frac{(\Delta M_K)^{Z^\prime}_{\rm VLL}}{(\Delta M_K)_\text{exp}}
\ee
 with $Z$ and $Z^\prime$ contributions given in (\ref{DM2a}) and  (\ref{DMZprime}), respectively.

In the case of $\kpn$ and $\klpn$ we simply have 

\be\label{H1a}
{\rm Re}\,X^{\rm NP}_{\rm eff}= {\rm Re}\,X_{\rm eff}(Z)+{\rm Re}\,X_{\rm eff}(Z^\prime)
\ee
and
\be\label{H2}
{\rm Im}\,X^{\rm NP}_{\rm eff}= {\rm Im}\,X_{\rm eff}(Z)+{\rm Im}\,X_{\rm eff}(Z^\prime)\,,
\ee
where different contributions can be found in (\ref{c2}), (\ref{c3}), (\ref{c7})
 and (\ref{c8}).

In order to get a rough idea about the relative size of $Z$ and $Z^\prime$ 
contributions to different observables we assume first that their 
contributions to  $\epe$ and $\varepsilon_K$ are related as follows
\be
\left(\frac{\varepsilon'}{\varepsilon}\right)_Z= a\,
\left(\frac{\varepsilon'}{\varepsilon}\right)_{Z^\prime}\,, \qquad 
(\varepsilon_K)^Z_{\rm VLL}=b\, (\varepsilon_K)^{Z^\prime}_{\rm VLL}
\ee
with $a$ and $b$ being real, positive and $\ord(1)$.

Proceeding as in the previous sections we find for $Z$ couplings now
\be\label{IMSDab}
 {\rm Im} \Delta_L^{s d}(Z)= -5.0\,\frac{a}{(1+a)}\kepe \left[\frac{0.76}{\bei}\right]\cdot 10^{-7}\,
\ee
and 

\be\label{RE1ab}
{\rm Re}\Delta_L^{s d}(Z)=4.7\,\frac{b(1+a)}{a(1+b)} \left[\frac{\keps}{\kepe}\right] \left[\frac{\bei}{0.76}\right] \cdot 10^{-5}\,,
\ee
which for $a\gg 1$ and $b\gg 1$ reduce to (\ref{IMSD}) and (\ref{RE1}), respectively.

For $Z^\prime$ scenario with QCDP dominance in $\epe$ we find
\be\label{IMSDPrime1}
 {\rm Im} \Delta_L^{s d}(Z^\prime)= \frac{2.1}{(1+a)}\, \left[\frac{\kepe}{\Delta_R^{q\bar q}(Z^\prime)} \right]\left[\frac{0.70}{\bsi}\right] \left[\frac{M_{Z^\prime}}{3\tev}\right]^2\cdot 10^{-3}\,
\ee
and 
\be\label{RESDPrime1}
{\rm Re} \Delta_L^{s d}(Z^\prime)=-1.4 \,\frac{(1+a)}{(1+b)} \keps\,
\left[\frac{\Delta_R^{q \bar q}(Z^\prime)}{\kepe}\right] \left[\frac{\bsi}{0.70}\right]  \cdot 10^{-5}\,,
\ee
which for $a=b=0$ reduce to (\ref{IMSDPrime}) and (\ref{RESDPrime}), respectively. 

Correspondingly for $Z^\prime$ scenario with EWP dominance in $\epe$ we find
\begin{align}
 {\rm Im} \Delta_L^{s d}(Z^\prime)&= \frac{3.5}{(1+a)}\,\left[\frac{\kepe}{\Delta_R^{q \bar q}(Z^\prime)}\right] \left[\frac{0.76}{\bei}\right]  \left[\frac{M_{Z^\prime}}{3\tev}\right]^2 \cdot 10^{-5}\,, \label{IMSDPrime81}\\
\label{IMSDPrime8a1}
{\rm Re} \Delta_L^{s d}(Z^\prime)&= -8.2\,\,\frac{(1+a)}{(1+b)} \keps\left[\frac{\Delta_R^{q \bar q}(Z^\prime)}{\kepe}\right] \left[\frac{\bei}{0.76}\right]  \cdot 10^{-4}\,,
\end{align}
which reduce for $a=b=0$ to (\ref{IMSDPrime8}) and (\ref{IMSDPrime8a1}), respectively.

The comparison of (\ref{RE1ab})  with  (\ref{RESD})  tells us that $b$ cannot 
be $\ord(1)$ but rather $b\le 0.05$. We conclude therefore that 
\begin{itemize}
\item
$Z^\prime$  dominates the contribution of NP to $\varepsilon_K$ which is consistent with previous general analysis \cite{Buras:2012jb}.
\end{itemize}

On the other hand assuming that $a=\ord(1)$ the inspection of the formulae 
for the quantities in (\ref{DM2aZZprime})-(\ref{H2}) implies the following 
pattern of $Z$ and $Z^\prime$ contributions.

In the QCDP scenario:
\begin{itemize}
\item
NP contribution to $\Delta M_K$ is dominated by $Z^\prime$.
\item
${\rm Re}\,X^{\rm NP}_{\rm eff}$ is dominated by $Z$
\item
${\rm Im}\,X^{\rm NP}_{\rm eff}$ is dominated by $Z^\prime$.
\end{itemize}

In the EWP scenario:
\begin{itemize}
\item
 $Z$ and $Z^\prime$ contributions to $\Delta M_K$ are of the same order.
\item
Contributions from $Z$ and $Z^\prime$ to 
${\rm Re}\,X^{\rm NP}_{\rm eff}$ are 
of the same order but as they have opposite signs for $\Delta_R^{q\bar q}(Z^\prime)\Delta_L^{\nu\bar\nu}(Z^\prime)>0$ the branching ratio for $\kpn$  can be enhanced or suppressed if 
necessary, dependently on the values of parameters involved.
\item
${\rm Im}\,X^{\rm NP}_{\rm eff}$ is dominated by $Z$.
\end{itemize}

Now, in many model constructions the full $Z^\prime$ and $Z$ flavour-violating 
couplings, both real and imaginary parts, are related by a common real factor
so that the ratio of real couplings of $Z^\prime$ and $Z$ equals the ratio of imaginary ones. Imposing this on the couplings obtained above we find the 
relations between the parameters  $a$ and $b$ and knowing already that $b\ll 1$  we can find out the size of $a$ in different scenarios.  In the case of QCDP scenario we obtain
\be
a^2=b \, \frac{1.4}{(\Delta_R^{q\bar q}(Z^\prime))^2} \cdot 10^4 \left[\frac{M_{Z^\prime}}{3\tev}\right]^2 \left[\frac{0.70}{\bsi}\right]^2\left[\frac{\bei}{0.76}\right]^2\,,
\qquad {(\rm QCDP)}
\ee
and for EWP one
\be
a^2=b \, \frac{4.0}{(\Delta_R^{q\bar q}(Z^\prime))^2} \left[\frac{M_{Z^\prime}}{3\tev}\right]^2 \,.
\qquad {(\rm EWP)}
\ee

For $b\le 0.05$ one has then in the QCDP scenario for $Z^\prime$
\be
a\le \frac{26.5}{\Delta_R^{q\bar q}(Z^\prime)}\left[\frac{M_{Z^\prime}}{3\tev}\right]\,,
\qquad {(\rm QCDP)}\,,
\ee
where we neglected  the difference between $\bsi$ and 
$\bei$. Evidently, unless the contribution of $Z$ to $\varepsilon_K$ is totally 
negligible, $Z$ generally dominates NP contribution to $\epe$ and therefore $Q_8$ operator wins over $Q_6$ as expected already from arguments given at the beginning of our paper.  This also implies 
that now, as opposed to the case of $a=\ord(1)$ discussed above, contributions from $Z$ and $Z^\prime$  to 
${\rm Im}\,X^{\rm NP}_{\rm eff}$ can be for sufficiently large $a$  of the same order. But, as they have  opposite signs for $\Delta_L^{\nu\bar\nu}(Z^\prime)>0$, the branching ratio for $\klpn$  can be enhanced or  suppressed if 
necessary, dependently on the values of parameters involved.

On the other hand in EWP
scenario  both contributions are dominated by $Q_8$ operator. We find then
\be
a\le \frac{0.45}{\Delta_R^{q\bar q}(Z^\prime)}\left[\frac{M_{Z^\prime}}{3\tev}\right]\,,
\qquad {(\rm EWP)}\,,
\ee
so that in this case $a=\ord(1)$ and  $Z$ contribution to $\epe$ can be comparable to the $Z^\prime$ one. Consequently the  pattern of NP effects listed for EWP above applies. Only for very suppressed $\Delta_R^{q\bar q}(Z^\prime)$ and large 
$M_{Z^\prime}$ the contribution from $Z$ can again dominate as in QCDP scenario.

Without a specific model it is not possible to make more concrete predictions 
but it is clear that the structure of NP contributions is more involved than 
in previous scenarios. One should also keep in mind that in certain models contributions from loop diagrams could play some role, in particular in models in 
which vector-like quarks and new heavy scalars are present.

\section{Selected Models}\label{sec:5}
\subsection{Preliminaries}
Here we will  briefly describe results in specific models as presented already in the literature. Some of these analyses have to be  updated but 
the pattern of NP effects in the described NP scenarios is known and consistent
with pattern found in previous sections.
\subsection{Models with Minimal Flavour Violation}
The recent analysis of simplified models, in particular those with minimal flavour 
violation and those with $U(2)^3$ symmetry shows that one should not expect a
solution to $\epe$ anomaly from such models \cite{Buras:2015yca}. This is 
also the case of the MSSM with MFV 
as already analyzed in \cite{Buras:2000qz} and 
 NP effects in this scenario must be presently even smaller due to the increase 
of the supersymmetry scale.

\subsection{A Model with a Universal Extra Dimension}
In this model NP contribution to $\epe$ depends on only one new parameter: 
the compactification radius. One finds $\epe$ to be smaller than 
its SM value independently of the compactification radius \cite{Buras:2003mk}.
Consequently this model  is disfavoured by $\epe$ and there is no need to discuss its implications for other observables.
\subsection{Littlest Higgs Model with T-Parity}
In this model NP contributions to $\kpn$, $\klpn$ and $\epe$ are governed by EWP and in particular  the ones in $\epe$ by the operator $Q_8$.
The model has the same operator structure as the SM and FCNC processes 
appear first at one loop level. But effectively for these three observables 
the model has the structure of $Z$ LH scenario with the coupling $\Delta_L^{s d}(Z)$ resulting from one-loop contributions involving new fermions and gauge 
bosons. Moreover NP contributions to $\varepsilon_K$ are governed by new box diagrams. Consequently the correlation 
with between $\kpn$, $\klpn$, $\epe$  is more involved than in simple models 
discussed by us. But the anticorrelation between $\epe$ and $\klpn$ is also 
valid here.

The most recent analysis in  \cite{Blanke:2015wba} shows that
\begin{itemize}
\item
The LHT model agrees well with the data on $\Delta F=2$ observables and is 
capable of removing some slight tensions between the SM predictions and the data. In particular $\varepsilon_K$ can be enhanced.
\item
If $\epe$ constraint is ignored the most interesting departures from SM predictions can be found for $\kpn$ and 
$\klpn$ decays.  {An enhancement} of the branching ratio for $\kpn$  by a factor of two relative to the SM prediction is {still} possible. An even 
larger enhancement in the case of $\klpn$ is allowed. But as expected from 
the properties of $Z$ LH scenario of Section~\ref{LHSZ}, when the $\epe$ constraint is taken into 
account the necessary enhancement of $\epe$ requires rather strong suppression 
of  $\klpn$. On the other hand  significant shifts of $\kpn$ with respect to SM are then no longer allowed. Figs.~6 and 7 in   \cite{Blanke:2015wba} show this behaviour in a spectacular manner.
\end{itemize}
\subsection{331 Models}\label{sec:331}
The 331 models are based on the gauge group $SU(3)_C\times SU(3)_L\times U(1)_X$. 
In these models new contributions to $\epe$ and other flavour observables are  dominated by tree-level exchanges of a $Z^\prime$ with non-negligible contributions from tree-level $Z$ exchanges generated through the $Z-Z^\prime$ mixing. The size of these NP effects depends not only on $M_{Z^\prime}$ but in particular on a parameter $\beta$, which distinguishes between various 331 
models, on fermion representations under the gauge group and a
parameter $\tan\bar\beta$ present in the $Z-Z^\prime$ mixing  \cite{Buras:2014yna}. The ranges of these parameters are restricted by electroweak precision 
tests and flavour data, in particular from $B$ physics. A recent updated 
analysis has been presented in \cite{Buras:2015kwd}.

The model belongs to the class of $Z^\prime$ models with LH flavour-violating 
couplings with only a small effect from $Z-Z^\prime$ mixing in  $\epe$ that is dominated by the operator $Q_8$.  But, in contrast to the general case analyzed 
in Section~\ref{Q8Zprime}, the diagonal couplings are known in a given 331 model
as functions of $\beta$. The new analysis in \cite{Buras:2015kwd} shows that the 
impact of  a required  enhancement of $\epe$  on other flavour observables 
is significant. The main findings of \cite{Buras:2015kwd} for $M_{Z^\prime}=3\tev$ are as follows:
\begin{itemize}
\item
Among seven 331 models singled out in \cite{Buras:2014yna} through electroweak precision study only three can provide significant shift of $\epe$ but for 
$M_{Z^\prime}=3\tev$ 
 not larger than $6\times 10^{-4}$, that is $\kepe\le 0.6$.
\item
Two of them can simultaneously suppress $B_s\to\mu^+\mu^-$ but do not offer 
the explanation of the suppression of the Wilson coefficient $C_9$ in $B\to K^*
\mu^+\mu^-$ (the so-called LHCb anomaly).
\item
On the contrary the third model offers partial explanation of this anomaly simultaneously enhancing $\epe$ but does not provide suppression of  $B_s\to\mu^+\mu^-$ 
which could be required when the data improves and the inclusive value of
$\vcb$ will be favoured.
\item
NP effects in $\kpn$, $\klpn$ and $B\to K(K^*)\nu\bar\nu$ are found to be  small. This could be challenged by NA62, KOPIO and Belle II experiments in this decade.
\end{itemize}

 Interestingly, the special flavour structure of 331 models implies that  even for  $M_{Z^\prime}=30\tev$ a shift of $\epe$ up to $8\times 10^{-4}$ and a significant shift in 
$\varepsilon_K$ can be obtained, while the effects in other flavour observables
are small. This makes these models appealing in view of the possibility of accessing masses of $M_{Z^\prime}$ far beyond the LHC reach. The increase in the 
maximal shift in $\epe$ is caused by RG effects summarized in Table~\ref{RG}.
But for  $M_{Z^\prime}>30\tev$ the $\Delta M_K$ constraint becomes important 
and NP effects in $\epe$ decrease as $1/M_{Z^\prime}$.

\subsection{More Complicated Models}
Clearly there are other possibilities involving new operators. In particular 
it has been pointed out that in general supersymmetric models $\epe$ 
can receive important contributions from chromomagnetic penguin operators
\cite{Masiero:1999ub,Buras:1999da}. In fact in 1999 this contribution could 
alone be responsible for experimental value of $\epe$ subject to very large uncertainties 
of the relevant hadronic matrix element. This assumed the masses of squarks 
and gluinos in the ballpark of $500\gev$. With the present lower bounds on 
these masses in the ballpark of few TeV, it is unlikely that these operators 
can still provide a significant contribution to $\epe$ when all constraints from other observables are taken into account. Similar comments apply to 
other models like the one in \cite{Davidson:2007si}, 
Randall-Sundrum models \cite{Gedalia:2009ws} and left-right symmetric models
\cite{Bertolini:2012pu}, where in the past $\epe$ could receive important contributions from chromomagnetic penguins. It would be interesting to update such analyses, in particular when
the value of $\bsi$ and the hadronic matrix elements of chromomagnetic penguins will be better known.

\boldmath
\section{New Physics in ${\rm Re} A_0$ and  ${\rm Re} A_2$}\label{DeltaI}
\unboldmath
The calculations of $K\to\pi\pi$ isospin amplitudes  ${\rm Re} A_0$ and  ${\rm Re} A_2$ within the SM, related to the $\Delta I=1/2$ rule in 
(\ref{IDI12}) have been the subject of many efforts in the last
40 years. Some aspects of these efforts have been recalled in \cite{Buras:2014maa}. Here we only note that  both the dual approach to QCD  \cite{Buras:2014maa} and lattice approach  \cite{Blum:2015ywa}
obtain satisfactory results for the amplitude  ${\rm Re}A_2$ within the SM leaving there only small room for NP contributions. 

On the other 
hand, whereas in the large $N$ approach one finds \cite{Buras:2014maa}
\be\label{DRULEN}
\left(\frac{{\rm Re}A_0}{{\rm Re}A_2}\right)_{{\rm dual~QCD}}=16.0\pm1.5 \,,
\ee
the most recent result from the RBC-UKQCD collaboration reads \cite{Bai:2015nea}
\be\label{DRULEL}
\left(\frac{{\rm Re}A_0}{{\rm Re}A_2}\right)_{{\rm lattice~QCD}}=31.0\pm6.6 \, .
\ee
Due to large error in the lattice result, both results are compatible with 
each other and both signal that this rule follows dominantly from the 
QCD dynamics related to current-current operators. In addition both leave 
room for sizable NP contributions. But, from the present perspective 
only lattice simulations 
can provide  precise value of  ${\rm Re}A_{0}$ one day, so that we will know 
whether some part of this rule at the level of $(20-30)\%$, as signalled 
by the result in (\ref{DRULEN}), originates in NP contributions. 

This issue has been addressed in \cite{Buras:2014sba}, where it has been 
demonstrated that a QCDP generated by a heavy $Z^\prime$ and in particular a heavy $G^\prime$ in the reach of the 
LHC could be responsible for the missing piece in ${\rm Re}A_0$ in 
(\ref{DRULEN}) but this requires a very large fine-tuning of parameters in 
order to satisfy the experimental bounds from $\Delta M_K$ and $\varepsilon_K$ 
 even in the absence of the $\epe$ anomaly, which was unknown at the time 
of the publication in \cite{Buras:2014sba}. 

The point is that a sizable contribution of $Q_6$ operator to ${\rm Re}A_0$ 
requires ${\rm Re}\Delta^{sd}_L(Z^\prime)=\ord(1)$ which as stressed in 
 \cite{Buras:2014sba} violates $\Delta M_K$ by many orders of magnitude if 
only LH flavour-violating currents are considered. In the presence of 
$\epe$ anomaly, which requires  ${\rm Im}\Delta^{sd}_L(Z^\prime)=\ord(10^{-3})$ the results of previous sections show that also $\varepsilon_K$ 
constraint is then violated by several orders of magnitude.

The only possible solution is the introduction of both LH and RH flavour 
violating currents with real and imaginary parts of both currents 
properly  chosen so that both $\Delta M_K$ and $\varepsilon_K$ constraints 
are satisfied and significant contribution to  ${\rm Re}A_0$  is obtained.  
The $\epe$ anomaly provides additional constraint but as seen in Fig.~4 of 
\cite{Buras:2014sba} in the case of $Z^\prime$ scenario and in Section~6 of 
that paper in the case of $G^\prime$ scenario,  satisfactory results for 
${\rm Re}A_0$, $\epe$, $\varepsilon_K$ and $\Delta M_K$ can be obtained.
But it should be kept in mind that such a solution requires very high 
fine-tuning of parameters and on the basis of the analysis in \cite{Buras:2014sba} the central value of lattice result in (\ref{DRULEL})  is too far away 
from the data that one could attribute this difference to any NP.

In summary, the future precise lattice calculations  will hopefully tell us whether there 
is some NP contributing significantly to ${\rm Re}A_0$. This would enrich the 
present analysis as one would have, together with  ${\rm Re}A_2$,  two 
additional constraints. But on the basis of \cite{Buras:2014sba} 
it is rather unlikely that this NP is represented 
by heavy $Z^\prime$ or $G^\prime$ unless the nature allows for very high fine-tunings.

\section{2018 Visions}\label{Vision}
With all these results at hand we can dream about the discovery of NP in 
$\kpn$  by the   NA62 experiment:
\be\label{NA62}
\mathcal{B}(\kpn)= (18.0\pm 2.0)\cdot 10^{-11},\qquad ({\rm NA62},~2018)\,.
\ee
Indeed, looking at the grey bands in several figures presented by us, such a result would be truly tantalizing with a big impact on our field.

We will next assume that the lattice values of $\bsi$ and $\bei$ will 
be  close to our central values
\be
\bsi\approx 0.70,\qquad \bei\approx 0.76,
\ee
and that the CKM parameters are such that $\kepe\approx 1.0$ will be required.

Concerning $\varepsilon_K$ we will consider two scenarios, one with 
$\keps=0.4$ and the other with $\keps=0$, that is no $\varepsilon_K$ anomaly.
\boldmath
\subsection{$\kepe=1.0$ and $\keps=0.4$}
\unboldmath
Inspecting the results of previous sections, we conclude the following
\begin{itemize}
\item
$Z$ scenarios with only LH and RH couplings will be ruled out as they cannot 
accommodate $\varepsilon_K$ anomaly with  $\keps=0.4$ unless at one loop level 
in the presence of new heavy fermions or scalars significant contributions to $\varepsilon_K$ would be generated. Then in principle the rates for $\kpn$ in LHS and RHS
could be made consistent with the result in (\ref{NA62}).
\item
It is clearly much easier to reproduce the data in the general $Z$ scenario. 
In fact as seen in Fig.~\ref{R4ab} both examples presented by us could 
accommodate the result in (\ref{NA62}), explain simultaneously $\epe$ and 
$\varepsilon_K$ anomalies and predict an enhancement of $\mathcal{B}(\klpn)$
by a factor of two to three in the first example and by an order of magnitude in 
the second example.
\item
As seen in Fig.~\ref{R5} the QCDP generated by $Z^\prime$ can reproduce the result in (\ref{NA62}) 
for $\Delta_L^{\nu\bar\nu}(Z^\prime)=0.5$ and $\kepe=1.0$. This then implies
the enhancement of the rate for $\klpn$ by a factor of $15-20$: good news 
for KOPIO. Moreover, $\varepsilon_K$ can be made consistent with the 
data independently of $\kepe$.
\item
Interestingly, as seen in Fig.~\ref{R6}, EWP generated by $Z^\prime$ will not be able to explain 
the result in (\ref{NA62}) unless the coupling $\Delta_R^{q\bar q}(Z^\prime)$ is very strongly suppressed below unity. Also NP effects in $\klpn$ are predicted to
be small.
\end{itemize}

\boldmath
\subsection{$\kepe=1.0$ and $\keps=0.0$}
\unboldmath
If $\epsilon_K$ can be explained within the SM the main modification 
relative to the case of $\keps\not=0$ is that in all scenarios the correlation between
$\klpn$ and $\kpn$ takes place on the branch parallel to GN bound in strict
correlation with $\epe$ or equivalently $\kepe$. Yet, there are differences
between various scenarios:
\begin{itemize}
\item
$Z$ scenarios with only LH or RH currents and EWP$(Z^\prime)$ scenario with 
$\Delta_R^{q\bar q}=\ord(1)$ imply SM-like values for $\mathcal{B}(\kpn)$, 
far below the result in (\ref{NA62}).
\item
For QCDP$(Z^\prime)$ nothing changes relative to the previous case and 
interesting 
results for both rare decay branching ratios can be obtained. Also the 
general $Z$ case can work in view of sufficient number of free parameters.
EWP scenario can also work provided $\Delta_R^{q\bar q}(Z^\prime)$ is very 
strongly suppressed below unity.
\end{itemize}

In summary we observe that a NA62 measurement of  $\mathcal{B}(\kpn)$ 
in the ballpark of the result in (\ref{NA62}) will be able to make 
reduction of possibilities with the simplest scenario being QCDP generated 
through a tree-level $Z^\prime$ exchange. But then the crucial question will 
be what is the value of $\Delta M_K$ in the SM.

\section{Outlook and Open Questions}\label{sec:6}
Our general analysis of $\epe$ and $\varepsilon_K$ in models with tree-level flavour-violating $Z$ and $Z^\prime$ exchanges shows that such dynamics could 
be responsible for the observed  $\epe$ anomaly with interesting implications 
for other flavour observables in the $K$ meson system. In particular it 
could shed some light on NP in $\varepsilon_K$ and $\Delta M_K$.
Our results are 
summarized in numerous plots and two tables which show that the inclusion 
of other observables can clearly distinguish between various possibilities.

Except for the case of $Z$ scenarios with only left-handed (LHS) and right-handed (RHS) flavour violating currents, where $K_L\to \mu^+\mu^-$ bound was the 
most important constraint on the real parts of flavour violating couplings, 
in the remaining scenarios the pattern of flavour violation was  governed in the large part of the parameter space entirely by CP-violating quantities: $\epe$ and $\varepsilon_K$. NP effects in them where described by two parameters 
$\kepe$ and $\keps$ as defined in (\ref{deltaeps}) and (\ref{DES}).

In LH and RH $Z^\prime$ scenarios the role of $\epe$ was to determine imaginary parts of flavour violating $Z^\prime$ couplings. Having them, the role of 
$\varepsilon_K$ was to determine the real parts of these couplings. These 
then had clear implications for other observables, in particular for the  branching ratios for $\kpn$ and $\klpn$ and for $\Delta M_K$. The case of general 
scenarios with LH and RH couplings is more involved but also here we could 
get a picture what is going on.

From our point of view the most interesting results of this work are as 
follows:
\begin{itemize}
\item
In LH and RH $Z$ scenarios the enhancement of $\epe$ implies uniquely suppression 
of $\klpn$. Moreover, NP effects in $\varepsilon_K$ and $\Delta M_K$ are very 
small.
\item
Simultaneous enhancements of $\epe$, $\varepsilon_K$ and of the branching 
ratios for $\kpn$ and $\klpn$ in $Z$ scenarios are only possible in the presence 
of both LH and RH flavour violating couplings. As far as $\epe$ and $\klpn$ 
are concerned this finding has already been reported in \cite{Buras:2015yca} but
our new analysis summarized in Figs.~\ref{R2a}-\ref{R4ab} extended this case 
significantly.
\item
If the enhancement of $\epe$ in $Z^\prime$ scenarios is governed by QCDP operator $Q_6$, the
branching ratios for $\klpn$ and $\kpn$ are strictly correlated, as seen in 
Fig.~\ref{R5}, along the branch parallel to the GN bound. They can be both enhanced or suppressed dependently on the signs of diagonal quark and neutrino couplings that are
relevant for $\epe$ and these rare decays, respectively. Various possibilities 
are summarized in Table~\ref{tab:ZPQ6}. There we see that in these scenarios $\Delta M_K$ is uniquely {\it suppressed} relative to its SM value. This is directly related to the dominance 
of {\it imaginary} parts of flavour violating couplings necessary to provide sufficient enhancement of $\epe$. The suppression of $\Delta M_K$ could turn out
to be a challenge for this scenario implying possibly an upper bound on $\kepe$ 
as we stressed in Section~\ref{BLHC} and illustrated in Fig.~\ref{R5} and in 
particular in Fig.~\ref{R7}.
On the other hand the role of $\varepsilon_K$ is 
smaller, even if solution to possible tensions there are offered. 
\item
But two  
messages on QCDP scenario from our analysis are clear. If $\mathcal{B}(\kpn)$ will turn out one day to be enhanced by NP relative to the SM prediction and
 $\mathcal{B}(\klpn)$ suppressed or vice versa, the QCDP scenario will not be 
able to describe it. This is also the case when $\Delta M_K$ in the SM will be
found below its experimental value.
\item
Rather different pattern of the implications of the $\epe$ anomaly are found 
in $Z^\prime$ scenarios in which the enhancement of $\epe$ is governed by EWP operator $Q_8$. In particular the correlation between 
$\kpn$ and $\klpn$ depends on the size and the sign of NP contribution to $\varepsilon_K$
which was not the case of QCDP scenario. Moreover, as seen in Fig.~\ref{R6},
the structure of this correlation is very different from the one in Fig.~\ref{R5}, although also in this case, for  $\keps>0$, both branching ratios are enhanced with respect their SM values. They can also be simultaneously suppressed for
different signs of diagonal quark and neutrino couplings. Various possibilities 
are summarized in Table~\ref{tab:ZPQ8}.
\item
But as we emphasized and shown in this table, for $\keps<0$ in the EWP scenario, the enhancement of $\kpn$  implies  simultaneous suppression of $\klpn$ or vice versa which is not 
possible in the QCDP scenario. Moreover, in this scenario
$\Delta M_K$ is uniquely {\it enhanced} relative to its SM value. This is directly related to the dominance 
of the {\it real} parts of flavour violating couplings necessary to provide sufficient contribution to $\varepsilon_K$ in the presence of an enhancement of $\epe$. But, as opposed to the QCDP case, this NP effect is small.
\end{itemize}

These results show that a good knowledge of $\Delta M_K$ within the SM would 
help a lot in distinguishing between QCDP and EWP scenarios. 
Presently the uncertainties in $\Delta M_K$ from both perturbative contributions \cite{Brod:2011ty} and long distance calculations both within 
large $N$ approach \cite{Buras:2014maa} and lattice simulations \cite{Bai:2014cva} are too large to be able to conclude whether positive or negative shift, if any, in    $\Delta M_K$ from NP is favoured.

The dominant part of our $Z^\prime$ study concerned $M_{Z^\prime}$ in the reach of 
the LHC but as we demonstrated in Section~\ref{BLHC}, $\epe$ will give us an 
insight into short distance dynamics even if $Z^\prime$ cannot be seen by ATLAS 
and CMS experiments. We also restricted our study to the $K$ meson system.
In concrete models there are correlations between observables in $K$ meson 
system and other meson systems. An example are models with minimal flavour 
violation. But as shown in \cite{Buras:2015yca}, in  such models NP effects in $\epe$, $\varepsilon_K$, $\kpn$ and $\klpn$ are small. Larger effects can be obtained in  LHT and 331 models  for which the most recent analyses can be found in 
 \cite{Blanke:2015wba} and \cite{Buras:2015kwd}, respectively. 

There is no doubt that in the coming years $K$ meson physics will strike back, 
in particular through improved estimates of SM predictions for $\epe$, 
$\varepsilon_K$, $\Delta M_K$ and $K_L\to\mu^+\mu^-$ and through crucial 
measurements of the branching ratios for $\kpn$ and $\klpn$. Correlations 
with other meson systems, lepton flavour physics, electric dipole moments 
and other rare processes should allow us to identify NP at very short distance 
scales \cite{Buras:2013ooa} and we should hope that this physics will also be directly seen at the LHC. 

Let us then end our paper by listing most pressing questions for the coming 
years. On the theoretical side we have:
\begin{itemize}
\item
{\bf What is the value of $\kepe$?} Here the answer will come not only from lattice QCD but also through improved values of the CKM parameters, NNLO QCD corrections and an improved understanding of FSI and isospin breaking effects. The NNLO QCD corrections should be available soon. 
 The recent analysis in the large $N$ approach in \cite{Buras:2016fys}
 indicates  that 
 FSI are likely to be important for the $\Delta I=1/2$  rule  in agreement with previous 
studies \cite{Pallante:1999qf,Pallante:2000hk,Buras:2000kx,Buchler:2001np,Buchler:2001nm,Pallante:2001he}, but much less relevant for $\epe$. 
\item
{\bf What is the value of $\keps$?} Here the reduction of CKM uncertainties is most important. But the most recent analysis in \cite{Blanke:2016bhf} indicates that if no NP is  
present in $\varepsilon_K$, it is expected to be found in $\Delta M_{s,d}$.
\item
{\bf What is the value of $\Delta M_K$ in the SM?} Here lattice QCD should provide
useful answers.
\item
{\bf What are the precise values of $\RE A_2$ and $\RE A_0$?} Again lattice QCD 
will play the crucial role here.
\end{itemize}

On the experimental side we have:
\begin{itemize}
\item
{\bf What is $\mathcal{B}(\kpn)$ from NA62?} We should know it in 2018.
\item
{\bf What is $\mathcal{B}(\klpn)$ from KOPIO?} We should know it around the year 2020.
\item
{\bf Do $Z^\prime$, $G^\prime$ or other new particles with masses in the reach of 
the LHC exist?} We could know it already this year.
\end{itemize}

Definitely there are exciting times ahead of us!

\section*{Acknowledgements}
First of all I would like to thank Robert Buras-Schnell for a very careful and 
critical reading of the manuscript and decisive help in numerical calculations, 
in particular for constructing all the plots present in this paper. I thank 
 Jean-Marc G{\'e}rard for illuminating discussions.
The collaboration with Fulvia De Fazio on $\epe$ in the context of 331 models
 and brief discussions with Christoph Bobeth are also highly appreciated.
This research was done and financed in the context of the ERC Advanced
Grant project ``FLAVOUR''(267104) and was partially supported by the DFG
cluster of excellence ``Origin and Structure of the Universe''.

\appendix
\boldmath
\section{More Information on Renormalization Group Evolution}\label{app:X}
\unboldmath
\subsection{QCD Penguins}
We follow here \cite{Buras:2014sba} and consider first the case of $Z^\prime$ 
with flavour universal diagonal quark couplings. In this case the QCDP 
$Q_5$ and $Q_6$ have to be considered. The mixing with other operators is 
neglected and we work in LO approximation.

Denoting then by $\vec C(M_{Z^\prime})$ the column vector with components given 
by the  Wilson coefficients $C_5$ and $C_6$ at $\mu=M_{Z^\prime}$ we 
find their values at $\mu=m_c$ by means of
\begin{equation}\label{cmub1}
 \vec C(m_c)=\hat U(m_c,M_{Z^\prime}) \vec C(M_{Z^\prime})
\end{equation}
where
\begin{equation}\label{cmub1a}
 \hat U(m_c,M_{Z^\prime})=\hat U^{(f=4)}(m_c,m_b)\hat U^{(f=5)}(m_b, m_t) \hat U^{(f=6)}(m_t, M_{Z^\prime})
\end{equation}
and \cite{Buras:1998raa}
\begin{equation}\label{u0vd0} 
\hat U^{(f)}(\mu_1,\mu_2)= \hat V
\left({\left[\frac{\alpha_s(\mu_2)}{\alpha_s(\mu_1)}
\right]}^{\frac{\vec\gamma^{(0)}}{2\beta_0}}
   \right)_D \hat V^{-1}.   \end{equation}

The relevant $2\times 2$  one-loop anomalous dimension matrix in the basis
 $(Q_5,Q_6)$ can  be extracted from the known  $6\times6$ matrix \cite{Gilman:1979bc} and is given as follows
\begin{equation}
\hat\gamma_s(\alpha_s)=\hat\gamma_s^{(0)}\frac{\alpha_s}{4\pi}, \qquad
\hat \gamma^{(0)}_s = 
\left(
\begin{array}{cc}
2 & -6  \\ \svs
-f\frac{2}{9} & -16
   + f\frac{2}{3}
\end{array}
\right)
\label{reduced}
\end{equation}
with $f$ being the number of quark flavours.

The matrix $\hat V$ diagonalizes ${\hat\gamma^{(0)T}}$
\begin{equation}\label{ga0d} 
\hat\gamma^{(0)}_D=\hat V^{-1} {\hat\gamma^{(0)T}} \hat V\,,
  \end{equation}
$\vec\gamma^{(0)}$ is the vector containing the diagonal elements of
the diagonal matrix :
\begin{equation}\label{g120d} \hat\gamma^{(0)}_D=
 \left(\begin{array}{cc} \gamma^{(0)}_+ & 0 \\
                          0 & \gamma^{(0)}_-
    \end{array}\right)  \end{equation}
 and
\be
\beta_0= \frac{33-2f}{3}\, .
\ee

For $\alpha_s(M_Z)=0.1185$,  $m_c=1.3\gev$ and $M_{Z^\prime}=3\tev$ we have
\begin{equation}\label{C5C6} 
\left[\begin{array}{c} C_5(m_c) \\
                         C_6(m_c)
    \end{array}\right]
 = \left[\begin{array}{cc} 0.86 & 0.19 \\
                          1.13 & 3.60
    \end{array}\right]  \left[\begin{array}{c} 1 \\
                         0
    \end{array}\right] \frac{\Delta_L^{s d}(Z^\prime)\Delta_R^{q q}(Z^\prime)}{4 M^2_{Z^\prime}}.
  \end{equation}
Consequently 
\be\label{LOC5C6}
 C_5(m_c)= 0.86  \frac{\Delta_L^{s d}(Z^\prime)\Delta_R^{q q}(Z^\prime)}{4 M^2_{Z^\prime}}\,,\qquad   C_6(m_c)= 1.13\frac{\Delta_L^{s d}(Z^\prime)\Delta_R^{q q}(Z^\prime)}{4 M^2_{Z^\prime}}.
\ee
Due to the large element $(1,2)$ in the matrix (\ref{reduced}) and 
the large anomalous dimension of the $Q_6$ operator represented by the $(2,2)$ 
element of this matrix, $C_6(m_c)$ is by a factor of $1.3$ larger than 
$C_5(m_c)$ even if $C_6(M_{Z^\prime})$ vanishes at LO. Moreover the matrix element $\langle Q_5\rangle_0$ is strongly colour suppressed \cite{Buras:2015xba}
which is not the case of  $\langle Q_6\rangle_0$ and within a good 
approximation we can neglect the contribution of $Q_5$. In the case of 
$(Q_5^\prime,Q_6^\prime)$ the formulae remain unchanged except that the 
value of $C_5^\prime(M_{Z^\prime})$ differs from $C_5(M_{Z^\prime})$.

In the case of $G^\prime$ the initial conditions for the Wilson coefficients $C_5$ and $C_6$ at $\mu=M_{G^\prime}$ are modified and given in (\ref{C5A})  and 
(\ref{C6A}). One finds then
\begin{equation}\label{C5C6c} 
\left[\begin{array}{c} C_5(m_c) \\
                         C_6(m_c)
    \end{array}\right]
 = \left[\begin{array}{cc} 0.86 & 0.19\\
                          1.13 & 3.60
    \end{array}\right]  \left[\begin{array}{c} -1/6 \\
                         1/2
    \end{array}\right] \frac{\Delta_L^{s d}(G^\prime)\Delta_R^{q q}(G^\prime)}{4 M^2_{G^\prime}}.
  \end{equation}
Consequently instead of (\ref{LOC5C6}) one has
\be
 C_5(m_c)= -0.05  \frac{\Delta_L^{s d}(G^\prime)\Delta_R^{q q}(G^\prime)}{4 M^2_{G^\prime}}\,,\qquad   C_6(m_c)= 1.61\frac{\Delta_L^{s 
d}(G^\prime)\Delta_R^{q q}(G^\prime)}{4 M^2_{G^\prime}}
\ee
so that now $Q_6$ operator is even more dominant over $Q_5$ than in the $Z^\prime$ 
scenario.

\subsection{Electroweak Penguins}
The basic equation for the RG evolution can also be used for $Z$ models 
except that
\begin{equation}\label{cmub1Z}
 \vec C(m_c)=\hat U(m_c,M_{Z}) \vec C(M_{Z})
\end{equation}
where
\begin{equation}\label{cmub1aZ}
 \hat U(m_c,M_{Z})=\hat U^{(f=4)}(m_c,m_b)\hat U^{(f=5)}(m_b, M_Z) 
\end{equation}
and  the relevant 
one-loop anomalous dimension matrix in the $(Q_7,Q_8)$ basis is very similar 
to the one in  (\ref{reduced})
\begin{equation}
\hat \gamma^{(0)}_s = 
\left(
\begin{array}{cc}
2 & -6  \\ \svs
0 & -16
   \end{array}
\right).
\label{reducedZ}
\end{equation}

Performing the renormalization group evolution from $M_Z$ to $m_c=1.3\gev$ 
we find \cite{Buras:2014sba}
\be\label{LOC7C8}
 C_7(m_c)= 0.87\,C_7(M_Z)\qquad   C_8(m_c)= 0.76\,C_7(M_Z).
\ee
Due to the large element $(1,2)$ in the matrix (\ref{reducedZ}) and 
the large anomalous dimension of the $Q_8$ operator represented by the $(2,2)$ 
element in (\ref{reducedZ}), the two coefficients are comparable in size.
But  the matrix element $\langle Q_7\rangle_{2}$ is colour suppressed 
which is not the case of  $\langle Q_8\rangle_{2}$ and within a good 
approximation we can neglect the contributions of $Q_7$. 
In the case of 
$(Q_7^\prime,Q_8^\prime)$ the formulae remain unchanged except that the 
value of $C_7^\prime(M_{Z})$ differs from $C_7(M_{Z})$.

If a $Z^\prime$ model has such flavour diagonal couplings that at the end 
 only the operators $(Q_7,Q_8)$ or  $(Q^\prime_7,Q^\prime_8)$  have to be considered, 
additional 
evolution from $M_Z$ to $M_{Z^\prime}$ has to be performed as in (\ref{cmub1a}) 
but the anomalous dimension matrix is as given in (\ref{reducedZ}). One 
finds then for  $\alpha_s(M_Z)=0.1185$,  $m_c=1.3\gev$ and $M_{Z^\prime}=3\tev$ 
\cite{Buras:2014yna}
\be
C_8(m_c)=1.35\, C_7(M_{Z^\prime})
\ee
with $1.35$ being RG factor. The longer RG evolution than in the case of $Z$
made this factor larger.

\subsection{Beyond the LHC Scales}
In the case of $Z^\prime$ we define for arbitrary $M_{Z^\prime}$ the factors $r_{65}$ and $r_{87}$ by
\be
C_6(m_c)=r_{65}\, C_5(M_{Z^\prime}), \qquad
C_8(m_c)=r_{87}\, C_7(M_{Z^\prime}).
\ee
In the case of $G^\prime$  we define the corresponding factor through
\be
 C_6(m_c)= r_{G^\prime}\frac{\Delta_L^{s 
d}(G^\prime)\Delta_R^{q q}(G^\prime)}{4 M^2_{G^\prime}}\,.
\ee
All these factors increase with increasing  $M_{Z^\prime}$. We show this dependence in Table~\ref{RG}\footnote{We thank Christoph Bobeth for checking this table.}.

\begin{table}[!htb]
\begin{center}
\begin{tabular}{|c|c|c|c|c|c|c|}
\hline
  $M_{Z^\prime}$ & $3\tev$ & $6\tev$& $10\tev$& $20\tev$ & $50\tev$ & $100\tev$\\
\hline
 $r_{65}$  & $1.13$ & $1.22$ & $1.28$ & $1.37$ &  $1.48$ & $1.56$ \\
 $r_{87}$  & $1.35$ & $1.48$ & $1.56$ & $1.69$ &  $1.85$ & $1.97$ \\
$r_{G^\prime}$  & $1.61$ & $1.70$ & $1.77$ & $1.85$ &  $1.96$ & $2.05$ \\
\hline
\end{tabular}
\end{center}
\caption{The $M_{Z^\prime}(M_{G^\prime})$ dependence of the RG factors $r_{65}$, $r_{87}$ and $r_{G^\prime}$ at 
LO with two-loop running of $\alpha_s$.
\label{RG}}
\end{table}

\boldmath
\section{$\varepsilon_K$  and $\Delta M_K$}\label{DF2}
\unboldmath
\subsection{General Formulae}
For the CP-violating parameter $\varepsilon_K$  and $\Delta M_K$ we have
respectively
\be
\varepsilon_K=\frac{\tilde\kappa_\eps e^{i\varphi_\eps}}{\sqrt{2}(\Delta M_K)_\text{exp}}\left[{\rm Im}\left(M_{12}^K\right)\right]\equiv e^{i\varphi_\eps}\, 
\left[\varepsilon_K^{\rm SM}+\varepsilon^{\rm NP}_K\right] \,,
\label{eq:3.35}
\ee
\be\label{eq:3.35a}
\Delta M_K=2{\rm Re}\left(M_{12}^K\right) = (\Delta M_K)^{\rm SM}+(\Delta M_K)^{\rm NP}
\ee
where
$\varphi_\eps = (43.51\pm0.05)^\circ$ and $\tilde\kappa_\eps=0.94\pm0.02$ \cite{Buras:2008nn,Buras:2010pza} takes into account
that $\varphi_\eps\ne \tfrac{\pi}{4}$ and includes long distance  effects in 
${\rm Im}(\Gamma_{12})$ and ${\rm Im}(M_{12})$. We have separated the overall 
phase factor so that $\varepsilon_K^{\rm SM}$ and $\varepsilon^{\rm NP}_K$ are real quantities with $\varepsilon^{\rm NP}_K$ representing NP contributions.

Generally we can write
\be
M_{12}^K= [M_{12}^K]_{\rm SM}+[M_{12}^K]_{\rm NP}\,,
\ee
where the first term is the SM contribution for which the explicit expression 
can be found e.g. in  \cite{Buras:2013ooa}. We decompose the NP part as follows
\be
[M_{12}^K]_{\rm NP}= [M_{12}^K]_{\rm VLL}+[M_{12}^K]_{\rm VRR}+[M_{12}^K]_{\rm LR}.
\ee
The first two contributions come from the operators
\be
{Q}_1^\text{VLL}=\left(\bar s\gamma_\mu P_L d\right)\left(\bar s\gamma^\mu P_L d\right)\,, \qquad
{Q}_1^\text{VRR}=\left(\bar s\gamma_\mu P_R d\right)\left(\bar s\gamma^\mu P_R d\right)\,
\ee
and the last one from 
\be
{Q}_1^\text{LR}=\left(\bar s\gamma_\mu P_L d\right)\left(\bar s\gamma^\mu P_R d\right)\,,\qquad
{Q}_2^\text{LR}=\left(\bar s P_L d\right)\left(\bar s P_R d\right)\,.
\ee

\boldmath
\subsection{$Z$ and $Z^\prime$ Cases}
\unboldmath
Using formulae in \cite{Buras:2012jb} we find then in the case of tree-level 
$Z$ contribution
\be\label{VLL}
 [M_{12}^K]^*_{\rm VLL}= \frac{1}{6}F_{K}^2\hat B_{K}m_{K} \eta_2 \tilde{r}
\left[\frac{\Delta_L^{sd}(Z)}{M_{Z}}\right]^2\,
\ee
where 
\be
\eta_2=0.576, \qquad \tilde r \approx 1.068, \qquad \hat B_K\approx 0.75\,.
\ee
 For VRR one should just replace L by R. We emphasize the complex conjugation in this formula.

For the LR contribution we simply have
\be
[M_{12}^K]^*_{\rm LR}=\frac{\Delta_L^{sd}(Z)\Delta_R^{sd}(Z)}{
 M_{Z}^2} \langle \hat Q_1^\text{LR}(M_{Z})\rangle^{sd}
\ee
where using the technology of \cite{Buras:2001ra,Buras:2012fs} we have  expressed the amplitude  in terms of the renormalisation scheme independent matrix 
element
\be
\langle \hat Q_1^\text{LR}(M_{Z})\rangle^{sd} =\langle  Q_1^\text{LR}(M_{Z})\rangle^{sd}\left(1-\frac{1}{6}\frac{\alpha_s(M_{Z})}{4\pi}\right) -\frac{\alpha_s(M_{Z})}{4\pi}\langle Q_2^\text{LR}(M_{Z})\rangle^{sd}\,.\label{Q1LR}
\ee

On the basis of \cite{Boyle:2012qb,Bertone:2012cu,Jang:2015sla} one finds for $M_Z$ and $M_{Z^\prime}=3\tev$ 
\be\label{LRM}
\langle \hat Q_1^\text{LR}(M_{Z})\rangle^{sd}\approx -0.09\,\gev^3\,, \qquad
\langle \hat Q_1^\text{LR}(M_{Z^\prime})\rangle^{sd}\approx -0.16\,\gev^3\,.
\ee
This matrix element increases with increasing $M_{Z^\prime}$. See Table~5 in \cite{Buras:2014zga}.

For $\varepsilon_K$ and $\Delta M_K$, inserting relevant contributions to $M_{12}$ into (\ref{eq:3.35}) and (\ref{eq:3.35a}),
 we get then in the case of $Z$ 
\be\label{c5}
\varepsilon^{\rm NP}_K= -4.26\cdot 10^7\left[{\rm Im} \Delta_{L}^{s d}(Z){\rm Re} \Delta_{L}^{s d}(Z)+{\rm Im} \Delta_{R}^{s d}(Z){\rm Re} \Delta_{R}^{s d}(Z)\right] +(\varepsilon_K)^Z_{\rm LR}
\ee
with
\be\label{c5a}
(\varepsilon_K)^Z_{\rm LR}= 2.07\cdot 10^9 \left[{\rm Im} \Delta_{L}^{s d}(Z){\rm Re} \Delta_{R}^{s d}(Z)+{\rm Im} \Delta_{R}^{s d}(Z){\rm Re} \Delta_{L}^{s d}(Z)\right]
\ee
and 
\be\label{c6}
\frac{(\Delta M_K)^{\rm NP}}{(\Delta M_K)_\text{exp}}=
6.43\cdot 10^7\sum_{P=L,R}\left[({\rm Re} \Delta_{P}^{s d}(Z))^2 -({\rm Im} \Delta_{P}^{s d}(Z))^2\right] + \frac{(\Delta M_K)^Z_{\rm LR}}{(\Delta M_K)_\text{exp}}
\ee
with 
\be\label{c6a}
\frac{(\Delta M_K)^Z_{\rm LR}}{(\Delta M_K)_\text{exp}}
= -6.21\cdot 10^9
\left[{\rm Re} \Delta_{L}^{s d}(Z){\rm Re} \Delta_{R}^{s d}(Z)-{\rm Im} \Delta_{L}^{s d}(Z){\rm Im} \Delta_{R}^{s d}(Z)\right].
\ee
The fact that the LR contributions in these expressions have opposite sign 
to the ones from VLL and VRR operators is related to the opposite signs 
in the relevant hadronic matrix elements. 

For the $Z^\prime$ tree-level exchanges, $M_{Z}$ should be replaced by $M_{Z^\prime}$, in VLL and VRR contributions $\tilde r=0.95$ should used and in LR contribution 
the value of the matrix element $\langle \hat Q_1^\text{LR}\rangle$ in (\ref{LRM}). See Section~\ref{sec:3} for explicit formulae.

\boldmath
\section{$\kpn$  and $\klpn$}\label{KPNN}
\unboldmath
\subsection{General Formulae}
The branching ratios for $\kpn$  and $\klpn$ in any extension of the SM in 
which light neutrinos couple only to left-handed currents  are given as 
follows
\begin{align}
\mathcal{B}(K^+\to\pi^+\nu\bar\nu)&=\kappa_+ \cdot
\left[\left(\frac{{\rm Im}\,X_{\rm eff}}{\lambda^5}\right)^2+
\left(\frac{{\rm Re}\,\lambda_c}{\lambda}P_c(X)+
\frac{{\rm Re}\,X_{\rm eff}}{\lambda^5}\right)^2\right]\,,\label{bkpnn}\\
\mathcal{B}(\klpn)&=\kappa_L\cdot
\left(\frac{{\rm Im}\,X_{\rm eff}}{\lambda^5}\right)^2\,,\label{bklpn}
\end{align}
where $\lambda=\vus$ and \cite{Mescia:2007kn}
\be
\kappa_+ = (5.173\pm 0.025 )\cdot 10^{-11}\left[\frac{\lambda}{0.225}\right]^8,\quad  
\kappa_L = (2.231\pm 0.013)\cdot 10^{-10}\left[\frac{\lambda}{0.225}\right]^8\,.
 \label{kapl}
\ee
For the charm contribution, represented by $P_c(X)$, the calculations in 
\cite{Buras:2005gr,Buras:2006gb,Brod:2008ss,Isidori:2005xm,Mescia:2007kn}
imply \cite{Buras:2015qea}
\be\label{PCFINAL}
P_c(X)= 0.404\pm 0.024,
\ee
where the error is dominated by the  long distance uncertainty estimated in \cite{Isidori:2005xm}. Next
\be\label{XK}
X_{\rm eff} = V_{ts}^* V_{td}\, \left[X_{\rm L} + X_{\rm R}\right]\,,
\ee
where the functions $X_{\rm L}$ and $X_{\rm R}$ summarise the contributions from 
left-handed and right-handed quark currents, respectively. $\lambda_i=V^*_{is}V_{id}$ are the CKM factors. In what follows we will set these factors to 
\be\label{CKM}
{\rm Re}\lambda_t=-3.0\cdot 10^{-4}, \qquad {\rm Im}\lambda_t=1.4\cdot 10^{-4}
\ee
which are in the ballpark of present best estimates  \cite{Bona:2006ah,Charles:2015gya}.
The Grossman-Nir (GN) bound on $\mathcal{B}(\klpn)$ reads \cite{Grossman:1997sk}
\be\label{GN}
\mathcal{B}(\klpn)\le \frac{\kappa_L}{\kappa_+}\,\mathcal{B}(\kpn)=4.31\, 
\mathcal{B}(\kpn),
\ee
where we have shown only the central value as it is never reached in the models 
considered by us. See Figs.~\ref{R4ab} and \ref{R5}.

 Experimentally we have \cite{Artamonov:2008qb}
\be\label{EXP1}
\mathcal{B}(\kpn)_\text{exp}=(17.3^{+11.5}_{-10.5})\cdot 10^{-11}\,,
\ee
and the $90\%$ C.L. upper bound \cite{Ahn:2009gb}
\be\label{EXP2}
\mathcal{B}(\klpn)_\text{exp}\le 2.6\cdot 10^{-8}\,.
\ee

\boldmath
\subsection{$Z$ and $Z^\prime$ Cases}
\unboldmath
In what follows we will give the expressions for $X_{\rm eff}$ in $Z$ and 
$Z^\prime$ models which inserted into (\ref{bkpnn}) and (\ref{bklpn}) give
the branching ratios for $\kpn$ and $\klpn$. It should be noted 
that the particular values of the CKM factors in (\ref{CKM}) enter only in 
the SM contributions and in their interferences with NP contributions.

In the case 
of tree-level $Z$ exchanges we have \cite{Buras:2012jb}
\be\label{XLK}
X_{\rm L}=X_L^{\rm SM}+\frac{\Delta_L^{\nu\bar\nu}(Z)}{g^2_{\rm SM}M_{Z}^2}
                                       \frac{\Delta_L^{sd}(Z)}{V_{ts}^* V_{td}},
\qquad
X_{\rm R}=\frac{\Delta_L^{\nu\bar\nu}(Z)}{g^2_{\rm SM}M_{Z}^2}
                                       \frac{\Delta_R^{sd}(Z)}{V_{ts}^* V_{td}}
\ee
where
\be\label{gsm}
g_{\text{SM}}^2=
4 \frac{M_W^2 G_F^2}{2 \pi^2} = 1.78137\times 10^{-7} \gev^{-2}\,.
\ee

In the SM only 
$X_{L}$ is non-vanishing and is given by \cite{Buchalla:1993bv,Misiak:1999yg,Buchalla:1998ba,Brod:2010hi}
\be\label{XSM}
X_L^{\rm SM}= 1.481 \pm 0.009
\ee
as extracted  in \cite{Buras:2015qea} from original papers.
With the known coupling $\Delta_L^{\nu\bar\nu}(Z)=0.372$ and the CKM factors 
in (\ref{CKM}) we have then
\begin{align}
{\rm Re}\,X_{\rm eff}(Z)&= -4.44\cdot 10^{-4}+2.51\cdot 10^2 [{\rm Re} \Delta_{L}^{s d}(Z)+ {\rm Re} \Delta_{R}^{s d}(Z)]\,,\label{c2}\\
{\rm Im}\,X_{\rm eff}(Z)& =2.07\cdot 10^{-4}+2.51\cdot 10^2 [{\rm Im} \Delta_{L}^{s d}(Z)+ {\rm Im} \Delta_{R}^{s d}(Z)]\,,\label{c3}
\end{align}
where the first terms on the r.h.s are SM contributions for CKM factors in 
(\ref{CKM}). Note that in $\klpn$ the enhancement of its branching 
ratio requires the sum of the imaginary parts of the couplings to be {\it positive}.  This enhances also $\kpn$ but could be compensated by the 
decrease of ${\rm Re}\,X_{\rm eff}$ unless the sum of the corresponding real parts is {\it negative}. 

In the case of tree-level $Z^\prime$ exchanges one should just replace everywhere
the index $Z$ by $Z^\prime$, in particular $M_Z$ by $M_{Z'}$, 
 and use $\Delta_L^{\nu\bar\nu}(Z^\prime)$.

The numerical factors in the NP parts in (\ref{c2}) and (\ref{c3}) above should then be multiplied by 
\be\label{RR}
R=\left[\frac{M_Z}{M_{Z^\prime}}\right]^2\frac{\Delta_L^{\nu\bar\nu}(Z^\prime)}{0.372}
=2.48\times 10^{-3}\left[\frac{3\tev}{M_{Z^\prime}}\right]^2\Delta_L^{\nu\bar\nu}(Z^\prime)\,.
\ee
Thus we get

\be\label{c7}
{\rm Re}\,X_{\rm eff}(Z^\prime)= -4.44\cdot 10^{-4}+0.62\,\left[\frac{3\tev}{M_{Z^\prime}}\right]^2\, [{\rm Re} \Delta_{L}^{s d}(Z^\prime)+ {\rm Re} \Delta_{R}^{s d}(Z^\prime)]  \Delta_L^{\nu\bar\nu}(Z^\prime)  \,,
\ee
\be\label{c8}
{\rm Im}\,X_{\rm eff}(Z^\prime)=2.07\cdot 10^{-4}+0.62\,\left[\frac{3\tev}{M_{Z^\prime}}\right]^2\, [{\rm Im} \Delta_{L}^{s d}(Z^\prime)+ {\rm Im} \Delta_{R}^{s d}(Z^\prime)]\Delta_L^{\nu\bar\nu}(Z^\prime)\,.
\ee

\boldmath
\section{$K_L\to \mu^+\mu^-$}\label{sec:KLmm}
\unboldmath
\subsection{General Formulae}
Only the so-called short distance (SD)
part of a dispersive contribution
to $K_L\to\mu^+\mu^-$ can be reliably calculated. It is given 
generally as follows {($\lambda=0.2252$)}
\be
\mathcal{B}(K_L\to\mu^+\mu^-)_{\rm SD} = 2.01\cdot 10^{-9} 
\left( \frac{{\rm Re}\,Y_{\rm eff}}{\lambda^5} + \frac{{\rm Re}\,\lambda_c}{\lambda} P_c(Y)  \right)^2\,,
\ee
where at NNLO \cite{Gorbahn:2006bm}
\begin{align}
P_c(Y) &= 0.115\pm 0.017.
\end{align}
The short distance contributions are described by
\be\label{YK}
Y_{\rm eff} = V_{ts}^* V_{td} \,\left[Y_{L}(K) - Y_{R}(K)\right],
\ee
where the functions $Y_{\rm L}$ and $Y_{\rm R}$ summarise the contributions from 
left-handed and right-handed quark currents, respectively.
Notice the minus sign in front of $Y_R$, as opposed to $X_R$ in (\ref{XK}), that results from the fact that only the axial-vector current contributes. 
This difference allows to be sensitive to right-handed couplings, which 
is not possible in the case of $K\to\pi\nu\bar\nu$ decays.

The extraction of the short distance
part from the data is subject to considerable uncertainties. The most recent
estimate gives \cite{Isidori:2003ts}
\be\label{eq:KLmm-bound}
\mathcal{B}(K_L\to\mu^+\mu^-)_{\rm SD} \le 2.5 \cdot 10^{-9}\,,
\ee
to be compared with $(0.8\pm0.1)\cdot 10^{-9}$ in the SM. With our choice of 
CKM parameters we find $0.72\cdot 10^{-9}$. It is important to improve this 
estimate as this would further increase the role of this decay in bounding 
NP contributions not only in $Z$ scenarios.
\boldmath
\subsection{$Z$ and $Z^\prime$ Cases}
\unboldmath
In the case 
of tree-level $Z$ exchanges we have \cite{Buras:2012jb}
\begin{align}
Y_L(K)&=Y_L^{\rm SM}(K)+\frac{\Delta_A^{\mu\bar\mu}(Z)}{g^2_{\rm SM}M_{Z}^2}
                                       \frac{\Delta_L^{sd}(Z)}{V_{ts}^* V_{td}}, & 
Y_R(K)&=\frac{\Delta_A^{\mu\bar\mu}(Z)}{g^2_{\rm SM}M_{Z}^2}
                                       \frac{\Delta_R^{sd}(Z)}{V_{ts}^* V_{td}},
\end{align} 
where \cite{Bobeth:2013tba}
\be
Y_L^{\rm SM}(K) = 0.942\,.
\ee
With the known coupling $\Delta_A^{\mu\bar\mu}(Z)=0.372$ and the CKM factors 
in (\ref{CKM}) we have then 
\be\label{c4}
{\rm Re}\,Y_{\rm eff}(Z)= -2.83\cdot 10^{-4}+2.51\cdot 10^2 [{\rm Re} \Delta_{L}^{s d}(Z)- {\rm Re} \Delta_{R}^{s d}(Z)]\,.
\ee

In the case of tree-level $Z^\prime$ exchanges one should just replace everywhere the index $Z$ by $Z^\prime$, in particular $M_Z$ by $M_{Z'}$, 
 and use $\Delta_A^{\mu\bar\mu}(Z^\prime)$. As $\Delta_A^{\mu\bar\mu}(Z)=\Delta_L^{\nu\bar\nu}(Z)$ also the same  numerical 
factor in (\ref{RR}) should multiply NP part in (\ref{c4}). Thus we have
\be\label{c9}
{\rm Re}\,Y_{\rm eff}(Z^\prime)= -2.83\cdot 10^{-4}+0.62\,\left[\frac{3\tev}{M_{Z^\prime}}\right]^2\, [{\rm Re} \Delta_{L}^{s d}(Z^\prime)- {\rm Re} \Delta_{R}^{s d}(Z^\prime)]  \Delta_A^{\mu\bar\mu}(Z^\prime) \,.
\ee

\boldmath
\section{$K_L\to\pi^0 \ell^+\ell^-$}\label{sec:KLpmm}
\unboldmath
The rare decays $K_L\to\pi^0e^+e^-$ and $K_L\to\pi^0\mu^+\mu^-$ are dominated
by CP-violating contributions. The indirect CP-violating
contributions are determined by the measured decays 
$K_S\to\pi^0 \ell^+\ell^-$ and the parameter $\varepsilon_K$ in 
a model independent manner. It is the dominant contribution within the SM 
with both branching being $\ord(10^{-11})$ \cite{Mescia:2006jd} 
and by one order of magnitude smaller than the present experimental bounds
\be
\mathcal{B}(K_L\to\pi^0e^+e^-)_\text{exp} <28\cdot10^{-11}\quad\text{\cite{AlaviHarati:2003mr}}\,,\qquad
\mathcal{B}(K_L\to\pi^0\mu^+\mu^-)_\text{exp} <38\cdot10^{-11}\quad\text{\cite{AlaviHarati:2000hs}}\,,
\ee
leaving thereby large room for NP contributions. In the models analyzed 
by us these bounds 
have no impact on $\kpn$ and $\klpn$ decays but the present data on 
$\kpn$ do not allow to reach the above bounds in the $Z'(Z)$ scenarios considered.

To our knowledge, there are no definite plans to measure these decays in 
the near future and we will not analyze them here. They are similar to $B\to K \ell^+\ell^-$ decays except 
that the dipole operator contributions turn out to be small in the SM and in 
many NP scenarios.  NP contributions 
 shift the values of the coefficients $C_{7V}$ and $C_{7A}$ which 
are sensitive to  $\Delta_V^{\mu\bar\mu}(Z^\prime)$ and  $\Delta_A^{\mu\bar\mu}(Z^\prime)$, respectively. Similar for $Z$. In the presence of right-handed 
flavour violating currents also $C^\prime_{7V}$ and $C^\prime_{7A}$ are generated.
This is the case of RS scenario with custodial protection  \cite{Blanke:2008yr}.
There are also recent 
efforts to improve SM prediction by means of lattice QCD \cite{Christ:2015aha}.
The importance of testing NP scenarios, in particular those involving right-handed currents, by means of these decays has been stressed 
in \cite{Mescia:2006jd}. Moreover, the measurement of both decays could disentangle the scalar/pseudoscalar from vector/axialvector contributions.
But from present perspective such tests will eventually become realistic only in the next decade.  
References to reach literature can be found in  \cite{Mescia:2006jd} and 
the analysis of these decays within general $Z$ and $Z^\prime$ models can be found in \cite{Buras:2012jb}. As seen in Figs. 12 and 33 of that paper there is a 
strong correlation between these decays and $\klpn$ in $Z$ and $Z^\prime$ 
scenarios so that the increase of 
$\mathcal{B}(\klpn)$ increases also the branching ratios for $K_L\to\pi^0 \ell^+\ell^-$. But the presence of the indirect CP-violating contributions in the latter decays, that are negligible in $\klpn$, shadows NP effects in them. Only 
when $\mathcal{B}(\klpn)$ is enhanced by an order of magnitude sizable 
enhancements of  $\mathcal{B}(K_L\to\pi^0 \ell^+\ell^-)$ are possible. Similar 
correlations are found in the LHT model \cite{Blanke:2006eb} and RSc \cite{Blanke:2008yr}.

In $Z$ scenarios due to the smallness of  $\Delta_V^{\mu\bar\mu}(Z)$ NP 
enters these decays predominantly through $C_{7A}$ and $C_{7A}^\prime$. More 
interesting is the NP pattern in $Z^\prime$ scenarios due to the $SU(2)_L$ 
relation
\be\label{SU2}
\Delta_{L}^{\nu\bar\nu}(Z')=\frac{\Delta_V^{\mu\bar\mu}(Z')-\Delta_A^{\mu\bar\mu}(Z')}{2}. 
\ee
This relation implies correlations between $Z^\prime$ contributions to 
$\kpn$, $\klpn$, $K_L\to \mu^+\mu^-$ 
and $K_L\to\pi^0\ell^+\ell^-$ analogous to the ones between 
$B\to K(K^*)\nu\bar\nu$, $B_d\to K(K^*)\mu^+\mu^-$ and $B_{s}\to\mu^+\mu^-$ that 
have been analyzed in detail in \cite{Buras:2014fpa}. In order for such 
relations to become vital in the $K$-meson system theoretical uncertainties 
in  $K_L\to \mu^+\mu^-$ and $K_L\to\pi^0\ell^+\ell^-$ have to be decreased 
by much. For the most recent analysis of $K_{L,S}\to\pi^0\ell^+\ell^-$ and 
$K^+\to\pi^+\ell^+\ell^-$ decays including correlations with LHCb anomalies 
see \cite{Crivellin:2016vjc}.

\renewcommand{\refname}{R\lowercase{eferences}}

\addcontentsline{toc}{section}{References}

\bibliographystyle{JHEP}
\bibliography{AJBeprefs}
\end{document}